\documentclass[
    sigplan,
    natbib=false 
]{acmart}

\usepackage[nolist]{acronym}
\usepackage[inline]{enumitem}
\usepackage{csquotes}
\usepackage[capitalise]{cleveref}
\Crefname{figure}{Fig.}{Figs.}
\crefname{subsection}{\S}{\S\S}
\crefname{part}{\S}{\S\S}
\crefname{chapter}{\S}{\S\S}
\crefname{section}{\S}{\S\S}
\usepackage{booktabs}
\usepackage{multirow}
\usepackage{subcaption}
\usepackage[most]{tcolorbox}
\usepackage{tikz}
\usetikzlibrary{patterns, arrows.meta, backgrounds, positioning, calc, backgrounds, fit, decorations.pathreplacing, pgfplots.groupplots}
\definecolor{vworange}{RGB}{230,120,40}
\definecolor{vwblue}{RGB}{40,100,190}
\definecolor{gcfill}   {RGB}{232,231,224}
\definecolor{gcedge}   {RGB}{110,108,100}
\definecolor{dagfill}  {RGB}{208,235,222}
\definecolor{dagedge}  {RGB}{13, 100, 76}
\definecolor{boundfill}{RGB}{160, 100,  10}
\definecolor{vfill}    {RGB}{245,196,179}
\definecolor{vedge}    {RGB}{153, 60, 29}
\definecolor{errred}   {RGB}{185,  40, 40}
\definecolor{amberfill}{RGB}{250,199,117}
\definecolor{amberedge}{RGB}{160,100, 10}
\definecolor{buffedge} {RGB}{130,128,120}
\definecolor{ColorOrange}{RGB}{230, 115, 0}   
\definecolor{ColorBlue}{RGB}{0, 100, 200}      
\definecolor{frameworkbg}{RGB}{223,231,241}
\definecolor{frameworklabel}{RGB}{110,152,206}
\definecolor{abg}{RGB}{210,220,188}
\definecolor{ablabel}{RGB}{112,182,95}

\usepackage{pgfplots}
\usepgfplotslibrary{statistics}
\usepackage{adjustbox}
\usepackage{xcolor}
\usepackage{todonotes}
\presetkeys{todonotes}{
  inline,
  color=yellow!20
}{}

\newcommand{\algoname}{\texorpdfstring{\textsc{Pangolin}}{Pangolin}}

\definecolor{darkgreen}{RGB}{0,128,0}

\usepackage[ruled]{algorithm}
\definecolor{commentgray}{RGB}{110,110,110}

\usepackage[
  noEnd,
  commentColor=commentgray,
  beginComment={$\triangleright$~},
  beginLComment={$\triangleright$~},
  endLComment=
]{algpseudocodex}
\algdef{SL}[REPLICASTATE]{ReplicaState}{0}[1]{\textbf{state} #1}
\algrenewcommand\algorithmicoutput{\textbf{abort}}
\algrenewcommand\algorithmicprocedure{\textbf{upon}}
\newcommand{\var}[1]{\mathit{#1}}
\newcommand{\fn}[2]{\text{#1}(#2)}
\newcommand{\DAG}[2]{DAG[#1][#2]}

\newcommand{\invisible}[1]{}

\setcopyright{none}

\RequirePackage[
  datamodel=acmdatamodel,
  style=acmnumeric,
]{biblatex}
\newcommand{\para}[1]{\smallskip\noindent\textbf{#1}}

\begin{document}

\begin{acronym}
\acro{SMR}{state-machine replication}
\acro{BAB}{Byzantine atomic broadcast}
\acro{BFT}{Byzantine fault-tolerant}
\acro{DAG}{directed acyclic graph}
\acro{TEE}{trusted execution environment}
\acro{GST}{global stabilization time}
\acro{PBFT}{Practical Byzantine Fault Tolerance}
\acro{GC}{garbage collection}
\acro{AB}{atomic broadcast}
\acro{RB}{reliable broadcast}
\end{acronym}

\title{\algoname: Efficient DAG-based BFT State-Machine Replication with TEE-Accelerated Commitment}

\author{Paul Franke-Bergmann}
\email{paul.bergmann@fau.de}
\affiliation{
    \institution{Friedrich-Alexander-Universität Erlangen-Nürnberg}
    \city{Erlangen}
    \country{Germany}
}

\author{Maya Kuhlmann}
\email{maya.kuhlmann@student.kit.edu}
\affiliation{
    \institution{Karlsruhe Institute of Technology (KIT)}
    \city{Karlsruhe}
    \country{Germany}
}

\author{Alexander Heß}
\email{alexander.hess@uni-ulm.de}
\affiliation{
    \institution{Ulm University}
    \city{Ulm}
    \country{Germany}
}

\author{Marc Leinweber}
\email{marc.leinweber@kit.edu}
\affiliation{
    \institution{Karlsruhe Institute of Technology (KIT)}
    \city{Karlsruhe}
    \country{Germany}
}

\author{Rüdiger Kapitza}
\email{ruediger.kapitza@fau.de}
\affiliation{
    \institution{Friedrich-Alexander-Universität Erlangen-Nürnberg}
    \city{Erlangen}
    \country{Germany}
}

\author{Hannes Hartenstein}
\email{hannes.hartenstein@kit.edu}
\affiliation{
    \institution{Karlsruhe Institute of Technology (KIT)}
    \city{Karlsruhe}
    \country{Germany}
}

\author{Franz J. Hauck}
\email{franz.hauck@uni-ulm.de}
\affiliation{
    \institution{Ulm University}
    \city{Ulm}
    \country{Germany}
}

\author{Christian Berger}
\email{christian.g.berger@fau.de}
\affiliation{
    \institution{Friedrich-Alexander-Universität Erlangen-Nürnberg}
    \city{Erlangen}
    \country{Germany}
}

\begin{abstract}
Currently many \ac{BFT} protocols use \acp{DAG} to let replicas propose blocks of requests in parallel for improved throughput.
Yet, the existing DAG-based designs mainly consider the atomic-broadcast layer: They typically do not describe an efficient client interaction layer, their garbage collection is either unspecified or relies on trusted external storage, and the lowest-latency designs rely on timing assumptions that may be violated in geo-distributed deployments.

In this paper, we propose the \algoname{} protocol: It introduces a novel low-latency \acp{DAG}-based atomic broadcast layer and completes it into an efficient \ac{SMR} framework. To preserve system throughput, clients issue requests to a single replica and verify replies efficiently via compact execution proofs.
Furthermore, \algoname{} introduces  
a new garbage-collection scheme for \acp{DAG} which bounds memory usage even under full asynchrony. 
\algoname{} leverages \aclp{TEE} to tolerate $f$ Byzantine faults with only $n\geq{}2f+1$ replicas and accelerates vertex commitments significantly: 
Broadcasting certified vertices requires only a single communication step, and commit rules are refined to commit both frequently and early without revealing the coin-elected anchor.
This way, \algoname{} overtakes the fastest partially synchronous DAG protocols while being live under asynchrony: In our evaluation, \algoname{} achieves an average $1.27\times$ (up to $1.52\times$) latency speedup over Mysticeti. 
\end{abstract}


\settopmatter{printacmref=false}
\renewcommand\footnotetextcopyrightpermission[1]{}
\maketitle
\pagestyle{plain}

\acresetall

\noindent
\begin{minipage}[c]{0.78\linewidth}
    \begin{tcolorbox}[
        enhanced,
        colback=gray!10,
        colframe=gray!50,
        boxrule=0.2pt,
        arc=4mm,
        boxsep=2mm,
        overlay={
            \fill[gray!10]
                ([xshift=-0.5mm,yshift=3mm]frame.east) --
                ([xshift=3mm]frame.east) --
                ([xshift=-0.5mm,yshift=-3mm]frame.east) -- cycle;
            \draw[gray!50,line width=0.2pt]
                ([yshift=3mm]frame.east) --
                ([xshift=3mm]frame.east) --
                ([yshift=-3mm]frame.east);
        }
    ]
    You can check out a visualization of the \algoname{} protocol here:
    \textcolor{blue}{\url{https://pangolin-protocol.netlify.app}}!
    \end{tcolorbox}
\end{minipage}\hfill
\begin{minipage}[c]{0.18\linewidth}
    \centering
    \includegraphics[width=\linewidth]{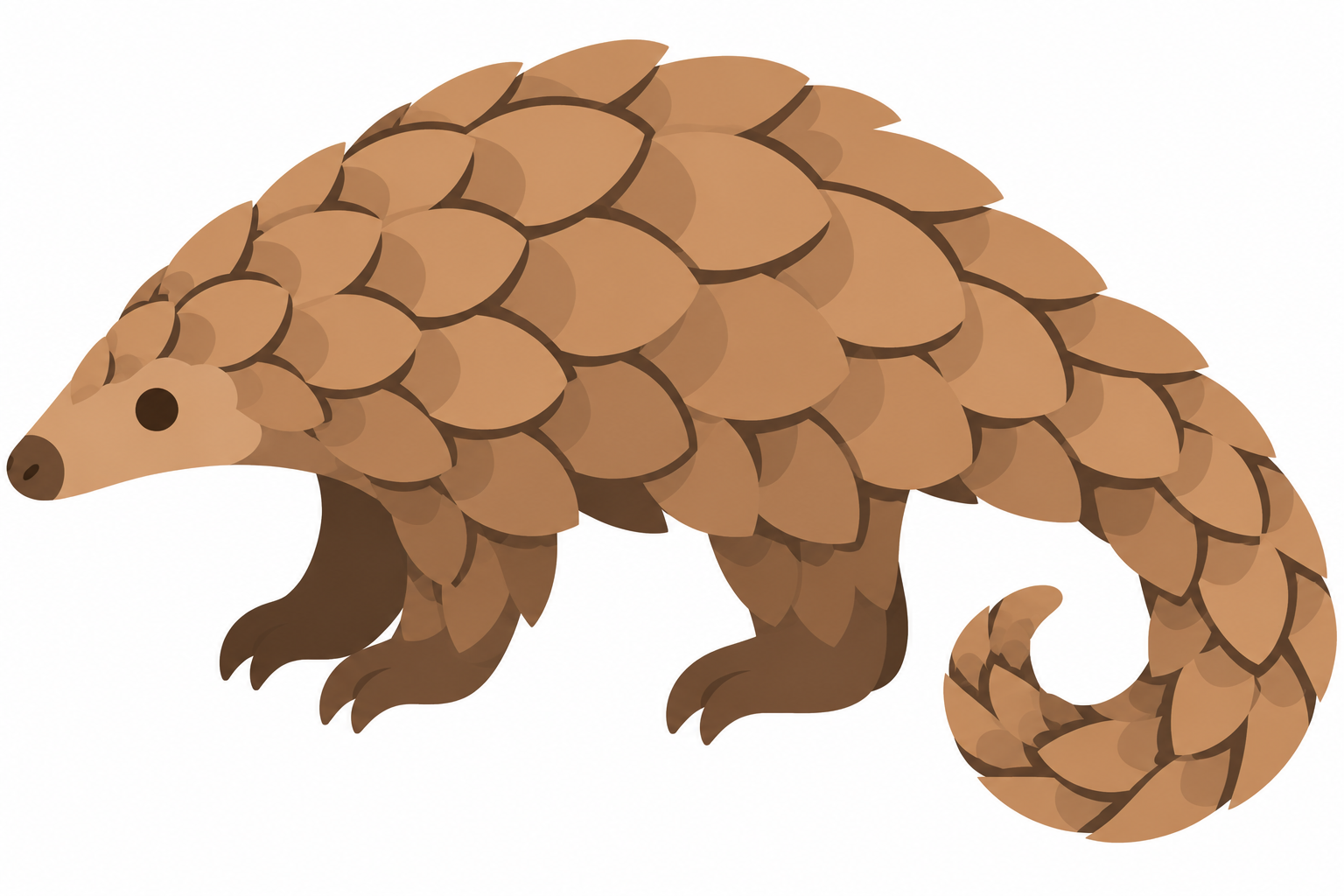}
\end{minipage}

\section{Introduction}
\Acf{BFT} \acf{SMR} provides clients with the illusion of a single, \textit{correct} service even if an adversary controls up to $f$ out of $n$ replicas~\cite{castro1999practical, lamport1982byzantine, schneider1990implementing}.
Moreover, BFT SMR is used in intrusion-tolerant systems such as firewalls~\cite{garcia2013intrusion, garcia2016sieveq} and critical infrastructure~\cite{babay2019deploying, bessani2008crutial, kirsch2013survivable}, and most prominently distributed ledgers~\cite{arun_2025_shoalpp, babel_2025_mysticeti, baudet2019state, buchman2018latest, gelashvili2022jolteon, gueta2019sbft, neiheiser2021kauri, hotstuff19}.
To support generic services, \ac{BFT} \ac{SMR} protocols must provide two fundamental guarantees~\cite{castro1999practical}:
\emph{safety}, which ensures that clients observe the replicated service as a centralized implementation that executes operations invoked by clients, and \emph{liveness}, which guarantees clients eventually receive valid responses even in the presence of faults.
Building correct \ac{BFT} \ac{SMR} systems is \textit{inherently difficult}:
agreement on a total order of requests, client-replica interaction and garbage collection must work correctly together~\cite{distler_byzantine_2021}.

\para{Solving scalability and the rise of DAG BFT protocols.}
Traditional BFT SMR relied on a dedicated leader and coupled request dissemination with request ordering~\cite{amir2010prime, bessani2014state, castro1999practical, clement2009upright,clement2009making, cowling2006hq, guerraoui2010next, kapitza2012cheapbft, wester2009tolerating}.
As the number of replicas $n$ grows, the leader's bandwidth is increasingly exhausted by forwarding request data, and becomes the dominating bottleneck for scalability~\cite{neiheiser2021kauri, stathakopoulou2022mirbft, danezis_narwhal_2022}. 
To overcome the leader bottleneck, a recent line of works proposes \ac{BFT} protocols based on \acp{DAG}~\cite{babel_2025_mysticeti, danezis_narwhal_2022, spiegelman2024shoal, spiegelman_bullshark_2022, jovanovic2025mahi, shrestha2025sailfish, polyanskii2025starfish}.
Rather than routing all requests through a single leader, these protocols achieve scalability by decoupling dissemination from ordering:
All replicas can propose blocks of client requests in parallel, referencing previously seen blocks to form a \ac{DAG} that encodes causal relationships~\cite{danezis_narwhal_2022, keidar_all_2021}.
The \ac{DAG} induces a partial order on requests.
Subsequently, the total order is derived by selecting shared anchor blocks (e.g., randomly, using a common coin) and then applying deterministic topological sorting with the anchor as starting point~\cite{keidar_all_2021}.
This parallelism allows \ac{DAG}-based \ac{BFT} protocols to achieve high throughput, and has driven their adoption in major blockchain platforms such as Aptos~\cite{aptos2025, spiegelman2024shoal} and Sui~\cite{babel_2025_mysticeti, sui2025}, with both combined over \${}880M total value locked~\cite{defillama_aptos_2026, defillama_sui_2026}.

\para{Completing atomic broadcast to SMR: The \textit{challenges}.}
Despite their adoption in large-scale blockchain deployments, existing \ac{DAG}-based designs primarily address the \ac{BAB} problem~\cite{babel_2025_mysticeti, danezis_narwhal_2022, keidar_all_2021, spiegelman2024shoal, spiegelman_bullshark_2022, tonkikh_raptr_2025} --- making all correct replicas agree on the same total order of requests --- while further components required for complete \ac{SMR} received considerably less attention.
We identify three concrete challenges that existing protocols fail to address:

\emph{Efficient client-replica interaction}.
Existing \ac{DAG}-based \ac{SMR} protocols do not specify a client interaction layer.
Their evaluations typically measure latency only until the moment the replicas locally commit the request, before the client has received any reply or can tell the request has completed~\cite{babel_2025_mysticeti, danezis_narwhal_2022, spiegelman_bullshark_2022}.
In classical \ac{BFT} \ac{SMR} systems, clients wait for $f+1$ matching replies to ensure at least one correct replica has executed the request~\cite{castro1999practical}.
Clients might also need to broadcast requests to prevent censorship~\cite{castro1999practical, distler_byzantine_2021}.
Naively applying this approach to \ac{DAG}-based protocols reduces the effective throughput by \textbf{up to 93\%}, as we show in \S\,\ref{sect:problems:client}.

\emph{Garbage Collection.}
Current \ac{DAG} protocols do not bound memory usage within the \ac{BFT} fault model.
Many protocols do not specify \ac{GC} at all~\cite{babel_2025_mysticeti, keidar_all_2021, spiegelman2024shoal, tonkikh_raptr_2025},
while those that do offload old vertices to external storage such as Amazon S3, thereby introducing a trusted component outside the \ac{BFT} fault model~\cite{danezis_narwhal_2022, spiegelman_bullshark_2022}.
Moreover, their \ac{GC} mechanism remains incomplete as the memory usage for \textit{buffering vertices} is unbounded during periods of asynchrony (see \S\,\ref{sect:problems:gc}).
However, the underlying problem is fundamental:
Recovering old vertices requires either external storage or complete \ac{SMR} (this work).

\emph{Low latency under asynchrony and $(2f+1)$-resilience.}
Obtaining low latency is a key challenge for asynchronous \ac{DAG}-based protocols~\cite{jovanovic2025mahi, fides_2026_arxiv}.
To avoid the problem, some approaches adopt partial synchrony to shorten their commit paths~\cite{babel_2025_mysticeti, spiegelman2024shoal, spiegelman_bullshark_2022}.
However, partial synchrony requires careful timeout configuration that can be violated under adversarial network conditions~\cite{tennage2023quepaxa}.
\algoname{} explores an alternative path, in which it leverages \acp{TEE} to improve both latency and resilience of the protocol while still operating in full asynchrony.

While they seem like isolated concerns, these three gaps -- client interaction, garbage collection and low latency under asynchrony -- share one root cause:
existing DAG protocols stop at the atomic-broadcast layer and never close the loop into full state-machine replication.
We therefore treat them as related and present a DAG-based BFT protocol that provides \textit{complete} SMR:
parallel atomic broadcast, efficient client interaction, garbage collection with bounded memory, and a TEE-accelerated commit path under full asynchrony.

\para{Completing DAG-based BFT towards SMR: \algoname.}
To address these challenges, we propose \algoname{}, a \ac{DAG}-based \ac{BFT} protocol designed explicitly as a complete \ac{SMR} system.
To make interactions with clients more efficient, \algoname{} introduces a \textit{client-handler} mechanism that exploits threshold signatures to preserve the throughput benefits of parallel request dissemination and satisfies \ac{SMR} execution guarantees:
The client validates a correctness proof of the execution result and enjoys exactly-once semantics.
Furthermore, \algoname{} presents a practical garbage-collection scheme that bounds resource usage under full asynchrony.
\algoname{} also shows that \acp{TEE} not only improve the resilience bound to $n \geq 2f+1$ but can also enable a substantially faster commit path in a DAG protocol.
On average, \algoname{} achieves latency gains of \(1.27\times\) over Mysticeti, the fastest partially synchronous DAG protocol, and \(1.64\times\) over Fides~\cite{fides_2026_arxiv}, an asynchronous TEE-based \ac{DAG} \ac{BFT} protocol.
Depending on the deployment, the gain over Mysticeti reaches up to \(1.52\times\).
To the best of our knowledge, this makes \algoname{} the lowest-latency DAG BFT protocol to date.
\textbf{Our contributions are as follows}:

\begin{itemize}
\item \textbf{Low latency and high resilience} -- We show how ($2f+1$)-resilience and low latency without relying on timing assumptions can be achieved using \acp{TEE}~(\S\,\ref{sect:design:buildingblocks} -- \S\,\ref{sect:design:rbc+consensus}).
\item \textbf{Completing BAB to SMR} -- We design and implement an efficient client interaction mechanism for \ac{DAG}-based \ac{BFT} to preserve benefits of parallel request dissemination while fulfilling SMR guarantees~(\S~\ref{sect:design:clienthandler}).
\item \textbf{GC under asynchrony} -- We introduce a novel garbage collection scheme that bounds resource usage under full asynchrony without any external storage
(\S\,\ref{sect:design:gc}).
\item \textbf{Extensive Evaluation} -- We conduct experiments that show that \algoname{} commits up to \(1.52\times\) faster than the fastest DAG-based protocols and can maintain high throughput 
(92–97\% of the maximum possible throughput) 
with full BFT SMR guarantees for clients.
\end{itemize}

\section{Analysis of Existing Approaches}
\label{sect:problems}
Despite achieving substantial throughput, existing DAG-based BFT protocols are designed around the \underline{B}yzantine-\underline{A}tomic-\underline{B}roadcast (BAB) layer and translating them into complete SMR systems raises three practical challenges.

\subsection{Efficient Client Interaction \& SMR Guarantees}
\label{sect:problems:client}
Practical \ac{SMR} systems must provide liveness, i.e., clients eventually receive replies to their requests~\cite{castro2001practical}.
However, many recent \ac{DAG}-based designs focus solely on \ac{BAB}.
Their evaluations typically observe a request only from the time of submission until it is locally committed by the replicas~\cite{babel_2025_mysticeti, danezis_narwhal_2022, spiegelman_bullshark_2022}, when the client has not yet received a reply.
In classical BFT SMR, clients wait for \(f+1\) matching replies, so that at least one correct replica confirms the result, and clients broadcast requests, directly or as a fallback, to prevent censorship~\cite{castro1999practical, distler_byzantine_2021, reiter1994replicate}.
Replicas cannot easily tell apart a correct client evading censorship from a spamming Byzantine one, so without intervention replicas include every request they receive.
Since every replica proposes blocks, this allows the same request to be included up to \(n\) times, each copy consuming ordering capacity~\cite{stathakopoulou2022mirbft}.

\begin{figure}[t]
    \centering
    \centering
\begin{tikzpicture}
    \begin{axis}[
        width=5.5cm,
        height=4cm,
        ybar,
        bar width=14pt,
        enlarge x limits=0.65,
        ymin=0,
        ymax=130,
        enlarge y limits=false,
        ylabel={Throughput [kOp/s]},
        xlabel={Number of replicas $n$},
        symbolic x coords={20,40},
        xtick=data,
        ytick distance=30,
        minor y tick num=1,
        ymajorgrids=true,
        yminorgrids=true,
        major grid style={dashed, gray!40},
        minor grid style={dashed, gray!35, line width=0.3pt},
        tick label style={font=\footnotesize},
        label style={font=\footnotesize},
        nodes near coords,
        nodes near coords style={
            font=\scriptsize,
            /pgf/number format/fixed,
            xshift=\pgfplotbarwidth/2,
            inner xsep=1.5pt,
        },
        legend style={
            font=\footnotesize,
            at={(1.4,1)},
            anchor=north east,
            xshift=2mm,
            yshift=2mm,
            cells={anchor=west},
        },
    ]

    \addplot[
        fill=blue!10,
        draw=black,
        postaction={
            pattern=north east lines,
            pattern color=blue!70,
        },
        nodes near coords style={anchor=north east, xshift=-\pgfplotbarwidth-1pt},
    ]
    coordinates {
        (20,120)
        (40,72.5)
    };

    \addplot[
        fill=black!8,
        draw=black,
        postaction={
            pattern=crosshatch,
            pattern color=black!60,
        },
        nodes near coords style={anchor=south, xshift=-\pgfplotbarwidth/2},
    ]
    coordinates {
        (20,13)
        (40,5)
    };

    \legend{No Client Guarantee, Classical BFT Client}

    \end{axis}
\end{tikzpicture}
    \caption{Sustained throughput of Narwhal/Tusk~\cite{danezis_narwhal_2022} with a baseline client without guarantees and a classical client that sends requests to all replicas and waits for \(f + 1\) replies.}
    \label{fig:client-perf}
\end{figure}
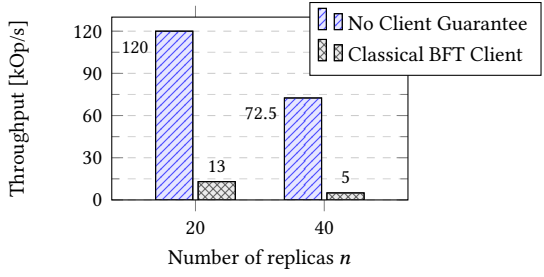

\para{Client interaction significantly impacts throughput.}
\Cref{fig:client-perf} measures the sustainable throughput of Narwhal~\cite{danezis_narwhal_2022} with a \textit{classical} client (requests to all replicas, \(f + 1\) matching replies).
Only $f+1$ designated replicas include each request, the minimum for one to be correct.
Compared to a \textit{baseline} client (one replica, one reply, no guarantees), the classical client loses 88--93\% of throughput (\(\S\,\ref{sec:eval:client}\)).

\para{Client interaction in \algoname's design.}
\algoname{} retains high throughput, by making a single contact replica sufficient while providing full \ac{SMR} guarantees (\S\,\ref{sect:design:clienthandler}).
The client waits for a single reply from its contact replica, which carries a threshold signature~\cite{boneh_short_2004} that certifies \(f+1\) replicas have executed the request.
If no reply arrives, the client resubmits to another replica, which answers immediately if the request was already executed (kept in a \textit{response storage}), or eventually re-injects it otherwise.
A \textit{round-based cooldown} keeps a spamming client from repeated inclusion this way, while still letting a censored one eventually get through.
Prior work also used threshold signatures to cut client communication, but falls back to broadcasting if no timely response arrives~\cite{gueta2019sbft, hess_consensus-agnostic_2024}.
Mir-BFT avoids multiple inclusions via consistent hashing, at the cost of forcing clients to contact a possibly distant replica~\cite{stathakopoulou_2022_mir_bft}.
\algoname{}'s mechanisms instead let clients choose the contact replica.

\subsection{Garbage Collection}
\label{sect:problems:gc}
\ac{GC} in \ac{DAG}-based protocols requires bounding memory usage while ideally providing \textit{fairness} by ensuring that all correct replicas' vertices 
are eventually totally ordered.

\para{Dilemma between fairness and garbage collection.}
To achieve ideal fairness, no unordered vertex can be deleted from the storage of all replicas because this vertex would never be totally ordered.
At the same time, DAG-based protocols do not provide an upper bound on the time required to order a vertex.
While, in expectation, it does not take more than a fixed number of rounds to order some vertices, slow replicas' vertices can take longer and can even be excluded under network partitions.
Therefore, a garbage-collection algorithm must either delete unordered vertices and continue, or wait for the network partition to pass.
Both options hinder fairness, as slow replicas are disadvantaged either by having their vertices deleted or by being forced to wait for better network conditions before proposing new ones.
Thus, providing fairness in asynchrony is effectively impossible.
Narwhal decides to remove unordered vertices from the DAG and thereby accepts fairness violations, which can occur even in a synchronous network~\cite[Sec. 1, 4th paragraph]{spiegelman_bullshark_2022}. Bullshark~\cite{spiegelman_bullshark_2022} addresses Narwhal's fairness problems by relying on weak edges combined with a timestamp-based garbage-collection scheme that guarantees \textit{timely fairness} (fairness after \ac{GST}~\cite{dwork1988consensus}).

\para{Garbage-collection limitations in existing works.}
DAG-based protocols have the advantage of decoupling request dissemination (reliable broadcast) from ordering (atomic broadcast).
However, while DAG-based atomic broadcasts are strictly round-based, reliable broadcasts are typically not.
Consequently, under unfortunate network conditions, the reliable broadcast might deliver vertices for future rounds that a slow replica has not yet reached.\footnote{This also affects the partially synchronous model (before GST is reached).} Neither
Narwhal's~\cite{danezis_narwhal_2022} nor Bullshark's~\cite{spiegelman_bullshark_2022} GC addresses the memory usage of these vertices that are stored in the so-called \textit{vertex buffer}, which makes their garbage collection mechanism unsafe in the general case.\footnote{An explicit attack against the above protocols can be found in Appendix~\ref{sec:vertexbufferoverflowattack}.}
Additionally, both algorithms offload \textit{old} (``garbage collected'') vertices to external storage such as Amazon S3
since memory must be freed, but old vertices may still be required to support other replicas during the reliable broadcast, a mechanism referred to as \textit{backfilling}.
Backfilling from external storage implicitly introduces a trusted third party that lies outside the BFT failure model.
The last two problems are inherent:
Bounding the vertex buffer requires discarding vertices whose causal history is still incomplete.
Since all correct replicas may have discarded the same vertices, recovering them later requires either external storage or a complete \ac{SMR} approach.

\para{\algoname's SMR approach to garbage collection.}
\algoname{} addresses these challenges by using an SMR execution layer instead of treating GC as a pure DAG-level concern.
First, we delete and thereby ignore the transactions of unordered vertices not when a DAG round is deemed old by some timestamp heuristic, but after a fixed number of DAG-rounds, which bounds memory consumption and provides timely fairness.
Second, \algoname{} employs the SMR approach to periodically build \textit{checkpoints}, which are certified snapshots of the replicated state that are signed by a quorum of replicas.
A vertex can be safely deleted once a stable checkpoint $ckpt_i$ covering the execution or ignorance of its transactions exists, as any replica that missed the vertex recovers by performing a state transfer to the most recent stable checkpoint $ckpt_j$, $j\geq i$, rather than relying on backfilling.
Third, to prevent unbounded growth of the vertex buffer during asynchrony,
\algoname{} enforces a \textit{bounded receiving window}, which can discard all incoming vertices whose rounds fall too far ahead of the local progress.
Importantly, if such a vertex is later needed, the same state-transfer mechanism covers its content.

As a result, both the DAG and the vertex buffer are bounded at all times, even under full asynchrony, with the \textit{timely fairness}~\cite{spiegelman_bullshark_2022} guarantee preserved and without the need for external storage outside the BFT failure model.

\subsection{Asynchrony, Low Latency and (\(2f+1\))-Resilience}
\label{sect:problems:latency}
Early asynchronous designs such as \ac{DAG}-Rider~\cite{keidar_all_2021} select leaders (\textit{anchor} vertices) retrospectively by requiring a shared random coin to circumvent the impossibility of deterministic, fault-tolerant consensus under asynchrony~\cite{fischer1985impossibility}.

\para{Low-latency DAG protocols.}
After Narwhal~\cite{danezis_narwhal_2022}, many research works aimed to reduce DAG protocol commit latency~\cite{spiegelman_bullshark_2022, arun_2025_shoalpp, shrestha2025sailfish, polyanskii2025starfish, keidar2023cordial, babel_2025_mysticeti, jovanovic2025mahi},
often assuming partial synchrony~\cite{dwork1988consensus}, which sidesteps this impossibility by bounding message delays after \ac{GST} and allows pre-defined leaders~\cite{spiegelman_bullshark_2022, arun_2025_shoalpp, shrestha2025sailfish, polyanskii2025starfish, babel_2025_mysticeti}. Mysticeti~\cite{babel_2025_mysticeti} achieves fast commit latency by using a novel commit rule and an uncertified DAG, an idea that Mahi-Mahi-5 transfers to the asynchronous network setting~\cite{jovanovic2025mahi}. Mahi-Mahi-5 needs additional rounds compared to Mysticeti at the cost of slightly increased commit latency.
Partial synchrony is a widely used assumption in practice~\cite{castro1999practical, hotstuff19}, but it requires careful timeout selection and misconfigured or temporarily violated timeouts can slow or stall progress under adversarial network conditions~\cite{tennage2023quepaxa}.
We show that the trade-off between low latency and a partial synchrony assumption is not fundamental:
Asynchronous \ac{DAG}-based protocols can be accelerated by \textit{cleverly leveraging \acp{TEE}} in the protocol design, while at the same time offering $n\geq2f+1$ resilience.

\begin{figure}[t]
    \centering
    \begin{tikzpicture}
\begin{axis}[
    ybar stacked,
    bar width=15pt,
    width=0.98\columnwidth,
    height=4cm,
    ymin=0,
    enlarge x limits=0.1,
    ylabel={Latency [ms]},
    symbolic x coords={DAGRider, Narwhal, Bullshark,  MahiMahi5, Shoal++, Fides, Mysticeti, Pangolin},
    xtick=data,
    xticklabels={
    DAGRider,
    Narwhal,
    Bullshark,
    Shoal++,
    Mahi-Mahi-5,
    Fides,
    Mysticeti,
    \textbf{Pangolin}
},
    x tick label style={font=\footnotesize},
    tick label style={font=\footnotesize},
    label style={font=\footnotesize},
    ymajorgrids=true,
    yminorgrids=true,
    grid style=dashed,
    minor y grid style={dotted},
    ytick distance=500,
    minor y tick num=4,
    extra y ticks={284},
    extra y tick labels={},
    extra y tick style={
        grid=major,
        major grid style={black!80, thick, solid}
    },
    x tick label style={
    font=\footnotesize,
    rotate=30, anchor=north east, inner sep=1pt
},
    legend style={
        at={(1,1)}, anchor=north east,
        font=\footnotesize,
        draw=black, fill=white,
        row sep=-2pt,
    },
    legend cell align=left,
]

\addplot[fill=blue!20, draw=blue, forget plot] coordinates {
    (DAGRider,284) (Narwhal,284) (Bullshark,0) (Shoal++, 0) (MahiMahi5, 284) (Fides, 284) (Mysticeti,0) (Pangolin,284)
};

\addplot[fill=red!20, draw=red, forget plot] coordinates {
    (DAGRider,0) (Narwhal,0)  (Bullshark,284)  (Shoal++, 284) (MahiMahi5, 0) (Fides, 0) (Mysticeti,284) (Pangolin,0)
};
\addplot[fill=blue!40, draw=blue, forget plot] coordinates {
    (DAGRider,1608) (Narwhal,1006) (Bullshark,0) (Shoal++, 0) (MahiMahi5, 232) (Fides, 220) (Mysticeti,0) (Pangolin,0)
};
\addplot[fill=red!40, draw=red, forget plot] coordinates {
    (DAGRider,0) (Narwhal,0) (Bullshark,656) (Shoal++, 233)  (MahiMahi5, 0) (Fides, 0) (Mysticeti,88) (Pangolin,0)
};

\addlegendimage{area legend, fill=blue!30, draw=blue}
\addlegendentry{asynchronous}
\addlegendimage{area legend, fill=red!30, draw=red}
\addlegendentry{partially synchronous}
\end{axis}
\end{tikzpicture} 
    \vskip-2mm
    \caption{Mean block decision latency across DAG-based BFT protocols.
     Although \algoname{} is fully asynchronous, it is even faster than the partially synchronous Mysticeti.
    }
    \label{fig:motivation-latency}
\end{figure}
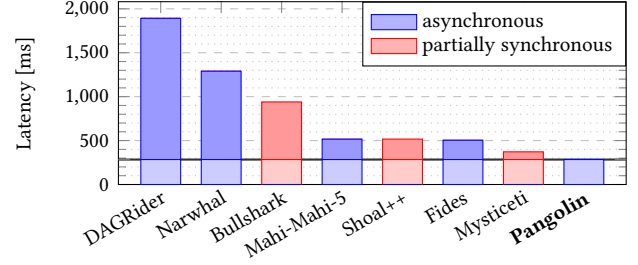

\para{\algoname{} attacks latency with \acp{TEE}.}
With \ac{TEE}s \algoname{} achieves the ($n\ge{}2f+1$)-resilience bound, but TEEs can also reduce the latency of atomic broadcast:
First, as TEEs can prevent equivocation, they can be used to realize a reliable broadcast primitive with fewer communication steps.
Second, the TEE assumption enables proportionally smaller quorums (of size $\lceil \frac{n+1}{2} \rceil$) which can be collected faster in a WAN than the usual Byzantine dissemination quorums.
Third, a TEE-based common coin can be used to check if commit rules apply earlier without revealing the randomly selected anchor.
We show how TEEs reduce latency in the asynchronous network model.
Moreover, in our evaluation \algoname{} is faster than the partially synchronous Mysticeti, the fastest DAG protocol known to us~\cite{babel_2025_mysticeti}.
Fig.~\ref{fig:motivation-latency} shows \algoname{} overtakes Mysticeti with a latency of 284~ms vs. 372~ms at \(n=13\) (and using the same 13 regions that the evaluation setup of Mysticeti employs~\cite{babel_2025_mysticeti}).
The details on our evaluation method for Fig.~\ref{fig:motivation-latency} are provided in Sect.\,\ref{sec:eval:generality} and App.~\ref{app:sim-mysticeti-deployment}.

\section{System Model}

\para{Adversary model.} We consider a system of $n$ replicas, of which a static adversary may control up to $f$ replicas chosen before execution and where $n \geq 2f+1$.
Furthermore, corrupted replicas are \textit{Byzantine}:
They may deviate from the protocol and coordinate with one another.
Replicas communicate over authenticated perfect point-to-point links~\cite{cachin2011introduction}.
The adversary cannot break cryptographic primitives such as employed hash functions and digital signature algorithms.

\para{Network model.}
The network is asynchronous:
messages may be reordered and delayed, but not dropped.
We make no timing assumptions and place no bounds on message delays.

\begin{table}[t]
  \footnotesize
  \begin{tabular}{@{}p{2.05cm}p{1.7cm}p{3.7cm}@{}}
    \toprule
    \textbf{Method} & \textbf{Trusted state} & \textbf{Guarantee} \\
    \midrule
     $\fn{setup}$ $\rightarrow (p_{k},s_{k}), cs$ & key pair, coin seed & attested identity; coin \emph{agreement} via the shared seed \\
    \addlinespace[3pt]
    $\fn{enclave\_sign}{v}$ $\rightarrow \sigma \mid \bot$ & $r_{\text{max}}$ & \emph{non-equivocation}: at most one well-formed vertex per round, rounds are strictly increasing \\
    \addlinespace[3pt]
    $\fn{check\_commit}{v}$ $\rightarrow u \mid \bot$ & $\var{sup}$, CPRNG & \emph{commit-gated coin reveal}: the wave's anchor is disclosed when the direct commit rule first holds, which requires sufficient support ($\var{sup}$) \\
    \bottomrule
  \end{tabular}
   \caption{\algoname{}'s abstract \ac{TEE} interface.
    Each method guards a piece of enclave state and provides one guarantee.
    }
  \label{tab:tee-interface}
\end{table}

\para{TEE model.}
We assume each replica is equipped with an enclave-based \ac{TEE} that provides integrity, confidentiality, and remote attestation.
Table~\ref{tab:tee-interface} shows the abstract interface of our assumed TEE, its enclave state and the guarantees that are given for the BFT protocol execution.

\para{TEE assumptions.}
We assume the \ac{TEE} fails only by crashing.
Upon a crash, its internal state is lost and is not recoverable.
The untrusted host, including the operating system and application layer, could be under adversary control:
it may omit or delay messages to and from the enclave, but cannot read or tamper with its internal state.
We assume the \ac{TEE} hardware itself is not compromised.
Furthermore, side-channel attacks and hardware-level exploits are outside our threat model.

\para{Threshold signatures.}
\label{sect:model:thresholdsig}
A $(t,n)$-threshold signature scheme involves $n$ participants, whereby each participant $i$ holds a private key share $sk_i$, and any subset of $t$ participants can collaborate to create a threshold signature, which is verifiable using a single public key.
Previous work has primarily used RSA-based threshold signatures~\cite{amir10steward, reiter1994replicate, shoup2000threshold}, or BLS-based threshold signatures~\cite{boneh_short_2004}, which are favored in recent systems~\cite{gueta2019sbft, hess_consensus-agnostic_2024, neiheiser2021kauri} due to their shorter signatures.

\para{TEE-based threshold signatures.}
In contrast to previous work, \algoname{} builds on the \emph{SpliTEE}~\cite{haller2026splitee} approach, which utilizes a trusted execution environment to produce efficient threshold signatures.
We make use of the \emph{common key} variant:
A common EdDSA key-pair~\cite{bernstein2011eddsa} is encapsulated by the TEEs of all participating nodes, and the public key is used as the common verification key.
In addition, each node holds an individual EdDSA key pair, whose public key is known in advance by the TEEs.
A TEE creates a threshold signature for some data $d$ using the common private key, if it is provided with valid signature shares for $d$ from $t$ distinct nodes.

\section{Design of \algoname}
\label{sect:design}

\algoname{}'s design follows our objectives of leaderless operation, low-latency commitment, and efficient client handling to maintain high \ac{SMR} throughput.
\label{sect:design:overview}
These goals are achieved by composing several components into a complete \ac{SMR} system:
two TEE-based building blocks (a reliable broadcast and a common coin (\S\,\ref{sect:design:buildingblocks})),
a novel, asynchronous DAG-based AB-layer that leverages them for fast vertex commitment (\S\,\ref{sect:design:rbc+consensus}),
a client handler that supports load-balancing of clients across several replicas via compact execution proofs (\S\,\ref{sect:design:clienthandler}),
and a checkpoint-driven garbage collection scheme that bounds memory usage under full asynchrony (\S\,\ref{sect:design:gc}).
We also provide a visualization of \algoname{} that may help in understanding its commit rules and comparing it with other DAG protocols\footnote{See \url{https://pangolin-protocol.netlify.app/} to visualize \algoname{}}.

  \begin{figure}[t]
   \centering
 \includegraphics[width=\columnwidth]{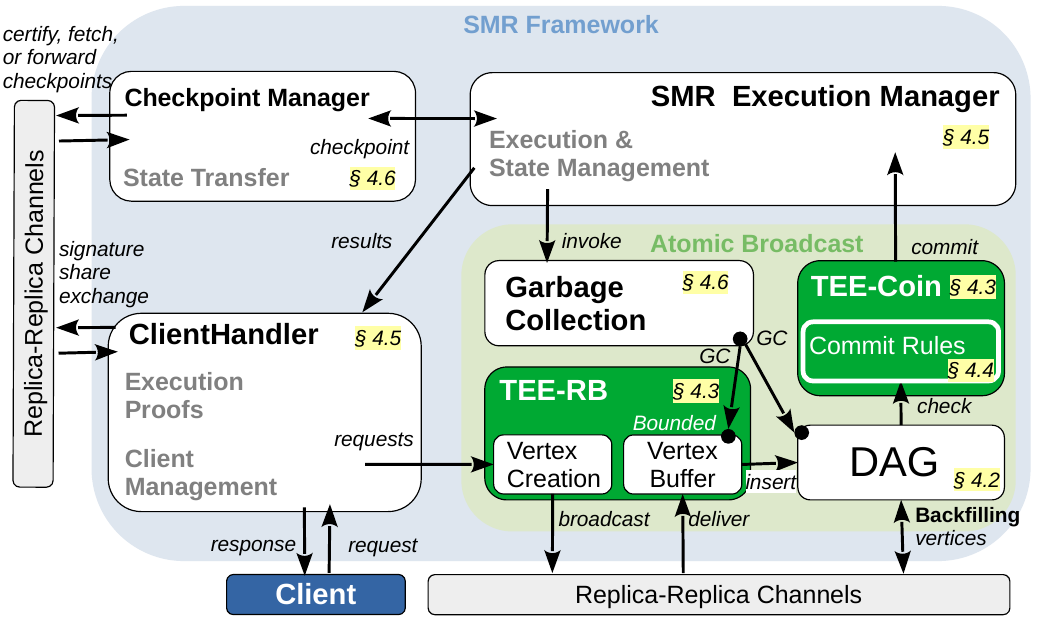}
    \caption{Architecture of \algoname{}.}
    \label{fig:architecture}
 \end{figure}

\subsection{Architectural Overview}
\Cref{fig:architecture} illustrates \algoname{} at a single replica.
Clients submit their requests to a contact replica of their choice, typically selecting one that is geographically close.
The client handler~(\S\,\ref{sect:design:clienthandler}) receives these requests, buffers them, and forwards them to the \ac{AB} layer, which periodically includes them in a newly created vertex, which is reliably broadcast to all replicas as soon as a new DAG-round can begin (details follow in \S\,\ref{sect:design:rbc+consensus}).
When the reliable broadcast layer delivers a vertex, it is placed in the \textit{vertex buffer};
If it corresponds to the current round and its causal history is available, it is inserted into the DAG.
If some references to a delivered vertex are missing, \textit{backfilling} is used to request the missing dependencies directly from other replicas.
Vertex insertion may trigger an evaluation of the commit rules (which we describe in \S\,\ref{sect:design:rbc+consensus}), which may commit several vertices at once.
The resulting sequence of committed vertices is then passed from the \ac{AB} layer to the execution and state management layer, where the requests contained in these vertices are filtered for duplicates before execution.
After execution, responses are passed back to the client handler, which attaches compact execution proofs based on threshold signatures (\S\,\ref{sect:design:clienthandler}) and stores the most recent response for each client.
Whenever a replica acts as a client's contact, it responds directly to that client.
In addition, the SMR execution manager periodically creates checkpoints that drive the garbage-collection scheme, enabling vertices to be removed from both the vertex buffer and the DAG (see \S\,\ref{sect:design:gc}).
To become stable, checkpoints need to be certified by $f+1$ replicas by collecting threshold signature shares and forming a valid $(f+1)$-threshold signature.
Stable checkpoints can then be forwarded to stale replicas, allowing them to recover by installing the most recent application state and invoking their garbage collection routine.

\subsection{DAG Construction}
\label{sect:dag-construction}
In this subsection, we briefly recap the core concepts of DAG construction and explain how they are applied in \algoname{}.

\begin{figure}[t]
    \centering
    \resizebox{\columnwidth}{!}{\definecolor{pcFill}{RGB}{168,224,196}  
\definecolor{pcEdge}{RGB}{34,140,100}
\definecolor{pcSlate}{RGB}{255,255,255} 
\definecolor{pcSlateE}{RGB}{140,148,160}
\definecolor{pcGold}{RGB}{240,176,32}   
\definecolor{pcAmber}{RGB}{198,158,70}  
\definecolor{pcStroke}{RGB}{150,157,168}\definecolor{pcHi}{RGB}{232,132,20}     
\definecolor{pcWeak}{RGB}{150,110,214}  

\begin{tikzpicture}[
  >={Stealth[length=2.1mm,width=1.8mm]},
  proc/.style={circle,draw=pcEdge,line width=0.7pt,fill=pcFill,minimum size=10mm,inner sep=0pt},
  frn/.style={circle,draw=pcSlateE,line width=0.7pt,fill=pcSlate,minimum size=10mm,inner sep=0pt},
  ldD/.style={circle,draw=pcGold,line width=2pt,fill=pcFill,minimum size=12.5mm,inner sep=0pt,
              font=\large,text=black},                                          
  ldR/.style={ldD,dash pattern=on 2.3pt off 1.7pt},                            
  se/.style={->,draw=pcStroke,line width=0.6pt,shorten >=0.5pt,shorten <=0.5pt},
  hl/.style={->,draw=pcHi,line width=1.6pt,shorten >=0.5pt,shorten <=0.5pt},
  we/.style={->,draw=pcWeak,line width=1.2pt,dashed,shorten >=0.5pt,shorten <=0.5pt},
]

\def\xI{0}\def\xII{1.8}\def\xIII{3.6}\def\xIV{5.4}
\def\xV{7.6}\def\xVI{9.4}\def\xVII{11.2}\def\xVIII{13.0}
\def\yA{4}\def\yB{2}\def\yC{0}

\node[proc] (n11) at (\xI,\yA){};   \node[proc] (n12) at (\xII,\yA){};
\node[proc] (n13) at (\xIII,\yA){}; \node[proc] (n14) at (\xIV,\yA){};
\node[ldD]  (n15) at (\xV,\yA){$v^1_{w{+}1}$};
\node[ldD]  (n16) at (\xVI,\yA){$v^2_{w{+}1}$};
\node[ldD]  (n17) at (\xVII,\yA){$v^3_{w{+}1}$};
\node[frn]  (n18) at (\xVIII,\yA){};
\node[proc] (n21) at (\xI,\yB){};   \node[proc] (n22) at (\xII,\yB){};
\node[proc] (n23) at (\xIII,\yB){}; \node[proc] (n24) at (\xIV,\yB){};
\node[proc] (n25) at (\xV,\yB){};   \node[proc] (n26) at (\xVI,\yB){};
\node[frn]  (n27) at (\xVII,\yB){}; \node[frn]  (n28) at (\xVIII,\yB){};
\node[ldD]  (n31) at (\xI,\yC){$v^1_w$};   \node[ldD] (n32) at (\xII,\yC){$v^2_w$};
\node[ldR]  (n33) at (\xIII,\yC){$v^3_w$}; \node[proc] (n34) at (\xIV,\yC){};
\node[proc] (n35) at (\xV,\yC){};   \node[frn]  (n36) at (\xVI,\yC){};
\node[frn]  (n37) at (\xVII,\yC){}; \node[frn]  (n38) at (\xVIII,\yC){};

\foreach \r in {2,3,5,6,7,8}{
  \pgfmathtruncatemacro{\pr}{\r-1}
  \foreach \p/\ta/\tb in {1/1/2, 2/2/3, 3/3/1}{
    \draw[se] (n\p\r) -- (n\ta\pr);
    \draw[se] (n\p\r) -- (n\tb\pr);
  }
}
\foreach \p/\ta/\tb in {1/1/2, 2/1/2, 3/3/1}{
  \draw[se] (n\p4) -- (n\ta3);
  \draw[se] (n\p4) -- (n\tb3);
}

\draw[hl] (n22) -- (n31);   
\draw[hl] (n32) -- (n31);   

\draw[we] (n15) to[out=215,in=52,looseness=0.8] (n33);

\node[font=\large,text=pcHi,align=left,anchor=south west] at (-1.0,5.05)
  {$f{+}1$ support in $(w,2)$\\ $\Rightarrow$ \textbf{direct commit} of $v^1_w$};
\node[font=\large,text=pcWeak,align=left,anchor=south west] at (4.0,4.7)
  {commit path traverses a \textbf{weak} edge\\ ($\mathrm{path}$, not $\mathrm{strongPath}$)
   $\;\Rightarrow\;$ \textbf{retrospective commit} of $v^3_w$};

\node[left=2.5mm,font=\LARGE] at (n11.west) {$p_1$};
\node[left=2.5mm,font=\LARGE] at (n21.west) {$p_2$};
\node[left=2.5mm,font=\LARGE] at (n31.west) {$p_3$};

\def\yi{-1.2}
\node[font=\LARGE] at (-1.35,\yi) {$k$};
\foreach \x/\lab in {\xI/1,\xII/2,\xIII/3,\xIV/4,\xV/1,\xVI/2,\xVII/3,\xVIII/4}
  \node[font=\LARGE] at (\x,\yi) {\lab};
\def\yba{-1.8}
\draw[decorate,decoration={brace,amplitude=7pt,mirror,raise=2pt}]
  ($(\xI,\yba)+(-0.55,0)$) -- ($(\xIV,\yba)+(0.55,0)$)
  node[midway,below=13pt,font=\LARGE] {wave $w$};
\draw[decorate,decoration={brace,amplitude=7pt,mirror,raise=2pt}]
  ($(\xV,\yba)+(-0.55,0)$) -- ($(\xVIII,\yba)+(0.55,0)$)
  node[midway,below=13pt,font=\LARGE] {wave $w{+}1$};

\begin{scope}[shift={(-1.1,-3.5)}]
  \node[proc,minimum size=7mm] (lg1) at (0,0){};
  \node[right=1.5mm,font=\large,anchor=west] at (lg1.east) {ordered};
  \node[ldD,minimum size=7mm,font=\tiny] (lg2) at (3.3,0){};
  \node[right=1.5mm,font=\large,anchor=west] at (lg2.east) {direct commit};
  \node[ldR,minimum size=7mm,font=\tiny] (lg3) at (7.0,0){};
  \node[right=1.5mm,font=\large,anchor=west] at (lg3.east) {retrospective};
  \node[frn,minimum size=7mm] (lg4) at (10.5,0){};
  \node[right=1.5mm,font=\large,anchor=west] at (lg4.east) {not yet ordered};
  \draw[se,line width=0.9pt] (0.0,-1.05) -- (1.1,-1.05);
  \node[right=1.5mm,font=\large,anchor=west] at (1.1,-1.05) {strong edge};
  \draw[we] (4.4,-1.05) -- (5.5,-1.05);
  \node[right=1.5mm,font=\large,anchor=west] at (5.5,-1.05) {weak edge};
  \draw[hl] (7.8,-1.05) -- (8.9,-1.05);
  \node[right=1.5mm,font=\large,anchor=west] at (8.9,-1.05) {support quorum for $v^1_w$};
\end{scope}

\end{tikzpicture}}
    \caption{
\algoname{}'s commit rules on an example DAG of $n=3$ replicas ($f=1$, quorum $\lceil \frac{n+1}{2} \rceil=2$) over two waves. Every vertex references $\lceil \frac{n+1}{2} \rceil$ vertices of the previous round, always including its own predecessor (solid \emph{strong} edges, the arrow pointing to the referenced older vertex); \emph{weak} edges (dashed) point to further-back rounds. Each wave has \textbf{three anchor slots} produced by a single coin-elected replica in every round of the wave but the last ($p_3$ in wave $w$, $p_1$ in wave $w{+}1$). \textbf{Direct commit:} an anchor is committed as soon as $\lceil \frac{n+1}{2} \rceil$ vertices of a later within-wave round have a path to it, and support is counted only until that quorum is first reached. $v^1_w$ already collects its $\lceil \frac{n+1}{2} \rceil$ support in the \emph{next} round $(w,2)$ (the two highlighted supporters), so the TEE uncovers the coin and commits it one round after it was proposed --- the support need not reach the wave's last round. $v^2_w$ likewise commits in $(w,3)$, as do all three slots of wave~$w{+}1$. \textbf{Retrospective commit:} $v^3_w$'s only later within-wave round is $(w,4)$, where fewer than $\lceil \frac{n+1}{2} \rceil$ vertices reach it, so it is not directly committed; it is committed retrospectively when $v^1_{w+1}$ commits, along a path that \emph{traverses the weak edge} $v^1_{w+1}\!\to\!v^3_w$. In contrast, DAGRider's commit rules traverse only strong edges. A vertex is colored green once it lies in a committed leader's causal history.}
    \label{fig:dag-sample}
\end{figure}
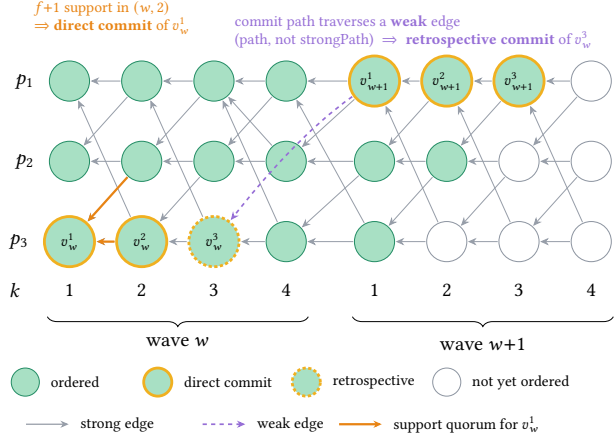

\para{Vertices, rounds, and edges.}
All replicas propose vertices in parallel, each containing a payload of client requests and references to previous vertices.
Each replica maintains a local view of the DAG as a two-dimensional array indexed by round and producer.
An example local view is shown in Fig.~\ref{fig:dag-sample}.
A vertex $v$ has a round $v.r \in \mathbb{N}$, a producer $v.p \in P$, a payload $v.\mathit{block}$ of client requests, a signature $v.\sigma$, and
a set of edges $v.\mathit{edges}$ that reference vertices from previous rounds. \textit{Strong edges} only reference
one vertex per producer from the last round $v.r{-}1$ that the creator of $v$ observed, while \textit{weak edges} include vertices from two or more rounds in the past.

\para{Round advancement.}
Edges capture causal dependencies and act as votes for the referenced vertices.
Because a TEE never signs more than one vertex per $(p,r)$-pair, Byzantine broadcasters cannot equivocate.
Two correct replicas therefore never observe conflicting vertices at any $(p,r)$.
A replica advances from round $r$ to $r{+}1$ once a quorum of $\lceil \frac{n+1}{2} \rceil$ vertices in round $r$ is in its local DAG, including its own.
The next vertex it produces will reference this quorum as edges.

\para{Waves, anchors and slots.}
Rounds are grouped into \emph{waves} of four consecutive rounds and  
$\mathit{round}(w, k) \triangleq 4w + k - 1$
denotes the $k$-th round of wave $w$.
For every wave $w$, the common coin toss randomly elects a single \textit{anchor replica}.
In the DAGRider protocol~\cite{keidar_all_2021}, the \emph{anchor slot} of wave $w$ is then only the $\mathit{round}(w, 1)$-vertex produced by the anchor replica.
In contrast, \algoname{} uses multiple anchor slots per wave.
The vertices the anchor replica produces in $w$'s first three rounds -- $(w,1), (w,2), (w,3)$ -- are the \textit{anchor slots}.

\subsection{Instantiation of the Building Blocks}
\label{sect:design:buildingblocks}
Next, we briefly explain how the two main building blocks of \algoname{} can be instantiated with the assistance of a TEE and how the TEE support helps to improve latency.

\subsubsection{The TEE-based Reliable Broadcast.}
\label{sect:design:buildingblocks:tee-broadcast}
\begin{figure}[t]
    \centering
    \resizebox{\columnwidth}{!}{\providecommand{\algoname}{\textsc{Porcupine}}\par
\providecolor{bcblue}  {RGB}{ 68,114,196}\providecolor{bcorange}{RGB}{237,125, 49}\providecolor{bcgreen} {RGB}{112,173, 71}\providecolor{bcpurple}{RGB}{152, 78,163}\par
\def\bcastFour#1#2{\foreach \s/\r/\c in {0/1/bcblue,   0/2/bcblue,   0/3/bcblue,
                        1/0/bcorange, 1/2/bcorange, 1/3/bcorange,
                        2/0/bcgreen,  2/1/bcgreen,  2/3/bcgreen,
                        3/0/bcpurple, 3/1/bcpurple, 3/2/bcpurple}
    \draw[msg, draw=\c] (#1,-\s) -- (#2,-\r);}\def\ackFour#1#2{\foreach \s/\r/\c in {1/0/bcblue,   2/0/bcblue,   3/0/bcblue,
                        0/1/bcorange, 2/1/bcorange, 3/1/bcorange,
                        0/2/bcgreen,  1/2/bcgreen,  3/2/bcgreen,
                        0/3/bcpurple, 1/3/bcpurple, 2/3/bcpurple}
    \draw[msg, draw=\c] (#1,-\s) -- (#2,-\r);}\par
\def\bcastThree#1#2{\foreach \s/\r/\c in {0/1/bcblue,   0/2/bcblue,
                        1/0/bcorange, 1/2/bcorange,
                        2/0/bcgreen,  2/1/bcgreen}
    \draw[msg, draw=\c] (#1,-\s) -- (#2,-\r);}\par
\def\lanes#1#2#3{\foreach \i/\k/\c in {#1} {\draw[lane]                                (-0.20,-\i) -- (#2,-\i);
    \draw[lane, dash pattern=on 1.6pt off 1pt] (   #2,-\i) -- (#3,-\i);
    \node[plabel, text=\c] at (-0.30,-\i) {$\boldsymbol{p_\k}$};}}\par
\begin{tikzpicture}[
    x=1.2513cm,           
    y=0.6882cm,           
    lane/.style       ={bcblue, line width=0.4pt},
    commstep/.style   ={bcblue, line width=0.4pt},
    bcastround/.style ={black,  line width=0.75pt},
    msg/.style        ={line width=0.4pt,
                        -{Triangle[length=1.5mm, width=1.35mm]}},
    plabel/.style     ={anchor=east, inner sep=1pt, font=\footnotesize},
    steplabel/.style  ={anchor=base, inner sep=0pt, font=\scriptsize},
    roundlabel/.style ={anchor=base, inner sep=0pt, font=\footnotesize},
    panelname/.style  ={anchor=base, inner sep=0pt, font=\footnotesize},
  ]

  \begin{scope}
    \lanes{0/1/bcblue, 1/2/bcorange, 2/3/bcgreen, 3/4/bcpurple}{3.00}{3.55}
    \foreach \x in {1,2} \draw[commstep]   (\x,0.40) -- (\x,-2.92);
    \foreach \x in {0,3} \draw[bcastround] (\x,0.40) -- (\x,-3.99);
    \bcastFour{0}{1}          
    \ackFour  {1}{2}          
    \bcastFour{2}{3}          
    \node[steplabel]  at (0.5,-3.42) {$n^2$ vertex};
    \node[steplabel]  at (1.5,-3.42) {$n^2$ ack};
    \node[steplabel]  at (2.5,-3.42) {$n^2$ cert};
    \node[roundlabel] at (1.500,-3.84) {$r$};
    \node[roundlabel] at (3.275,-3.84) {$r+1$};
    \node[panelname]  at (1.482,-4.40) {(a) Narwhal: $3n^2$};
  \end{scope}

  \begin{scope}[shift={(4.48,0)}]
    \lanes{0/1/bcblue, 1/2/bcorange, 2/3/bcgreen}{1.00}{1.55}
    \foreach \x in {0,1} \draw[bcastround] (\x,0.40) -- (\x,-2.99);
    \bcastThree{0}{1}         
    \node[steplabel]  at (0.500,-2.42) {$n^2$ vertex};
    \node[roundlabel] at (0.500,-2.84) {$r$};
    \node[roundlabel] at (1.275,-2.84) {$r+1$};
    \node[panelname]  at (0.482,-4.40) {(b) \algoname{}: $n^2$};
  \end{scope}

\end{tikzpicture}}
    \caption{Common-case communication of one broadcast round in the DAG, shown for $f=1$.
    Black lines delimit broadcast rounds, blue lines the communication steps within a round and an arrow's color is the originator of its payload.
 Narwhal's quorum-based reliable broadcast needs \textit{three} steps and $3n^2$ messages to complete a round, certifying $n$ vertices whereas \algoname{}'s \ac{TEE}-based reliable broadcast finishes a broadcast round in a \textit{single} step with $n^2$ messages only.}
    \label{fig:broadcast}
\end{figure}
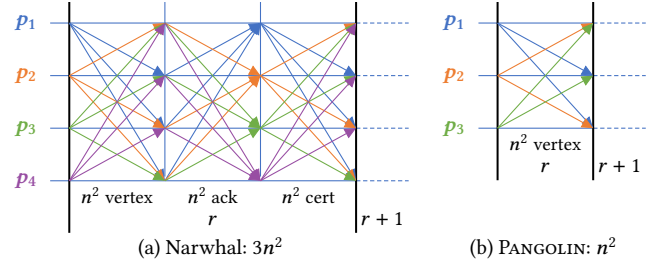

The RB-layer of \algoname{} broadcasts one vertex per replica and round.
We instantiate it as a \textit{Byzantine Consistent Broadcast}~\cite{cachin2011introduction} that is completed by a \textit{backfilling} mechanism:
The enclave prevents equivocation, so that dissemination itself only costs a single communication step, and causal history that a replica is missing is fetched reactively from others.
\algoname{} integrates the TEE-based RB from~\cite[\S\,4.2]{leinweber_2026_tee_ledgers}.
We compare \algoname{}'s communication pattern with the quorum-based certification of Narwhal~\cite{danezis_narwhal_2022} in Figure~\ref{fig:broadcast}.

\para{Ensuring \textit{non-equivocation} via the enclave.}
Before broadcasting a vertex $v$, a replica $p_i$ invokes the enclave to get a signature $v.\sigma$~(Alg.~\ref{alg:porcupinetee}, ll. 5--11).
The enclave guards $r_{max}$, the highest round it has signed a vertex for so far, and refuses to sign a vertex if its round $v.r$ is not higher than $r_{max}$.
Thus, every replica, correct or faulty, can receive at most one valid signature per round, and the rounds of the signed vertices are strictly monotonically increasing.
As a consequence, two correct replicas never see conflicting vertices for the same $(producer, round)$-slot, and the misbehavior a faulty broadcaster can cause is reduced from Byzantine to just omission.

\para{From RB delivery to DAG insertion.}
Verifying a vertex $v$ does not make it immediately usable, because the commit rules (see \S\,\ref{sect:design:rbc+consensus}) reason over paths, so $v$ can be included in the DAG only when its \textit{causal history} is available.
Thus, a replica puts a verified vertex $v$ into the \emph{vertex buffer} (Alg.~\ref{alg:porcupinerb}, l.~5) and inserts it in the DAG once (1)~it has reached round $v.r$ \textit{and} (2)~every vertex referenced by $v.\mathit{edges}$ is either present in the DAG or already stale (Alg.~\ref{alg:porcupineab}, ll.~10--12).
Vertices of rounds ahead of the local round can arrive at any time and remain buffered until they become insertable.
Note that the insertion rule is applied transitively, so the ancestry of an inserted vertex must be part of the DAG down to the garbage-collection boundary. 
We detail how \algoname{}'s garbage collection scheme can bound memory in \S\,\ref{sect:design:gc}.

\para{Backfilling.}
When a replica receives a vertex $v$ whose referenced vertices are neither in the DAG nor in the vertex buffer, then it requests the missing vertices from other replicas (Alg.~\ref{alg:porcupinerb}, l.~3 and ll.~14--17).
\algoname{} adopts a pull-based recovery like Narwhal~\cite{danezis_narwhal_2022}:
Retrieving the history of a vertex only costs extra messages when a vertex overtakes it or when omissions occur, so the common case is still a single communication step.
In case a requested vertex was already garbage collected, then the request is replied to with a \textit{state transfer} (Alg.~\ref{alg:porcupinerb}, ll.~9--10) that updates the asking replica to a state in which the vertex is no longer needed.
Because a vertex can be garbage collected before every correct replica has added it, our instantiation does not provide \textit{RB-Totality} in its unrestricted form at the broadcast layer.
Instead, we prove the weaker property that a vertex added by one correct replica is eventually added by every correct replica or becomes irrelevant to it (Theorem~\ref{the:rbagreement}), and show in \S\,\ref{sect:design:gc} that this suffices for the atomic broadcast layer.

\subsubsection{The TEE-assisted Common Coin.}
The coin has one task:
Elect one anchor replica in every wave and let all correct replicas observe the same result, yet the result needs to be unpredictable to the adversary and disclosed only \textit{after} the wave has passed or an anchor slot of that wave can be committed.
Commonly, asynchronous protocols use a coin based on threshold signatures that is revealed \textit{after} the wave has passed, which requires both an initial distributed key setup and an exchange of signature shares per coin toss.
Because \algoname{} already leverages the TEE for the broadcast, we might as well use it to derive the coin from a cryptographically secure pseudo-random generator (CPRNG) that runs inside the enclave of each replica.
Enclaves are seeded with the same value during setup (see Table~\ref{tab:tee-interface}), so each enclave outputs the same anchor for the same wave and thus replicas agree on it without a single message exchange.

\para{Commit-gated disclosure.}
If an anchor replica is disclosed too early by the common coin, then an adversarial scheduler can prevent the anchor from being committed:
The adversary would delay the anchor replica's vertices so that they are not referenced by sufficiently many other replicas and then fail to collect the quorum support.
For this reason, other asynchronous protocols hide the anchor replica until the corresponding wave has passed, implying that commits can be checked in the subsequent wave at the earliest.
To prevent such delay, the TEE evaluates the commit rules itself:
Once a replica inserts a vertex $v$ in the DAG, it invokes $\fn{check\_commit}{v}$ (Alg.~\ref{alg:porcupineab}, l.~13).
The enclave verifies $v.\sigma$, computes the anchor replica of wave $v.w$ from the CPRNG \textit{without returning it}, and tests for each of the wave's three anchor slots whether $v$ supports that slot, either by referencing it directly or by referencing a vertex that already supports it (Alg.~\ref{alg:porcupinetee}, ll.~12--20).
The commit rules follow in \S\,\ref{sect:design:rbc+consensus};
here it is only relevant that the supporting vertices accumulate in the enclave state $\var{sup}$.
Only when an anchor slot commits does the enclave return that slot and unveil it to the outside.
At the latest, the anchor is revealed once the next wave starts.

\begin{figure}[t]
    \centering
    \resizebox{\columnwidth}{!}{\providecolor{cnGold} {RGB}{240,176, 32}\providecolor{cnHi}   {RGB}{232,132, 20}\providecolor{cnFill} {RGB}{168,224,196}\providecolor{cnSlate}{RGB}{140,148,160}\par
\begin{tikzpicture}[
    x=1.8344cm,          
    y=0.8000cm,
    font=\footnotesize,
    encl/.style     ={draw=black, line width=0.7pt},
    divider/.style  ={draw=black!65, line width=0.6pt,
                      dash pattern=on 1.6pt off 1.2pt},
    lane/.style     ={draw=black!40, line width=0.4pt},
    guide/.style    ={draw=black!22, line width=0.4pt},
    edge/.style     ={draw=cnHi, line width=0.7pt, shorten >=1.2pt,
                      -{Triangle[length=1.5mm, width=1.3mm]}},
    trigger/.style  ={draw=black!65, line width=0.6pt,
                      -{Triangle[length=1.4mm, width=1.25mm]}},
    anchorv/.style  ={circle, draw=cnGold, line width=1.1pt, fill=cnFill,
                      inner sep=0pt, minimum size=6.0pt},
    supportv/.style ={circle, draw=cnHi, line width=0.8pt, fill=white,
                      inner sep=0pt, minimum size=5.0pt},
    plainv/.style   ={circle, draw=cnSlate, line width=0.7pt, fill=white,
                      inner sep=0pt, minimum size=5.0pt},
    dagedge/.style  ={draw=cnSlate, line width=0.8pt, shorten >=1.2pt,
                      -{Triangle[length=1.4mm, width=1.2mm]}},
    boxtext/.style  ={anchor=base, inner sep=0pt, font=\scriptsize},
    tag/.style      ={anchor=west, inner sep=1pt, fill=white, font=\scriptsize},
    plabel/.style   ={anchor=east, inner sep=1pt, font=\scriptsize},
    rlabel/.style   ={anchor=base, inner sep=0pt, font=\scriptsize},
    vlabel/.style   ={anchor=east, inner sep=1pt, font=\scriptsize},
    note/.style     ={anchor=base, inner sep=0pt, font=\scriptsize},
  ]

  \fill[black!8] (-0.15,1.62) rectangle (1.85,2.55);
  \draw[encl]    (-0.15,1.62) rectangle (5.15,2.55);
  \draw[divider] (1.85,-0.45) -- (1.85,2.55);
  \draw[divider] (4,-0.45) -- (4,2.55);
  \node[tag] at (0.02,2.55) {TEE enclave};

  \node[boxtext] at (0.85,1.79) {$p \gets \fn{CPRNG}{w}$};
  \node[boxtext] at (0.85,2.17) {\textbf{anchor hidden}};

  \node[boxtext] at (3.00,2.17) {\textbf{anchor revealed}};
  \node[boxtext] at (3.00,1.79) {due to commit of $p_2$};

  \node[boxtext] at (4.55,2.17) {\textbf{anchor revealed}};
  \node[boxtext] at (4.55,1.79) {due to next wave};

  \foreach \x in {0,1,2,3,4,5} \draw[guide] (\x,-0.06) -- (\x,1.44);
  \foreach \y/\k in {1.20/1, 0.70/2, 0.20/3} {
    \draw[lane] (-0.12,\y) -- (5.15,\y);
    \node[plabel] at (-0.20,\y) {$p_\k$};
  }
  \foreach \x/\k in {0.5/1, 1.5/2, 2.5/3, 3.5/4} {
    \node[rlabel] at (\x,-0.32) {$(w,\k)$};
  }
  \node[rlabel] at (4.5,-0.32) {$(w+1,1)$};
  \node[plainv]   (b1) at (0.50,1.20) {};
  \node[anchorv]  (a)  at (0.50,0.70) {};
  \node[plainv]   (b3) at (0.50,0.20) {};
  \node[vlabel] at (0.38,0.70) {$v^1_w$};

  \node[supportv] (s1) at (1.30,0.20) {};
  \node[supportv] (s2) at (1.58,1.20) {};

  \draw[dagedge] (s2) -- (b1);
  \draw[dagedge] (s1) -- (b3);
  \draw[edge]    (s1) -- (a);
  \draw[edge]    (s2) -- (a);

  \draw[trigger] (1.63,1.32) -- (1.85,1.59);

  \node[note] at (2.5,-0.78)
       {$\lceil \frac{n+1}{2} \rceil$ support for $v^1_w$ $\Rightarrow$ commit $\Rightarrow$ $p_2$ revealed};

\end{tikzpicture}}
    \caption{Commit-gated disclosure of the common coin over one wave $w$.
    Each enclave computes the wave's anchor replica from its CPRNG immediately, but \textit{withholds it} until an anchor slot commits or the next wave begins.
    Inserting the $(\lceil \frac{n+1}{2} \rceil)$-th vertex of round $(w,2)$ that supports the anchor slot $v^1_w$ fulfills the commit rule: Now the enclave returns the anchor replica and unveils the coin results of decided waves.}
    \label{fig:coin}
\end{figure}
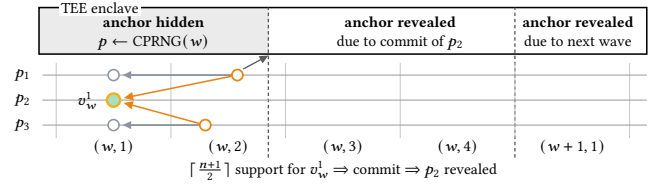

\para{Probing the enclave.}
One might think that since the method $\fn{check\_commit}{v}$ accepts any well-formed vertex, Byzantine replicas could probe their own enclave with candidate vertex sets to learn whether a slot is already committable, thereby trying to narrow down the anchor replica before the protocol commits it.
In the Appendix, we prove in \textbf{Lemma~\ref{lem:hiddencoin}} \textit{Hidden Common Coin} that such a strategy is not useful to break liveness:
The expected number of waves until a correct replica observes the direct commit rule to be fulfilled remains below two.

\subsection{DAG-based Consensus}
\label{sect:design:rbc+consensus}
\algoname{}'s atomic broadcast layer is a DAG-based BAB in the lineage of DAGRider~\cite{keidar_all_2021}, but leverages the \ac{TEE} to re-design DAGRider's commit rules to improve latency.

\para{Committing in DAGRider.}
In DAGRider, a replica \emph{directly commits} the anchor slot $v_w$ of wave $w$ once its local DAG contains a quorum of vertices in $\mathit{round}(w, 4)$ with a \textbf{strong path} to $v_w$.
It then \emph{retrospectively commits} every earlier anchor reachable from \(v_w\) by a \textbf{strong path}, walking back from wave \(w{-}1\) to the last decided wave.
Earlier anchors are ordered first.
By quorum intersection, an anchor that any replica directly committed in an earlier wave has a path to $v_w$ and is ordered before it.
Thus all correct replicas commit the same anchors in the same order.

\para{\algoname{} commit rules.}
\algoname{} keeps DAGRider's wave structure, but commits both \textit{more frequently} and \textit{earlier}, and exploits smaller quorums the \ac{TEE} permits ($f+1$ out of $n=2f+1$ or $\lceil \frac{n+1}{2} \rceil$ in general).
Three things change.
First, all three vertices of the anchor replica in rounds $(w,1)-(w,3)$ are potential anchor slots:
$v^1_{w}, v^2_{w}, v^3_{w}$, so almost every round offers an opportunity to commit, instead of only one anchor slot per wave as in DAGRider.
Second, the \textit{direct commit} rule broadens the support window:
an anchor slot $(w, k)$ is directly committed as soon as a quorum of $\lceil \frac{n+1}{2} \rceil$ vertices of \textit{any} later round of the same wave have a \textbf{path} to it, not only the last round as in DAGRider.
Support is counted round by round until this quorum is first reached.
This improvement makes use of the commit-gated disclosure provided by the TEE to preserve liveness.
Third, support is counted along \textit{all} edges, strong and weak, and propagates transitively:
A vertex supports an anchor slot if one of its edges references the anchor slot itself, or if it references any vertex already supporting it.
Each change makes commitment easier to reach than in DAGRider, while preserving correctness (as we prove in the Appendix).
If a new vertex completes the quorum for several slots of its wave at once, only the highest commits directly;
the lower slots resolve retrospectively as described next.

\para{Retrospective commits and total order.}
As in DAGRider, upon directly committing anchor slot $(w,k)$, the replica revisits all undecided anchor slots below it:
First the lower-round slots of the same wave $w$, from $(w,k{-}1)$ down to $(w,1)$, then the slots of earlier waves, down to the last decided anchor slot.
If an anchor slot is present in the local DAG and reachable over a \textbf{path} from the most recently committed anchor, it is committed as well.
Otherwise, the slot is skipped.
Both outcomes are final because a decided slot is never revisited.
This procedure creates a total order on all reachable anchor slots.
The committed anchors are then delivered in ascending slot order, and each anchor deterministically delivers the vertices of its causal history that no earlier anchor has delivered, so all correct replicas can deterministically linearize all vertices referenced by anchor slots.
Vertices older than the GC offset are ignored during this ordering step and must be handled by the checkpointing scheme instead (\S\,\ref{sect:design:gc}).

\para{Correctness.}
We sketch an intuition for safety and liveness here, while providing full proof in the Appendix.
Two of the above changes could plausibly break safety:
\begin{enumerate*}[label=(\arabic*)]
    \item Commits now originate from three anchor slots per wave instead of one, and
    \item the support is counted along \textit{all} edges of \textit{any} later within-wave round instead of just the wave's last round.
\end{enumerate*}
Neither breaks the property safety relies on:
any two $\lceil \frac{n+1}{2} \rceil$ quorums intersect, and since $\lceil \frac{n+1}{2} \rceil \geq f+1$ replicas support a direct commit of an anchor, every vertex after the support round must indirectly reference the anchor.
Together with the self-reference that the enclave enforces, this yields \textit{anchor connectivity} (Lemma~\ref{lem:anchorconnectivity}):
Once an anchor slot is directly committed, every anchor slot that may be committed later (in this or a future wave) still has a path to it.
Hence correct replicas make identical commit-or-skip decisions during the retrospective walk and deliver the same \textit{total order} (see Theorem~\ref{theorem:totalorder}).

Liveness rests on the coin:
Because the enclave reveals a wave anchor only after an anchor slot commits, an adversary probing check\_commit() cannot prevent the commit and a correct replica observes the direct commit rule fulfilled after fewer than two waves in expectation (Lemma~\ref{lem:hiddencoin}).
\subsection{Client Interaction Model}
\label{sect:design:clienthandler}
\para{Contact Replicas.}
A client sends its request to a single \textit{contact replica}, which buffers it and includes it in the payload of its next vertex ($v.block$).
However, a crashed contact replica might never propose the request and a Byzantine one could censor it.
In both cases, the client receives no timely reply and re-issues its request to the next replica of its \textit{preference list} of all \(n \geq 2f+1\) replicas (e.g., based on geographical proximity/latency~\cite{archer_2018_eischer}).
As at most $f$ replicas are faulty, each faulty contact replica can delay a request, but not suppress it, so the order of the list is a latency choice rather than a correctness one and can be dynamically adjusted.

Since a replica cannot tell such a failover from a spamming client (\S\,\ref{sect:problems:client}), replica $p_i$ applies a \textit{round-based cooldown}:
it rejects a request of client $c$ while a vertex of another replica from the last $\kappa$ rounds carries a not-yet-ordered request of $c$.
For censored clients, no such vertex exists, since their request was never included, and $\kappa$ ensures that a vertex which is never ordered cannot block $c$ forever.
Multi-inclusion nonetheless remains briefly possible, since concurrently proposed vertices may not yet reference each other;
but as each round requires a quorum of vertices from the previous one, later vertices are quickly forced to (indirectly) reference every vertex already carrying the request, closing the window within a few rounds.
Selecting $\kappa$ thus trades off this residual multi-inclusion against how quickly a censored client can rejoin.
Since $\kappa$ counts logical \ac{DAG} rounds rather than time, it further needs no network-specific tuning.

\para{Execution and response storage.}
After committing vertices from the DAG, replicas forward the requests, included as payload, to their local state machine in \textit{total order} (see §\ref{sect:design:rbc+consensus}) for execution.
Each request includes an identifier $rid$ that is unique per client, and the execution manager discards already executed requests.
This way, \algoname{} provides \textit{exactly-once} semantics, despite re-submission attempts resulting from the client-failover mechanism.
Replicas hold the most recent response (and associated proof) per client in a \textit{response storage} (Alg.~\ref{alg:porcupinesmr}, l.~4), which is queried to answer re-submitted requests.
The response storage becomes part of the replica state, so it must be included into checkpoints (Alg.~\ref{alg:porcupineab}, l.~40) so that a replica which recovered through a state transfer can still serve re-issued requests.

\begin{figure}[t]
    \centering
    \resizebox{\columnwidth}{!}{\providecolor{pfGold}  {RGB}{240,176, 32}\providecolor{pfSlate} {RGB}{140,148,160}\providecolor{pfB1}    {RGB}{ 68,114,196}\providecolor{pfB1f}   {RGB}{203,217,240}\providecolor{pfB2}    {RGB}{112,173, 71}\providecolor{pfB2f}   {RGB}{211,232,198}\providecolor{pfB3}    {RGB}{199, 48, 62}\providecolor{pfB3f}   {RGB}{247,208,212}\providecolor{pfHi}    {RGB}{232,132, 20}\providecolor{pfHif}   {RGB}{250,222,193}\par
\begin{tikzpicture}[
    x=1.2969cm, y=1.2969cm,
    font=\footnotesize,
    lane/.style     ={draw=black!40, line width=0.4pt},
    delivedge/.style={draw=pfGold, line width=0.9pt, shorten >=2.4pt,
                      -{Triangle[length=1.5mm, width=1.3mm]}},
    plainedge/.style={draw=black!30, line width=0.5pt, shorten >=2.2pt,
                      -{Triangle[length=1.1mm, width=0.95mm]}},
    vtx/.style      ={circle, draw=pfSlate, line width=0.6pt, fill=white,
                      inner sep=0pt, minimum size=7.6pt, font=\tiny},
    anchorring/.style={line width=1.0pt, draw=pfGold, minimum size=9.8pt},
    hnode/.style    ={rounded corners=1pt, draw=pfSlate, line width=0.5pt,
                      fill=white, inner sep=0pt, font=\tiny,
                      minimum width=0.38cm, minimum height=0.26cm},
    tedge/.style    ={draw=pfSlate, line width=0.5pt},
    pedge/.style    ={draw=pfHi, line width=1.0pt},
    sig/.style      ={rounded corners=1.5pt, draw=black, line width=0.6pt,
                      fill=black!6, inner sep=1.5pt, font=\tiny},
    plabel/.style   ={anchor=east, inner sep=1pt, font=\scriptsize},
    rlabel/.style   ={anchor=base, inner sep=0pt, font=\scriptsize},
    vlabel/.style   ={anchor=base, inner sep=0pt, font=\tiny},
    note/.style     ={anchor=base, inner sep=0pt, font=\scriptsize},
    panelname/.style={anchor=base, inner sep=0pt, font=\scriptsize},
  ]

  \foreach \y/\k in {2.55/1, 1.95/2, 1.35/3} {
    \draw[lane] (0.62,\y) -- (3.30,\y);
    \node[plabel] at (0.56,\y) {$p_\k$};
  }
  \foreach \x/\k in {0.85/1, 1.95/2, 3.05/3}
    \node[rlabel] at (\x,0.92) {$(w,\k)$};

  \node[vtx, draw=pfB2, fill=pfB2f] (a1) at (0.85,2.55) {2};
  \node[vtx, draw=pfB1, fill=pfB1f] (v1) at (0.85,1.95) {1};
  \node[vtx, draw=pfB3, fill=pfB3f] (a3) at (0.85,1.35) {3};
  \node[vtx, draw=pfB3, fill=pfB3f] (b1) at (1.95,2.55) {3};
  \node[vtx, draw=pfB2, fill=pfB2f] (v2) at (1.95,1.95) {2};
  \node[vtx]                        (b3) at (1.95,1.35) {};
  \node[vtx]                        (c1) at (3.05,2.55) {};
  \node[vtx, draw=pfB3, fill=pfB3f] (v3) at (3.05,1.95) {3};
  \node[vtx]                        (c3) at (3.05,1.35) {};

  \foreach \n in {v1,v2,v3} \node[circle, anchorring] at (\n) {};
  \node[vlabel] at (0.85,1.63) {$v^1_w$};
  \node[vlabel] at (1.95,1.63) {$v^2_w$};
  \node[vlabel] at (3.05,1.63) {$v^3_w$};

  \draw[plainedge] (b3) -- (a3);   \draw[plainedge] (b3) -- (v1);
  \draw[plainedge] (c1) -- (b1);   \draw[plainedge] (c1) -- (v2);
  \draw[plainedge] (c3) -- (b3);   \draw[plainedge] (c3) -- (v2);

  \draw[delivedge] (v2) -- (v1);
  \draw[delivedge] (v2) -- (a1);
  \draw[delivedge] (v3) -- (v2);
  \draw[delivedge] (v3) -- (b1);
  \draw[delivedge] (b1) -- (a1);
  \draw[delivedge] (b1) -- (a3);

  \node[panelname] at (1.86,0.48) {(a) batches 2 anchor slots};

  \node[hnode, draw=pfB3, fill=pfHif, text=black] (h1) at (4.45,1.35) {$h_1$};
  \node[hnode, draw=pfHi, line width=0.9pt, fill=pfHi, text=white]
                                                  (h2) at (5.15,1.35) {$h_2$};
  \node[hnode, draw=pfB3] (h3) at (5.85,1.35) {$h_3$};
  \node[hnode, draw=pfB3] (h4) at (6.55,1.35) {$h_4$};

  \node[hnode, draw=pfSlate]                  (h12) at (4.80,1.75) {$h_{12}$};
  \node[hnode, draw=pfSlate, fill=pfHif]      (h34) at (6.20,1.75) {$h_{34}$};
  \node[hnode, draw=pfSlate, minimum width=0.44cm] (hr) at (5.50,2.15) {$h_r$};

  \draw[tedge] (h3) -- (h34);   \draw[tedge] (h4) -- (h34);
  \draw[tedge] (h34) -- (hr);   \draw[tedge] (h1) -- (h12);
  \draw[pedge] (h2) -- (h12);   \draw[pedge] (h12) -- (hr);

  \node[sig] (sg) at (5.50,2.55) {$\sigma$ over $h_r$, $f{+}1$ shares};
  \draw[tedge, -{Triangle[length=1.2mm, width=1.0mm]}] (hr) -- (sg);

  \node[note] at (5.50,0.92) {proof: $\sigma$, $h_1$, $h_{34}$};
  \node[panelname] at (5.50,0.48) {(b) proof from $v^3_w$'s batch};

\end{tikzpicture}}
    \caption{From DAG commits to execution proofs, for $n=3$, $f=1$ and anchor replica $p_2$.
    (a)~Anchor slots carry a gold ring: Every anchor delivers the part of its causal history that no earlier anchor slot committed, so batches are disjoint.
    A vertex carries the number and color of the anchor that committed it. White means uncommitted.
    Edges along which vertices commit are gold and the remaining ones grey. Edges into later rounds are omitted.
    (b)~The $m$ requests of $v^3_w$'s batch form the leaves $h_i = \textsf{hash}(\var{c}, \var{rid}_i, \var{request}_i, \var{response}_i)$ of a Merkle tree whose root is certified by a threshold signature $\sigma$.
    The client receives a response, and a \textit{proof} ($\sigma$ and the shaded siblings on its path to the root, $\log_2 m = 2$ hashes).}
    \label{fig:proofs}
\end{figure}
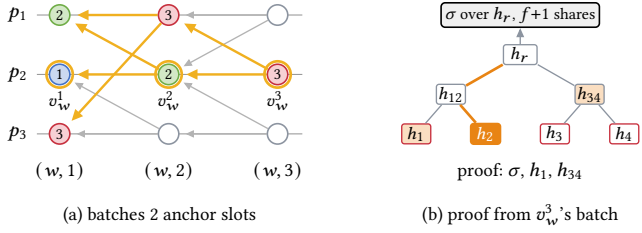

\para{Execution proofs.}
When an anchor slot commits, it delivers the part of its causal history that no earlier anchor slot has delivered in a deterministic order (Alg.~\ref{alg:porcupineab}, ll.~33--35), shown in \cref{fig:proofs}(a).
\algoname{} batches exactly this set (all vertices committed by the same anchor slot) for proof generation.
This means that up to three batches are delivered per wave.
After executing a batch, the \textit{client handler} constructs a Merkle tree~\cite{merkle_signatures_1989} whose leaves are the hashes of the tuples $(\var{c}, \var{rid}, \var{request}, \var{response})$, and certifies its root with the threshold scheme of \S\,\ref{sect:model:thresholdsig}:
\begin{enumerate*}[label=(\arabic*)]
    \item Each replica creates a signature share,
    \item and distributes it in a single all-to-all broadcast, subsequently
    \item the enclave can use $t=f{+}1$ valid shares to create a threshold signature over the root, which can be verified using a single common public key.
\end{enumerate*}

\para{Verification.}
The proof a client receives comprises the threshold signature and the sibling hashes on the path from its leaf to the root of the Merkle tree $(\sigma, h_i, \ldots{}, h_j)$ (\cref{fig:proofs}(b)).
A client verifies the proof in three steps:
\begin{enumerate*}[label=(\arabic*)]
    \item It employs the received response to hash its own tuple $(\var{c}, \var{rid}, \var{request}, \var{response})$ into a leaf, then
    \item re-computes the path in the Merkle tree from the leaf up to the root using the received sibling hashes (in the order given by the position in the batch), and
    \item checks $\sigma$ against the computed root with the common public key.
\end{enumerate*}
If the client's verification succeeds, the result is guaranteed to be correct, as \textit{at least one} out of the $f{+}1$ replicas that contributed to the threshold signature has to be benign.

\para{Why no broadcast is needed.}
In \algoname{}, a client whose contact replica is faulty never needs to broadcast its request, but can simply retry at another contact replica.
If the re-issued request was already ordered (and executed), it is served from the response storage.
Otherwise, the client keeps retrying until the cooldown passes, at which point any replica can simply propose it, since \textit{every replica} can propose vertices in the DAG.
Because all correct replicas execute identical batches, build the same Merkle trees, and collect signature shares over the same results, any of them can serve the client a valid response and proof.

\subsection{Garbage Collection}
\label{sect:design:gc}
DAG-based protocols produce two data structures that may grow without bound:
the \textit{DAG} itself and the \textit{vertex buffer}.

\para{Bounding the DAG with a GC barrier.}
To restrict DAG memory usage, vertices that could not be ordered before a configured number of rounds must be deleted.
However, agreement requires that if one correct replica deletes a vertex instead of ordering it, every other correct replica ignores it during ordering as well.
Since \ac{BAB} inherently ensures that commits occur in the same order at every correct replica, one can enforce a GC offset $g$ that moves with each commit, ensuring that it has the same effect on every replica (Theorem \ref{the:abagreement}).
More precisely, after a commit (direct or retrospective) in wave $w$ the GC barrier is increased to wave $w-g$, which determines that every unordered vertex before wave $w-g$ will never be ordered and hence can be ignored (Alg.~\ref{alg:porcupineab}, l.~33).

\para{Backfilling without external storage.}
When a vertex is ignored or delivered, the content of the vertex is no longer required by the AB-layer of the replica that already decided the vertex's fate.
Furthermore, references to the vertex can be ignored and are therefore irrelevant (Alg.~\ref{alg:porcupineab}, l.~11).
However, the vertex's content is still required by the RB-layer to achieve a memory-bounded and performant request dissemination.
Such a reliable broadcast requires replicas to help and recover \textbf{other} replicas, a mechanism we refer to as \textit{backfilling}.
For instance, a replica that delivered a vertex $v$ must still be able to provide the vertex to a replica that has fallen behind, such that this replica can also deliver $v$.
While Narwhal~\cite{danezis_narwhal_2022} can still access garbage collected vertices from the external storage for backfilling purposes, \algoname{} leverages the overlying SMR layer to build consecutive \textit{checkpoints} every $I_{\text{ckpt}}^{th}$ wave, which are certified snapshots of the RSM state that are signed by $f+1$ replicas.
A state transfer to a certified checkpoint can replace the answer to a backfilling request for a vertex that was already deleted (Alg.~\ref{alg:porcupinerb}, l.~10).
Consequently, to answer backfilling requests, a vertex $v$ is not deleted already when it is ignored or delivered, but only when a checkpoint covering the ignorance or delivery of the vertex gets stable.
This stable checkpoint therefore must be at least in wave $v.w + g$ such that the checkpoint proves that the GC barrier already caught up to the vertex, which implies that either the vertex transactions are applied to the state of the stable checkpoint, or the vertex is ignored.
Both options render the vertex's content irrelevant to the requester once the state transfer is complete (see Lemma~\ref{lem:backfillingresponse}).

\para{Checkpoint stabilization and active waiting.}
If a replica made more than $g$ waves of progress without stabilizing a checkpoint, it must `carefully' wait for validation signatures from correct replicas, as otherwise, faulty replicas could prevent all checkpoint stabilization (Alg.~\ref{alg:porcupineab}, ll. 10--14).
These signatures eventually arrive, because a replica continues to check for commits whenever a new vertex is added to the DAG (active waiting).
Hence, commits and checkpoint validation can occur even during active waiting, ensuring that the procedure cannot lead to deadlocks.
Moreover, if we set $g$ proportional to the expected latency ratio between the $f+1$ fastest replicas and the slowest, replica progress differs by less than $g$ waves and active waiting never takes place (see Theorem \ref{the:checkpointstabilization} and Lemma~\ref{lem:progressdifference}).

\para{Bounding the vertex buffer.}
\algoname{} bounds the memory consumption of the vertex buffer by limiting it to vertices that lie at most $g$ waves in the future, creating a bounded receiving window (Alg.~\ref{alg:porcupinerb}, ll.~4--5).
If a vertex does not fit into the receiving window, it is discarded and backfilled by other replicas later, which ensures the agreement of the RB-layer (Alg.~\ref{alg:porcupinerb}, ll.~6--7).
The receiving window algorithm is sufficient (cf.~Lemma~\ref{lem:slidingwindowsufficiency}) and does not affect \algoname{}'s performance, since replica progress is at most $g$ waves apart.

\para{Resulting memory bound.}
With GC and checkpointing, \algoname{} consumes a fixed amount of memory even during asynchrony, on the order of $O((n|v| + \frac{|S|}{I_{ckpt}})(g+I_{ckpt}))$ where $|v|$ and $|S|$ denote the sizes of a vertex and a checkpoint, respectively (for the derivation and constraints on the choice of $g$ and $I_{ckpt}$ see Prop.~\ref{obs:checkpointstabilization}).
Regarding timely fairness~\cite{spiegelman_bullshark_2022}:
When $g$ is set proportionally to the relative latencies of replicas that hold during periods of synchrony, then during those periods vertices are neither ignored nor out of the receiving window, and no active waiting for checkpoint stabilization is required (see Prop. \ref{lem:progressdifference}, \ref{the:timelyfairness}, \ref{obs:checkpointstabilization}).

\para{\algoname{} without \ac{TEE}s.}
Two of the three main contributions we make 
do not depend on the availability of TEEs:
\begin{itemize*}
    \item[\textbf{(1)}]The efficient client interaction layer which enables our DAG protocol to maintain high throughput (\S~\ref{sect:design:clienthandler})
      \item[\textbf{(2)}] Our garbage collection scheme that bounds memory consumption under asynchrony (\S\,\ref{sect:design:gc}).
\end{itemize*}
Latency gains and ($n \geq 2f+1$)-resilience rely on the TEE assumption.
In Appendix~§~\ref{app:sect:porcupine-without-tees} we explain how to realize \algoname{} without \acp{TEE}.

\section{Evaluation}
\label{sec:evaluation}

We implement \algoname{} in Rust on top of ABCperf~\cite{spannagel_abcperf_2023}, using Tokio for asynchronous networking, TCP and TLS for communication, Intel SGX with the Teaclave SDK~\cite{teaclave_repo} for the \ac{TEE}, and SpliTEE common-key threshold signatures~\cite{haller2026splitee} for the client mechanism.
\algoname{}'s trusted computing base consists of 238 LOC and the protocol-agnostic client interaction layer of 505 LOC.

\subsection{Setup and Claims}
\label{sec:eval:setup}

We evaluate two claims:
\begin{enumerate*}[label=\textbf{(C\arabic*)}, ref=\textbf{C\arabic*}]
  \item\label{claim:client} a naive \ac{BFT} client interaction significantly lowers the maximum achievable throughput of a \ac{DAG} protocol, whereas our single-contact threshold-signature client interaction sustains near-baseline throughput with full \ac{BFT} \ac{SMR} guarantees
  \item\label{claim:speed} \algoname{} commits substantially faster than Mysticeti~\cite{babel_2025_mysticeti}, the fastest \ac{DAG} \ac{BFT} consensus protocol we are aware of, with and without faults.
\end{enumerate*}

\para{Setup.}
We use 20 machines with 64\,GB of memory and 10\,Gbps links running Ubuntu 22.04, ten of them with an Intel Xeon E-2288G and ten with an Intel Xeon E-2388G CPU.
We emulate the geo-distributed AWS deployment of Mysticeti~\cite{babel_2025_mysticeti_preprint} with \texttt{netem}~\cite{netem} across the same 13 regions~\cite{awsregions}, halving the $p_{50}$ round-trip times reported by Cloudping~\cite{cloudping} to obtain one-way delays between 6\,ms and 140\,ms (full config in \S\,\ref{sec:evalapp:deployment}).
Replicas are spread evenly across regions, with at most one replica per machine for $n \leq 20$ and as even a colocation as possible beyond that.
Clients run on 4 additional machines, without additional emulated latency.
Requests are 64\,B.
Each configuration runs 4 times for 5\,min and we exclude 30\,s of warm-up and cooldown from each run.

\para{Baselines.}
To demonstrate generality of our client interaction layer and evaluate it with our implementation, we re-implement Narwhal/Tusk~\cite{danezis_narwhal_2022} in ABCperf.
For the commit speed comparison, we use Mysticeti's open-source implementation~\cite{mysticeti_repo} with the default configuration, which designates two proposer slots per round.
We test with equal $n$ and varying $f$, i.e. \algoname{} tolerates $f \leq \lfloor\frac{n-1}{2}\rfloor$ faults while both baselines tolerate only $f \leq \lfloor\frac{n-1}{3}\rfloor$.

\subsection{Client Interaction}
\label{sec:eval:client}

\begin{figure}[t]
  \centering
  \begin{tikzpicture}
    \begin{groupplot}[
        group style={
            group size=2 by 1,
            horizontal sep=0.4cm,
            ylabels at=edge left,
            yticklabels at=edge left,
        },
        width=4.3cm,
        height=4cm,
        ybar,
        /pgf/bar width=6pt,
        enlarge x limits=0.45,
        ymin=0,
        ymax=130,
        enlarge y limits=false,
        ylabel={Throughput [kOp/s]},
        symbolic x coords={20,40},
        xtick=data,
        xticklabels={$n{=}20$, $n{=}40$},
        ytick distance=30,
        minor y tick num=1,
        ymajorgrids=true,
        yminorgrids=true,
        major grid style={dashed, gray!40},
        minor grid style={dashed, gray!35, line width=0.3pt},
        tick label style={font=\footnotesize},
        label style={font=\footnotesize},
        title style={font=\footnotesize, yshift=-3pt},
    ]

    \nextgroupplot[
        title={\algoname{}},
        legend style={
            font=\footnotesize,
            at={(1.05,-0.17)}, anchor=north,
            legend columns=3,
            draw=none, fill=none,
            /tikz/every even column/.append style={column sep=6pt},
        },
    ]

    \addplot[
        fill=blue!10,
        draw=black,
        postaction={
            pattern=north east lines,
            pattern color=blue!70,
        },
    ]
    coordinates {
        (20,122)
        (40,72.5)
    };

    \addplot[
        fill=black!8,
        draw=black,
        postaction={
            pattern=crosshatch,
            pattern color=black!60,
        },
    ]
    coordinates {
        (20,15)
        (40,5)
    };

    \addplot[
        fill=vworange!15,
        draw=black,
        postaction={
            pattern=north west lines,
            pattern color=vworange,
        },
    ]
    coordinates {
        (20,115)
        (40,70)
    };

    \legend{No Client Guarantee (Baseline), Classical Client, Ours}
    
    \nextgroupplot[
        title={Narwhal/Tusk},
    ]

    \addplot[
        fill=blue!10,
        draw=black,
        postaction={
            pattern=north east lines,
            pattern color=blue!70,
        },
    ]
    coordinates {
        (20,120)
        (40,72.5)
    };

    \addplot[
        fill=black!8,
        draw=black,
        postaction={
            pattern=crosshatch,
            pattern color=black!60,
        },
    ]
    coordinates {
        (20,13)
        (40,5)
    };

    \addplot[
        fill=vworange!15,
        draw=black,
        postaction={
            pattern=north west lines,
            pattern color=vworange,
        },
    ]
    coordinates {
        (20,110)
        (40,70)
    };

    \end{groupplot}
\end{tikzpicture}
  \caption{Maximum sustained throughput of \algoname{} and Narwhal/Tusk for the three client mechanisms.}
  \label{fig:client}
\end{figure}
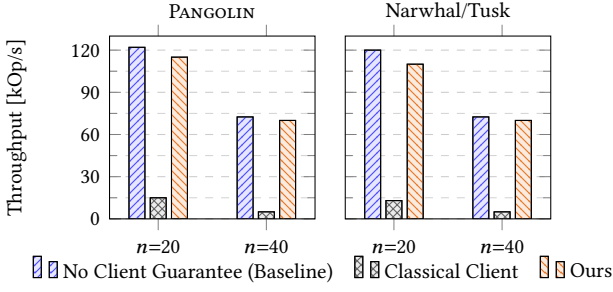

\begin{table}[t]
  \centering
  \small
  \caption{\(p_{50}\) end-to-end latency [ms] at $n=20$ for the three client mechanisms of \cref{fig:client}. $\pm$ intervals are t-based 95\% confidence intervals across runs.}
  \label{tab:client}
  \begin{tabular}{lrr}
    \toprule
    Client & \algoname{} & Narwhal/Tusk \\
    \midrule
    No Client Guarantee (Baseline)        & 349.29\(\pm\)61.69 & 1189.88\(\pm\)8.37\\
    Classical Client    & 272.62\(\pm\)59.77 & 1063.83\(\pm\)27.59 \\
    Ours            & 394.01\(\pm\)31.97 & 1260.78\(\pm\)22.16 \\
    \bottomrule
  \end{tabular}
\end{table}

We compare three client mechanisms over \algoname{} and Narwhal/Tusk at committee sizes \(n \in \{20, 40\}\).
Classically, clients submit requests to all \(n\) replicas and wait for \(f+1\) matching replies.
For each request, \(f+1\) of \(n\) replicas are deterministically designated responsible, the minimum needed for at least one to be honest.
Replicas include any request they are responsible for, since they cannot tell whether a client is recovering from censorship or spamming.
Our mechanism lets a replica decide this using the round-based cooldown of \S\,\ref{sect:design:clienthandler}, so only a single replica needs to include the request;
a single threshold-signed reply then suffices as proof, and the client contacts only that replica and waits for its reply.
As a reference for the maximum achievable throughput, we add a client that submits to a single replica and accepts a single reply without any client-side guarantee.
Replicas propose transactions as soon as they transition rounds, to exclude batching effects.
Throughput and end-to-end latency are both measured at the clients once they accept a reply.
Requests proposed by multiple replicas are counted once.

\Cref{fig:client} reports the highest load at which the system keeps pace with the offered request rate (within 5\%) while median end-to-end latency stays within $1.5\times$ its minimum, both required at that load and at every lower one.
With the naive client, both systems sustain 7--12\%{} of their baseline throughput:
15\,kOp/s against 122\,kOp/s for \algoname{} and 13\,kOp/s against 120\,kOp/s for Narwhal/Tusk at $n=20$, and 5\,kOp/s against 72.5\,kOp/s for both at $n=40$.
Our client instead reaches 92--97\% of the baseline (115\,kOp/s and 110\,kOp/s at $n=20$, 70\,kOp/s for both at $n=40$).
This confirms \ref{claim:client}.
\Cref{tab:client} reports the corresponding end-to-end latencies for $n=20$.
The threshold client is 45\,ms slower than the baseline for \algoname{} and 71\,ms slower for Narwhal/Tusk, reflecting the cost of the signature share exchange and the Merkle tree generation.
With clients directly connected at near-zero latency, this signature exchange dominates the \(f+1\) replica-client paths.
Waiting for the fastest \(f+1\) of \(n\) replies beats not only the threshold client, by 121\,ms (\algoname{}) and 197\,ms (Narwhal/Tusk), but also the baseline.

\subsection{Commit Latency}
\label{sec:eval:latency}

In this experiment, we compare \emph{commit latency}, the time between a request being received and it being committed by the replicas.
We do not compare end-to-end latency, since the open-source Mysticeti implementation has no real clients:
load is generated synthetically at the replica, so there is no client path.
We also do not compare throughput, which is not meaningful across codebases given differences in optimizations and implementation choices.
Each replica calculates the mean and percentiles of commit latency across the transactions it commits, which are then averaged across replicas to yield one of each statistic per configuration.
Following Mysticeti, we run both systems at 12\,500 requests/s, below either system's maximum and roughly ten times the combined rate of current real-world blockchains~\cite{babel_2025_mysticeti}.
We evaluate committee sizes $n \in \{10, 40, 50\}$.

\begin{figure}[t]
  \centering
  \input{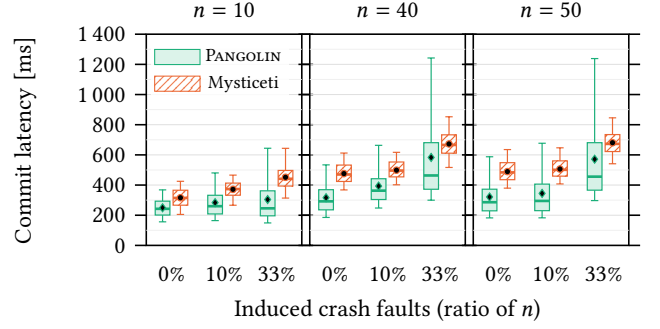}
  \caption{Commit latency [ms] at 12\,500 requests/s. Boxes span
  $p_{25}$--$p_{75}$, whiskers $p_{5}$--$p_{95}$; line is $p_{50}$,
  marker is the mean.
  Crashed replicas: $0$, $10\%\cdot{}n$ ($1$/$4$/$5$) and $33\%\cdot{}n$
  ($3$/$13$/$16$)}
  \label{fig:commit-latency}
\end{figure}

\para{Without faults.}
\algoname{}'s mean commit latency is 249\,ms at $n=10$, 316\,ms at $n=40$ and 321\,ms at $n=50$, against 316\,ms, 476\,ms and 489\,ms for Mysticeti, a speedup of $1.27\times$, $1.51\times$ and $1.52\times$ (\cref{fig:commit-latency}).
We attribute the difference to three mechanisms:
the single-step \ac{TEE}-based reliable broadcast (\cref{fig:broadcast}), the commit rules of \cref{sect:design:rbc+consensus}, which let an anchor slot commit one round after it was proposed, and the smaller $\lceil \frac{n{+}1}{2} \rceil$ quorum.
At $p_{95}$ the margin is $1.15\times$, $1.15\times$ and $1.08\times$.

\para{Under crash faults.}
We evaluate two failure scenarios, crashing $10\,\%$ and $33\,\%$ of the replicas in both systems before the measurement window opens, so the measurement covers the faulty steady state; $33\,\%$ is $f = \lfloor{}\frac{n-1}{3}\rfloor{}$, the most Mysticeti tolerates.
The set of crashed replicas is selected at random, but the same for both algorithms.
\algoname{} keeps its mean advantage at every point: $1.31\times$, $1.27\times$ and $1.47\times$ with $10\,\%$ crashed and $1.49\times$, $1.15\times$ and $1.19\times$ with $33\,\%$ crashed, at $n=10$, $40$ and $50$.
With a significant speedup both with and without faults, we can confirm \ref{claim:speed}.
With $33\,\%$ crashed, \algoname{}'s $p_{95}$ reaches 1242\,ms at $n=40$ and 1238\,ms at $n=50$, against 853\,ms and 846\,ms for Mysticeti.
\algoname{} elects its anchors by coin toss, and with fewer correct replicas the coin is more likely to land on a crashed one, causing commits to be postponed to the next wave, whereas Mysticeti's fixed leader schedule avoids this at the cost of a timing assumption.

\subsection{Generality across Deployments}
\label{sec:eval:generality}

\begin{figure}[t]
  \centering
  \includegraphics[width=\columnwidth]{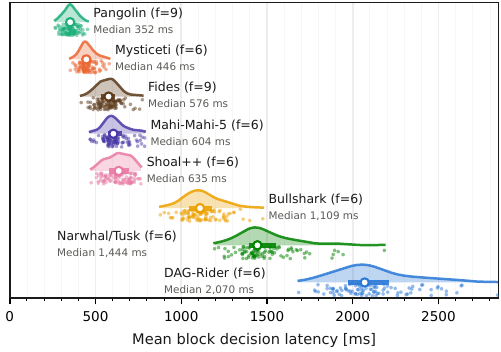}
  \caption{Distribution of mean block decision latency across 100 deployments ($n=20$, $600$\,s each). For each protocol, the curve shows the distribution over deployments; each dot marks one deployment's mean latency. The bar spans the interquartile range, and the white circle marks the median.}
  \label{fig:sim-violin}
\end{figure}

The metrics of \cref{sec:eval:client,sec:eval:latency} assume a single deployment setting, using $13$ hand-selected AWS regions (the regions used by Mysticeti's evaluation~\cite{babel_2025_mysticeti_preprint}).
To show that \algoname{}'s latency improvements do not depend on this concrete deployment scenario, we use a discrete-event network simulator to evaluate \algoname{} across \textbf{100 independently sampled deployments}, and extend our evaluation to more DAG protocols.
The simulator abstracts cryptography and models the network using replica-replica one-way delays, so it reproduces the protocol structure rather than an implementation;
in particular, it does not model transaction submission or queueing at a replica.
We therefore report \emph{block decision latency}, the time from a block being proposed to it being committed, rather than the commit latency of \cref{sec:eval:latency}.
Commit latency additionally includes the time a transaction spends queued until it is included in a vertex.
Block decision latency isolates the protocol's message-passing cost, which dominates commit latency in a real WAN.
Appendix~\ref{sec:evalapp:simulation} details the simulator for the interested reader.

\para{Setup.}
We simulate \algoname{} against seven baselines:
DAGRider~\cite{keidar_all_2021}, Narwhal/Tusk~\cite{danezis_narwhal_2022}, Bullshark~\cite{spiegelman_bullshark_2022}, Shoal++~\cite{arun_2025_shoalpp}, Mahi-Mahi-5~\cite{jovanovic2025mahi}, Mysticeti~\cite{babel_2025_mysticeti} and the TEE-based Fides~\cite{fides_2026_arxiv}.
A deployment sampler creates 100 deployments, reused across all protocols.
Each deployment places \(n = 20\) replicas in regions drawn independently, and with replacement, from a 34-region AWS latency matrix (see \cref{fig:heatmap}), so a deployment may co-locate replicas.
Runs are fault-free and cover $600$\,s of simulated time each.
Each replica computes the mean, $p_{50}$, $p_{90}$ and $p_{99}$ block decision latency over the blocks it commits.
We average each statistic separately across replicas, yielding one deployment-level mean, $p_{50}$, $p_{90}$ and $p_{99}$.

\para{Observations.}
\Cref{fig:sim-violin} shows, for each protocol, the distribution across deployments of this deployment-level mean.
In a direct, paired comparison, \algoname{} has lower block decision latency than every baseline in \textit{all 100 deployments}.
Taking the median of the deployment-level means, \algoname{} decides blocks in $352$\,ms against $446$\,ms for Mysticeti, $576$\,ms for Fides, $604$\,ms for Mahi-Mahi-5, $635$\,ms for Shoal++, $1{,}109$\,ms for Bullshark, $1{,}444$\,ms for Narwhal/Tusk and $2{,}070$\,ms for DAGRider,
which means speedups of $1.27\times$, $1.64\times$, $1.72\times$, $1.80\times$, $3.15\times$, $4.10\times$ and $5.88\times$, further confirming \ref{claim:speed}.
The speedup ratio depends considerably on the concrete deployment chosen:
for \algoname{} against Mysticeti, the speedup never drops below $1.16\times$ but can also reach $1.51\times$.
The speedup also narrows in the tail ($p_{90}$: $1.12\times$, $p_{99}$: $1.02\times$ for \algoname{} vs. Mysticeti), consistent with our finding in \cref{sec:eval:latency}, that attributes tail latency to the randomized leader election necessary in the asynchronous setting.

Compared to Fides~\cite{fides_2026_arxiv}, another asynchronous TEE-DAG protocol with \(2f + 1\) replicas and \(f + 1\) previous-round references, \algoname{} achieves an average speedup of \(1.64\times\) (\(p_{90}\): \(1.58\times\), \(p_{99}\): \(1.76\times\)).
Fides' trusted coin releases one first-round anchor per four-round wave upon presentation of a fourth-round certificate.
Direct commitment requires support from \(f+1\) fourth-round vertices.
Both protocols share the same reduced quorum size, but \algoname{}'s three anchor slots per wave and relaxed commit rules yield far more commitment opportunities.
Appendix~\ref{sec:evalapp:simulation} additionally reports paired comparisons and cumulative latency distributions.

\section{Related Work}
\label{sec:eval:rw}

\para{Scalability.}
SBFT~\cite{gueta2019sbft} also uses Merkle proofs and threshold signatures to reduce client communication to a single reply in the common case, although here the application state is used to construct the proof, rather than a per-batch tree of responses.
In SBFT, a replica sends its signature share only to a small designated group of replicas called \textit{collectors}, which combine the shares into the threshold signature before distributing it but if no collector's reply arrives in time, then the client must fall back to broadcasting its request to all replicas and collecting $f{+}1$ responses.
Mir-BFT and ISS~\cite{stathakopoulou2022state, stathakopoulou2022mirbft} parallelize the proposing of batches of requests among several leaders.
Raptr~\cite{tonkikh_2025_raptr} combines votes for replicas' locally available proposal prefixes to enable commitment without retrieving the entire proposal.
Autobahn~\cite{giridharan_autobahn_2024} disseminates batches along per-replica chains and orders their prefixes through compact proposals containing certified chain tips.
Raptr and Autobahn scale leader-driven ordering under partial synchrony, whereas \algoname{} uses \acp{TEE} to order DAG causal histories with randomized asynchronous progress. 
Shrestha and Kate~\cite{shrestha_2026_dag_bft_scalability} use sampled committees to reduce payload replication and partition transaction execution in a DAG protocol.
This reduces dissemination and execution work, complementary to \algoname{}'s faster ordering.
Finally, the Chop Chop~\cite{camaioni_2024_chop_chop} architecture uses untrusted brokers to amortize input authentication and ordering.
Its delivery certificates concern broadcast completion, whereas \algoname{} certifies execution results and does not necessitate brokers.

\para{Progress and resource management.}
Aardvark~\cite{clement_2009_aardvark} isolates communication resources and schedules request processing to prevent Byzantine replicas from starving useful work.
\algoname{} addresses a related problem of reclaiming DAG history while preserving recovery for lagging replicas.
DispersedLedger~\cite{yang_2022_dispersedledger} lets asynchronous agreement advance ahead of payload retrieval and discusses suspending proposals when retrieval falls too far behind.
This is similar to \algoname{}'s use of progress-dependent throttling, but \algoname{} binds throttling to checkpoint certification and uses certified snapshots to recover replicas that require history.

\para{Trusted components.}
MinBFT~\cite{veronese_2013_minbft} introduced a trusted counter to prevent equivocation and tolerate $f$ faults with only $2f+1$ replicas.
CheapBFT~\cite{kapitza2012cheapbft} further reduced normal-case work by only requiring $f+1$ active replicas to agree on and execute requests while $f$ passive replicas receive state updates.
Damysus~\cite{decouchant2022damysus} extends HotStuff~\cite{hotstuff19} with trusted services, requires $2f+1$ replicas, saves one protocol phase compared to HotStuff and also assumes partial synchrony.
OneShot~\cite{decouchant2024oneshot} improves on Damysus by shortening it by another protocol phase in the normal case while needing a smaller trusted computing base. Its key idea is to adapt each view to what its leader knows about previously prepared blocks, which allows OneShot to decide in just four instead of six communication steps in the normal case.
Furthermore, the leaderless set Byzantine consensus protocol of RedBelly~\cite{crain2021redbelly} was adapted with trusted components to $n \geq 2f{+}1$~\cite{zhao2024hybridset}. This approach assumes partial synchrony and employs a trusted counter and a trusted collector service: Each replica broadcasts a proposal through a two-step hybrid verified reliable broadcast, after which one binary agreement per proposal decides if the proposal can be included.

In contrast, \algoname{} orders all proposals through a single DAG without per-proposal agreement and remains live under asynchrony.
\algoname{} uses the TEE-based reliable broadcast from~\cite[\S\,4.2]{leinweber_2026_tee_ledgers} as a building block to construct a fast DAG-based SMR protocol that operates under asynchrony:
The trusted non-equivocation allows fast vertex certification, while backfilling retrieves missing causal dependencies.

Fides~\cite{fides_2026_arxiv} is a close asynchronous TEE-assisted DAG protocol with $2f+1$ replicas and $f+1$ previous-round references.
\algoname{} provides three anchor slots per wave, more support opportunities, and checks commitment inside the enclave, revealing an anchor once sufficient support exists.
This permits commitment before the wave ends and significantly accelerates it without losing liveness.

Pallas and Aegis~\cite{decouchant_2026_pallas_aegis} address
trusted-state rollback after crashes or cloning, which can complement
\algoname's application-state recovery.

\section{Conclusion}
\algoname{} is an asynchronous \ac{DAG}-based \ac{BFT} SMR protocol that combines low-latency commitment with efficient client interaction and garbage collection.
It requires only $n\geq2f+1$ replicas and is probabilistically live under asynchrony.
To speed up commitments, \algoname{} uses a single broadcast step per round that leverages trusted non-equivocation and can commit early via enclave-protected checks.
Furthermore, \algoname{} integrates batched execution proofs to let clients verify execution results through a single reply, including after retrying at another replica.
\algoname{} limits protocol state through the use of checkpoint-driven garbage collection and bounded receiving windows.
Certified checkpoints let lagging replicas recover even when they miss the DAG history.

In an emulated 13-region WAN, \algoname{} commits up to $1.52\times$ faster than Mysticeti and keeps its lead with a third of the replicas crashed;
in simulation, it decides blocks faster than seven DAG protocols in all 100 sampled deployments.
Clients retain 92--97\% of the maximum possible throughput, whereas a classical client that sends to all replicas and waits for $f{+}1$ replies achieves only 7--12\%.
\algoname{}'s architecture shows that trusted components and coordinated design across ordering, client interaction, and recovery can make asynchronous DAG-based replication practical for \ac{BFT} \ac{SMR}.

\begin{acks}
This work is funded by Deutsche Forschungsgemeinschaft (DFG, German Research Foundation) – 446811880 (BFT2Chain). This work was supported by funding from the topic Engineering Secure Systems of the Helmholtz Association (HGF).
\end{acks}

\printbibliography
\appendix

\section{Algorithms}
\label{app:algorithms}
\begin{algorithm*}
\begin{algorithmic}[1]
\caption{\algoname{} Byzantine Atomic Broadcast: Main Loop for Process $p_i$}
\label{alg:porcupineab}
\ReplicaState{$P \colon$ set of processes}
\ReplicaState{$\var{DAG} \colon$ array of array of vertices, $\var{DAG}[0]$ initialized with genesis vertices}
\ReplicaState{$r \colon \mathbb{N}$, initialized with 0}
\ReplicaState{$\var{decidedRound} \colon \mathbb{N}_0$, initialized with 0} 
\ReplicaState{$\var{messageBuffer} \colon$ queue of messages, initialized empty}
\ReplicaState{$\var{vertexBuffer} \colon$ set of vertices per process of size $m_b$, initialized empty}
\ReplicaState{$\var{delivered} \colon$ array of round of last delivered vertex per process, initialized with 0}
\ReplicaState{$added \colon$ array of round of last vertex added per process, initialized with 0}
\While{true}
    \For{$v \in \var{vertexBuffer}$}
        \If{$v.r \leq r \land (\forall h \in v.\var{edges}\colon \var{DAG}[h.r][h.p] \neq \bot \lor \fn{irrelevant}{h.r, h.p})$}
            \State $\DAG{v.r}{v.p} \gets v; \var{vertexBuffer}.\fn{remove}{v}; $$\var{added[v.p] \gets v.r}$
            \If{$\var{decidedRound} < v.r \land u \coloneq \fn{check\_commit}{v} \neq \bot $} $\fn{waveReady}{u}$
            \EndIf
        \EndIf
    \EndFor
    \If{$ 4 \cdot (\var{pendingCkpts}.\fn{earliestUnstable}{}.w + g) < r$} \textbf{continue} 
    \Comment{stable checkpoint should have been built by now}\EndIf
    \If{$|\var{DAG}[r]| < f+1 \lor \var{DAG}[r][p_i] = \bot$} \textbf{continue} \EndIf 
    \State $r \gets r +1$
    \State \textbf{wait until} $\neg \var{messageBuffer}.\fn{isEmpty}$  \Comment{allow inserting empty vertices for progress when there is no continuous load}
    \State $v \gets$ new vertex$(\var{block} \coloneq \var{messageBuffer}.\fn{pop}{}, r \coloneq r, p \coloneq p_i, w \coloneq \lfloor \frac{r}{4} \rfloor, \var{edges} \coloneq \var{added})$
    \State $v.\sigma \gets \fn{enclave\_sign}{v}$
    \For{\textbf{all} $p_j \in P$} \textbf{send} $\langle VERTEX, v\rangle$ to $p_j$\EndFor
\EndWhile
\Function{a\_broadcast}{$m$}
    \State $\var{messageBuffer}.\fn{insert}{m}$
\EndFunction
\Function{waveReady}{$u$}
    \State $u_r \gets \DAG{u.r}{u.p}$ 
    \State $\var{committedAnchors} \gets$ new stack; $\var{committedAnchors}.\fn{push}{u_r}$
    \For{$r' \gets u.r-1$ down to $\var{decidedRound} + 1$}
        \If{$r' \text{ mod }4 =3$} \textbf{continue}\EndIf
        \State $u_{r'} \gets \DAG{r'}{\fn{CPRNG}{\lfloor\frac{r'}{4}\rfloor}}$
        \Comment{Coin is already revealed}
        \If{$u_{r'} \not = \bot \land \fn{path}{u_r, u_{r'}}$}
            \State $\var{committedAnchors}.\fn{push}{u_{r'}}; u_r \gets u_{r'}$
        \EndIf
    \EndFor
    \While{$\neg \var{committedAnchors}.\fn{isEmpty}{}$}
    \State $u_{r'} \gets \var{committedAnchors}.\fn{pop}{}$
    \State $\var{verticesToDeliver} \gets \{v \mid v \in \var{DAG} \colon \fn{path}{u_{r'}, v} \land \var{delivered}[v.p] < v.r \land \neg \fn{ignored}{v.r}\}$
    \For{$v \in \var{verticesToDeliver}$ in deterministic order}
    \State \textbf{a\_deliver} $(v.p, v.r, v.\var{block}); \var{delivered}[v.p] \gets v.r$
    \EndFor
    \State $\var{decidedRound} \gets r'$
    \If{$r'\mod4\neq0 $} \textbf{continue}\EndIf 
    \Comment{build new checkpoint over commit of the first anchor in a wave}
    \State $w' \gets \frac{r'}{4};\fn{respondCkptRequests}{w'}$
    \If{$\var{pendingCkpts}.\fn{front}{}.w + I_{\var{ckpt}} \leq w'$}
    \Comment{two consecutive checkpoints are at least $I_{ckpt}$ waves apart}
    \State $\var{ckpt_{new}} \gets \text{new checkpoint}(w', \var{state}, \var{responseStorage}, \var{delivered}); \var{pendingCkpts}.\fn{enqueue}{\var{ckpt_{new}}}$
    \State $\fn{certify}{\var{ckpt_{new}}}$
    \EndIf
    \EndWhile
\EndFunction
\end{algorithmic}
\end{algorithm*}

\begin{algorithm*}
\begin{algorithmic}[1]
    \caption{\algoname{} SMR: Checkpoint Certification Logic for Process $p_i$}
    \label{alg:porcupinesmr}
    \ReplicaState{$\var{ckpt_{stable}} \colon$ last certified checkpoint, initialized by genesis}
    \ReplicaState{$\var{pendingCkpts} \colon$ queue of uncertified checkpoints, initialized by genesis}
    \ReplicaState{$\var{state} \colon$ SMR state}
    \ReplicaState{$\var{responseStorage} \colon$ buffer of last response to client requests per client}
    \ReplicaState{$\var{ckptRequests}\colon$ buffer of last $\lceil\frac{g}{I_{\text{ckpt}}}\rceil$ received checkpoint requests per replica}
    \ReplicaState{$I_{ckpt}\colon \mathbb{N}$, checkpoint interval}
    \ReplicaState{$g \colon \mathbb{N}_0$, garbage collection offset}
    \Function{certify}{$\var{\var{ckpt}}$}
  \For{\textbf{all} $p_j \in P$}
    \textbf{send} $\langle CKPT\_SIGN\_REQ, \fn{h}{\var{ckpt}}, \var{ckpt}.w \rangle$ to $p_j$
  \EndFor
\EndFunction
\Procedure{ckpt\_sign\_req}{$p_k, h, w$}
    \If{$\fn{h}{\var{pendingCkpts}.\fn{get}{w}} = h$}
     \State \textbf{send} $\langle CKPT\_SIGN\_RESP, h, \fn{sign}{h}, w \rangle$ to $p_k$
    \ElsIf{$w \leq \var{ckpt_{stable}}.w$}
         \textbf{send} $\langle STATE\_TRANSFER, \var{ckpt_{stable}}\rangle$ to $p_k$
         \Comment{$\var{ckpt}$ is in the past and cannot be verified}
    \Else $\text{ } \var{ckptRequests}.\fn{add}{h, w, p_k}$
    \EndIf
\EndProcedure
\Procedure{ckpt\_sign\_resp}{$p_k,h, \sigma_k, w$}
    \State $\var{ckpt} \gets \var{pendingCkpts}.\fn{get}{w}$
    \If{$w \leq \var{ckpt_{stable}}.w \lor \neg\fn{verify}{\fn{h}{ckpt}, \sigma_k, p_k}$}
    \textbf{abort}
    \EndIf
    \State $\var{ckpt}.\var{signatureShares}.\fn{push}{\sigma_k}$
    \If{$|\var{ckpt}.\var{signatureShares}|\geq f+1$} 
    \State $\var{ckpt}.\var{cert} \gets \fn{global\_sign}{\fn{h}{\var{ckpt}}, \var{ckpt}.\var{signatureShares}}$
    \State $\fn{pruneUpTo}{ckpt}; \var{ckpt_{stable}} \gets \var{ckpt}$
    \EndIf
\EndProcedure
\Function{respondCkptRequests}{$w$}
\For{$(h, w', p_k) \in \var{ckptRequests}$}
    \If{$w = w'$} 
    \State $h \gets h($new checkpoint$(w', \var{state}, \var{responseStorage}, \var{delivered}))$
    \State \textbf{send} $\langle CKPT\_SIGN\_RESP, h, \fn{sign}{h}, w \rangle$ to $p_k$
    \State $\var{ckptRequests}.\fn{remove}{h, w', p_k}$\EndIf
\EndFor
\EndFunction
\Procedure{state\_transfer}{$p_k, \var{ckpt}$}
        \If{$\var{ckpt}.w \leq \var{ckpt_{stable}}.w \lor \neg \fn{global\_verify} {\var{ckpt}.\var{cert},\var{ckpt}}$} \textbf{abort} \EndIf
        \State $\fn{pruneUpTo}{ckpt}; \var{ckpt_{stable}} \gets \var{ckpt}$
        \If{$\var{decidedRound} < 4\cdot\var{ckpt}.w$}
            \State $added \gets \fn{max}{\var{added}, \var{ckpt}.\var{delivered}};\var{delivered} \gets  \var{ckpt}.\var{delivered}$
            \State $\var{state} \gets \var{ckpt}.\var{state}; r\gets \max (4\cdot \var{ckpt}.w, r)$
        \EndIf
\EndProcedure
\Function{pruneUpTo}{ckpt}
\State \textbf{clean} $\var{DAG}, \var{vertexBuffer}, \var{pendingCkpts}, \var{sup}$ \textbf{up to} $\var{ckpt}.w-g$
\State \textbf{clean} $\var{ckptRequests}$ \textbf{up to} $\var{ckpt}.w$
\EndFunction
\end{algorithmic}
\end{algorithm*}
\begin{algorithm*}
\caption{\algoname{} Reliable Broadcast: Broadcast Logic for Process $p_i$}
\label{alg:porcupinerb}
\begin{algorithmic}[1]
\Procedure{vertex}{$p_k, v$}
    \If{$ \DAG{v.r}{v.p} \neq \bot \lor \neg \fn{verify}{v.\sigma, v, v.p}$} \textbf{abort}\EndIf
    \For{$p_j \in P$} $\fn{requestHistoryFrom}{v, p_j}$\EndFor
    \If{$\neg(\fn{irrelevant}{v.r, v.p} \lor \fn{outOfWindow}{v.r})$}
    \State $\var{vertexBuffer}[v.p][v.r] \gets v$
    \ElsIf{$\neg \fn{irrelevant}{v.r, v.p}$}
    \State \textbf{send}  $\langle BACKFILL\_REQ, v.r, v.p\rangle$ to $p_k$
    \Comment{$v$ fits into the receiving window when it is delivered by the RB again}
    \EndIf
\EndProcedure
\Procedure{backfill\_req}{$p_k, r', p_v$}
    \If{$\fn{deleted}{r'}$} \State \textbf{send} $\langle STATE\_TRANSFER, \var\var{ckpt_{stable}}\rangle$ to $p_k$ 
    \Comment{the requested content can only be provided by a state transfer}
    \Else 
        \State $v \gets \DAG{r'}{p_v}$
        \If{$v \neq \bot$} \textbf{send} $\langle VERTEX, v\rangle$ to $p_k$
        \EndIf
    \EndIf
\EndProcedure
\Function{requestHistoryFrom}{$v, p_j$}
    \For{\textbf{all} $h \in v.\var{edges}$}
    \If{$\neg \fn{irrelevant}{h.r, h.p} \land \DAG{h.r}{h.p} = \bot \land h \notin \var{vertexBuffer}$}
    \State \textbf{send} $\langle BACKFILL\_REQ, h.r, h.p\rangle$ to $p_j$
    \EndIf
    \EndFor
\EndFunction
\Function{outOfWindow}{$r'$}
    \State \textbf{return} $r+ 4\cdot g < r'$
\EndFunction
\Function{irrelevant}{$r', p_k$}
    \State \textbf{return} $\fn{deleted}{r'} \lor r' \leq \var{ckpt_{stable}}.\var{delivered}[p_k]$
\EndFunction
\Function{ignored}{$r'$}
    \State \textbf{return} $r' \leq decidedRound - 4 \cdot g$
\EndFunction
\Function{deleted}{$r'$}
    \State  \textbf{return} $r' \leq 4 \cdot(\var{ckpt_{stable}}.w - g)$
\EndFunction
\end{algorithmic}
\end{algorithm*}
\begin{algorithm*}
\begin{algorithmic}[1]
    \caption{\algoname{} TEE: TEE Secret and Verification for Process $p_i$}
    \label{alg:porcupinetee}
    \ReplicaState{$sk_i \colon$ private key of the TEE}
    \ReplicaState{$r_{\text{max}} \colon \mathbb{N}_0$ maximum round of a vertex the TEE signed, initialized with $0$}
    \ReplicaState{$\var{CPRNG} \colon$ random number generator}
    \ReplicaState{$\var{sup} \colon$ set of supporters per anchor}
    \Function{enclave\_sign}{$v$}
        \If{$v.r \leq r_{\text{max}}$} \textbf{abort} \EndIf
        \If{$|\{(p, r')\in v.edges|r' = v.r-1\}|<f+1 \lor \neg \forall h\in v.\var{edges}\colon \fn{verify}{h.\sigma, h, h.p}$} \textbf{abort} \EndIf
        \If{$\exists h \in v.edges\colon h.p = p_i \land h.r = r_{max}$} 
        \State $r_{max}\gets v.r$
        \If{$r_\text{max} \mod 4 = 0$} \textbf{reveal} $\var{sup}[ : r_\text{max}], \fn{CPRNG}{\frac{r_\text{max}}{4}-1}$\Comment{reveals commit objects from previous waves}\EndIf
        \State \textbf{return} $\fn{sign}{v, sk_i}$\EndIf
    \EndFunction
    \Function{check\_commit}{$v$}
        \Comment{is calculated out of enclave if commit objects operated on are already revealed}
        \If{$\neg \fn{verify}{v.\sigma,v, v.p}$} \textbf{return} $\bot$ 
        \EndIf
        \State $p_u \gets \fn{CPRNG}{v.w}$
        \For{$i=2 ...0$}
            \State $\var{anchorRound} \gets 4 \cdot v.w + i ; u \gets (p_u, \var{anchorRound})$
            \If{$(\{u\} \cup \var{sup}[\var{anchorRound}]) \cap v.edges = \emptyset$} \textbf{continue}
            \EndIf
            \State $\var{sup}[\var{anchorRound}] \gets \var{sup}[anchorRound] \cup {v}$
            \If{$|\{s\in \var{sup}[\var{anchorRound}] | s.r=v.r\}|\geq f+1$} 
            \State \textbf{reveal} $\var{sup}[: 4\cdot (v.w+1)], \fn{CPRNG}{v.w};$ \textbf{return} $u$ 
            \Comment{reveals commit objects up to current committable wave}\EndIf
        \EndFor
    \EndFunction
\end{algorithmic}
\end{algorithm*}

In this Appendix we present the logic of \algoname{} split into four snippets of pseudo-code. The first snippet (Alg.~\ref{alg:porcupineab}) contains the logic of the atomic broadcast which is: receiving requests from the client handler (ll. 21--22), putting them into new vertices (ll. 17--20), receiving vertices from the RB-layer and adding them to the DAG (ll. 10--16), committing anchors once the commit rules are fulfilled (ll. 23--36) and invoking checkpoint creation (ll. 37--41). The second snippet (Alg.~\ref{alg:porcupinesmr}) covers the checkpoint related tasks of the SMR execution manager which are: certifying checkpoints (ll. 8--27), performing state transfers at replicas that have fallen behind (ll. 28--33), and carrying out garbage collection whenever a new stable checkpoint is available (ll. 34--36).
Next, the RB-layer of \algoname{} makes use of a TEE-based reliable broadcast that is contrasted to the broadcast pattern of Narwhal in \Cref{fig:broadcast}. The functionality of the RB-layer is depicted in the third snippet (Alg. \ref{alg:porcupinerb}). The reliable broadcast receives vertices, checks them for correctness (ll. 1--2), and carries out backfilling (ll. 3--17) with regards to the current progress of the garbage collection (ll. 18--25).
The last code snippet describes the logic of the TEE (Alg.~\ref{alg:porcupinetee}) which is: signing valid vertices (ll. 5--11), checking for commits without revealing the anchor (ll. 12--18) and revealing the anchor once it is harmless, such that commits can be checked outside the TEE (ll. 10 and 19--20).

Throughout this appendix, we assume that the number of nodes $n$ is exactly $n=2f+1$.

\section{Proofs}
In the following we prove the correctness of \algoname{}'s commit rules and garbage collection respectively. We further provide a theoretical analysis of the impact of garbage collection on the performance of \algoname{} and construct an attack against Narwhal's existing approach of garbage collection.
\subsection{\algoname{}'s BAB}
We first show that \algoname{} without garbage collection solves the Byzantine Atomic Broadcast problem (BAB problem) by deriving its correctness from DAG-Rider~\cite{keidar_all_2021} and TEE-Rider~\cite{leinweber_2026_tee_ledgers}. 
TEE-Rider adapts DAGRider to $n \geq 2f{+}1$ by replacing its reliable broadcast with a TEE-based one and shrinking its quorums to $\lceil\frac{n+1}{2}\rceil$, while leaving the commit rules unchanged.
A Byzantine Atomic Broadcast fulfills the following properties~\cite{keidar_all_2021}:
\begin{itemize}[label={}, leftmargin=\parindent]
    \item \textbf{Validity}: If a correct replica broadcasts $v$, then every correct replica eventually calls $\fn{a\_deliver}{v.p, v.r, v.block}$, where $v.p$ denotes the producer, $v.r$ the round, and $v.block$ the content of vertex $v$.
    \item \textbf{Agreement}: If a correct replica calls a\_deliver($v.p$, $v.r$, $v.block$), then every other correct replica calls a\_deliver($v.p$, $v.r$, $v.block$) eventually.
    \item \textbf{Integrity}: For each vertex $v$ uniquely identified by $(v.p, v.r)$, a correct replica calls $\fn{a\_deliver}{v.p, v.r, \cdot}$ at most once.
    \item \textbf{Total Order}: If a correct process calls a\_deliver($v.p$, $v.r$, $v.block$) before $\fn{a\_deliver}{v'.p, v'.r, v'.block}$, then no correct process calls $\fn{a\_deliver}{v'.p, v'.r, v'.block}$ without first calling $\fn{a\_deliver}{v.p, v.r, v.block}$.
\end{itemize}
The changes \algoname{} introduces to TEE-Rider are (1) \textit{three} potential anchor slots per wave, (2) checking commit rules in \textit{all later} rounds within the same wave, and (3) counting support along \textit{all} edges.
Because \textit{integrity} follows directly from TEE-Rider, we prove \textit{validity}, \textit{agreement}, and \textit{total order}. DAG-based protocols fulfill these guaranties by relying on two properties: (a) anchor connectivity (two consecutive anchors are always connected with a path), and (b) eventual commits (commits happen eventually).
We now prove that \algoname{} satisfies those two properties.
\begin{lemma}
    [Anchor Connectivity] \label{lem:anchorconnectivity} If a process directly commits an anchor $u$, then there is a path from $u$ to every anchor $u'$ that might be committed in a future round.
\end{lemma}
\begin{proof}
    Let $u$ be an anchor that is directly committed by a correct process $p_c$.
    Let $u'$ be an anchor of a future round.\\
    \textbf{Case 1:} $u.w = u'.w$. Then $u$ and $u'$ are connected with a path, because $u$ and $u'$ were proposed by the same replica and the TEE guarantees that a replica always references its own vertices (\Cref{alg:porcupinetee}, l. 9).\\
    \textbf{Case 2:} $u.w < u'.w$. Because $u$ is directly committed, there is a round $(u.w, i)$ in which at least $f+1$ vertices have a path to $u$. Therefore, because replicas reference their own vertices, starting in round $(u.w, 4)$, there are $f+1$ replicas that have a path to $u$. A vertex from a future round must reference at least $f+1$ replicas. By quorum intersection, at least one of these replicas references $u$. Therefore, every vertex of future waves, including $u'$ has a path to $u$.
\end{proof}
Based on anchor connectivity, we can show that the \textit{total order} property holds. Moreover, if commits happen eventually, DAG-Rider's \cite{keidar_all_2021} proofs of \textit{validity} and \textit{agreement} in Propositions 3 and 4 apply directly. Before we show the former, we prove the latter. We therefore have to show that \algoname{} commits occur with finite expectation.
In TEE-Rider~\cite{leinweber_2026_tee_ledgers}, commits occur with probability $\frac{1}{2}$ because the common coin of a wave $w$ is completely hidden until the end of round $(w,4)$. However, in \algoname{} a malicious replica may try to gain knowledge about the anchor in rounds $(w,2)$ and $(w,3)$ of a wave $w$, respectively, by querying the TEE to check whether the commit rule is already satisfied. Therefore, we have to show that the common coin is sufficiently `hidden' from a Byzantine attacker. The following Lemma proves that the TEE can hide the common coin sufficiently, such that a commit happens at least with the same probability as in TEE-Rider. The proof follows a game-based security approach in which an attacker tries to prevent a commit by providing the TEE a vertex and learning some information about the anchor of this wave. The attacker faces a fundamental dilemma: if it successfully learns that a replica $p$ is the anchor, it inherently loses the ability to halt the commit. Alternatively, if it learns that a replica $p$ is not the anchor, the actions taken to gain this information carry a high risk of inadvertently committing the correct anchor, ensuring a successful commit with probability larger than a half.
\begin{lemma}
    [Hidden Common Coin] \label{lem:hiddencoin}
    The expected number of waves until a correct process observes the direct commit rule to be fulfilled is below 2.
\end{lemma}
\begin{proof}
    Let $\mathcal{A}$ be an attacker that tries to prevent the direct commit of an anchor $u$.
    Assume that $\mathcal{A}$ lets the TEE check if the direct commit rule in round $(w,i), i\in\{2,3\}$ is fulfilled, by providing a set of validated vertices $S_i$ from round $(w,i)$. More explicitly, $S_i, i\in\{2,3\}$ is the set of vertices the attacker shows to the TEE using check\_commit (Alg.~\ref{alg:porcupinetee}, l.~12) and which could possibly support an anchor in an earlier round.

    \textbf{Case 1:} The TEE responds with ``$u$ is the anchor''. Then $S_i$ contains at least $f+1$ vertices that have a path to $u$. By quorum intersection, each vertex in round $(w,i+1)\leq (w,4)$ will have a path to $u$ and hence $u$ will be committed in round $(w,4)$ at the latest. Therefore, $\mathcal{A}$ does not succeed in preventing the commit.

    \textbf{Case 2:} The TEE responds with ``none''. Then $\mathcal{A}$ learns that a set of vertices $\bar{U}, u \notin \bar{U}$ from round $(w, 1)$, which would have sufficient support in $S_i$ to fulfill the direct commit rule, are not the anchor. We call this operation requesting $\bar{U}$.

    We now show that each time $t\in\mathbb{N}$ the attacker $\mathcal{A}$ requests the TEE using check\_commit with a set $S_t=S_2^t\cup S_3^t$ of vertices that provide sufficient support to directly commit a set of vertices $\bar{U}_t$, the probability that $\mathcal{A}$ loses by accidentally hitting case 1 instead of case 2, is sufficiently high to ensure a commit with probability $>\frac{1}{2}$. The probability that $\mathcal{A}$ loses at the $t^\text{th}$ request can be modeled by a binomially distributed random variable $X_t \sim B(\frac{|\bar{U}_t - \bigcup_{t'<t}\bar{U}_{t'}|}{2f+1-|\bigcup_{t'<t}\bar{U}_{t'}|})$ with a hit probability, which is the number of requested vertices $\bar{U}$ not covered in previous requests divided by the number of vertices that could possibly still be the anchor. Combining the requests, the probability for $\mathcal{A}$ to lose the game is: $\sum_{t}\mathbb{P}[X_t=1]= \sum_t\mathbb{P}[X_t=1|X_1=...=X_{t-1}=0]\prod_{t'=1}^{t-1}\mathbb{P}[X_{t'}=0] = \text{expression }\star \text{below}.$ Wlog. $\mathcal{A}$ provides support $S_t$ such that at each TEE request a single vertex is requested, hence $|\bar{U}_t - \bigcup_{t'<t}\bar{U}_{t'}|=1$ since adding multiple vertices to the set of committable vertices at the same time would be worse than requesting them successively.
    Then if $\mathcal{A}$ requests $k\leq 2f+1$ vertices, it loses with probability {\small\begin{align*}
        \star &= \sum_{t=1}^k\left(\frac{1}{2f+1-(t-1)}\prod_{t'=1}^{t-1}(1-\frac{1}{2f+1-(t'-1)})\right)\\&=\sum_{t=1}^k\left(\frac{1}{2f+1-(t-1)}\prod_{t'=1}^{t-1}\frac{2f+1-t'}{2f+1-(t'-1)}\right)\\&=\sum_{t=1}^k\left(\frac{1}{2f+1-(t-1)}\frac{\prod_{t'=1}^{t-1}2f+1-t'}{\prod_{t'=0}^{t-2}2f+1-t'}\right)\\&=\sum_{t=1}^k\left(\frac{1}{2f+1-(t-1)}\frac{2f+1-(t-1)}{2f+1}\right)=\sum_{t=1}^k\frac{1}{2f+1}=\frac{k}{2f+1}
    \end{align*}} Therefore, every time the attacker $\mathcal{A}$ checks whether a vertex is the anchor by querying the TEE, $\mathcal{A}$'s probability to lose the game increases. However, by Lemma 4.3 in TEE-Rider~\cite{leinweber_2026_tee_ledgers} regarding the common core, at least $f+1$ vertices fulfill the direct commit rule with respect to vertices in round $(w,4)$. This fact is also true for \algoname{} using the same quorum intersection argument. It follows that if $\mathcal{A}$ did not lose during requests to the TEE, the probability of $u$ to be one of the $f+1$ committable vertices is at least $\max(\frac{f+1-k}{2f+1-k},0)$.
    Therefore, the probability that $\mathcal{A}$ will lose the game is: $\frac{k}{2f+1} + (1-\frac{k}{2f+1})\max(\frac{f+1-k}{2f+1-k},0) = \begin{cases} k < f+1 \colon \frac{k}{2f+1} +\frac{2f+1-k}{2f+1-k}\frac{f+1-k}{2f+1}=\frac{f+1}{2f+1} \\
    k\geq f+1\colon \frac{k}{2f+1}\end{cases} >\frac{1}{2}$

    As a result, in each wave a commit happens with probability $>\frac{1}{2}$. Thus, the expected number of waves until a correct process observes the direct commit rule to be fulfilled is below 2.
\end{proof}
Finally, we will show that \algoname{} satisfies the \textit{total order} property of the BAB problem.
\begin{theorem}
    [Total Order] \label{theorem:totalorder}
    If a correct process calls a\_deliver $(v.p, v.r, v.block)$ before $\fn{a\_deliver}{v'.p, v'.r, v'.block}$, then no correct process calls $\fn{a\_deliver}{v'.p, v'.r, v'.block}$ without first calling $\fn{a\_deliver}{v.p, v.r, v.block}$.
\end{theorem}
\begin{proof}
    By the agreement of the common coin, each wave has only one anchor replica, hence, three anchor vertices.
    By Lemma~\ref{lem:anchorconnectivity} if a correct process directly commits an anchor~$u$, then there is a path to every anchor $u'$ that might be committed in a future round. Therefore, by the retrospective commit rule, anchors are committed in the same order. As the anchors' causal histories are delivered in some deterministic order, which is identical for all correct processes, all correct processes that deliver vertices do so in the same order.
\end{proof}
\begin{figure*}[]
    \centering
    \resizebox{.85\linewidth}{!}{\definecolor{pcFill}{RGB}{168,224,196}  
\definecolor{pcEdge}{RGB}{34,140,100}
\definecolor{pcSlate}{RGB}{255,255,255} 
\definecolor{pcSlateE}{RGB}{140,148,160}
\definecolor{pcGold}{RGB}{240,176,32}   
\definecolor{pcAmber}{RGB}{198,158,70}  
\definecolor{pcStroke}{RGB}{150,157,168}\definecolor{pcHi}{RGB}{232,132,20}     
\definecolor{pcWeak}{RGB}{150,110,214}  
\begin{tikzpicture}[
    x=1cm, y=1cm,
    vertex/.style={circle, draw=pcEdge, thick, fill=pcFill, minimum size=0.4cm, inner sep=0pt},
    direct_commit/.style={circle, draw=pcGold, line width=1.5pt, fill=pcFill, minimum size=0.4cm, inner sep=0pt},
    empty_vertex/.style={circle, draw=gray!80, thick, fill=white, minimum size=0.4cm, inner sep=0pt},
    dashed_vertex/.style={circle, draw=gray, very thick, dashed, fill=none, minimum size=0.4cm, inner sep=0pt},
    edge/.style={->, >={Stealth[length=2.5mm, width=2.5mm]}, very thick, draw=pcGold},
    font=\sffamily
]

\begin{scope}[on background layer]

    \fill[green!30, rounded corners=3pt] (11.9, 0.6) rectangle (15.2, 7.5);
    \fill[brown!40, rounded corners=3pt]  (1.05, 3.6) rectangle (5.3, 7.5);
    \fill[brown!40, rounded corners=3pt]  (2.45, 2.55) rectangle (5.3, 3.7);
    \fill[brown!40, rounded corners=3pt]  (2.4, 0.6) rectangle (5.3, 1.45);
    \fill[brown!40, rounded corners=3pt]  (3.7, 1.35) rectangle (5.3, 2.7);
    \fill[brown!40, rounded corners=3pt]  (5.1, 1.55) rectangle (9.55, 3.7);
    \fill[brown!40, rounded corners=3pt]  (5.1, 0.6) rectangle (7.6, 1.7);
    \fill[brown!40, rounded corners=3pt] (9.4, 1.55) rectangle (10.55, 2.45);
    \fill[pink!90, rounded corners=3pt] (0.1, 0.6) rectangle (0.75, 7.5);
    \fill[pink!90, rounded corners=3pt] (0.4, 0.6) rectangle (2.2, 3.4);
    \fill[pink!90, rounded corners=3pt] (2.0, 1.6) rectangle (3.5, 2.4);
\end{scope}

\draw[thick, decorate, decoration={zigzag, segment length=4mm, amplitude=0.6mm}] (0.9, 7.8) -- (0.9, 0.3);
\draw[thick, dashed] (5.5, 7.8) -- (5.5, 0.3);
\draw[thick] (11.5, 7.6) -- (11.5, 0.3);
\draw[thick] (13.0, 4.0) -- (13.0, 0.3);
\draw[thick] (13.1, 4.0) -- (13.1, 0.3);
\draw[thick, decorate, decoration={zigzag, segment length=4mm, amplitude=0.6mm}] (15.5, 7.8) -- (15.5, 0.3);

\draw[thick] (0.5, 4.0) -- (16.0, 4.0);

\node[rotate=90] at (-0.7, 6.0) {$\var{vertexBuffer}$};
\node[rotate=90] at (-0.7, 2.0) {DAG};

\foreach \y/\l in {7/p_1, 6/p_2, 5/p_3, 3/p_1, 2/p_2, 1/p_3} {
    \node[] at (-0.2, \y) {$\l$};
}

\node[rotate=90] at (0.45, 5.5) {$\fn{deleted}{}$};
\node[text=orange!80!black, font=\sffamily\large\bfseries] at (3.0, 5.3) {decided};
\node[text=red!80!black, font=\sffamily\large\bfseries] at (1.9, 2.0) {irrelevant};
\node[rotate=90] at (15.8, 5.8) {$\fn{outOfWindow}{}$};
\node[text=green!50!black, font=\sffamily\Large\bfseries] at (13.7, 5.3) {future};
\node[rotate=90] at (5.2, 2.35) {$\fn{ignored}{}$};
\node[text=black!50, font=\sffamily\large\bfseries] at (7.1, 5.3) {undecided};

\node[vertex] (p11) at (0.45, 3) {};
\node[vertex] (p12) at (0.45, 2) {};
\node[vertex] (p13) at (0.45, 1) {};

\node[font=\Large, text=black!70] at (1.7, 3) {\dots};
\node[font=\Large, text=black!70] at (1.7, 1) {\dots};

\node[vertex] (c31) at (3.0, 3) {};
\node[direct_commit] (c32) at (3.0, 2) {};
\node[vertex] (c33) at (3.0, 1) {};

\node[vertex] (c41) at (4.2, 3) {};
\node[vertex] (c42) at (4.2, 2) {};
\node[vertex] (c43) at (4.2, 1) {};

\draw[edge] (c41) -- (c32);
\draw[edge] (c42) -- (c32);

\node[font=\Large, text=black!70] at (6.0, 3) {\dots};
\node[font=\Large, text=black!70] at (6.0, 2) {\dots};
\node[font=\Large, text=black!70] at (6.0, 1) {\dots};

\node[direct_commit] (u61) at (7.0, 3) {};
\node[vertex] (u62) at (7.0, 2) {};
\node[vertex] (u63) at (7.0, 1) {};

\node[vertex] (u71) at (8.2, 3) {};
\node[vertex] (u72) at (8.2, 2) {};
\node[dashed_vertex] (u73) at (8.2, 1) {};

\draw[edge] (u71) -- (u61);
\draw[edge] (u72) -- (u61);

\node[font=\Large, text=black!70] at (9.1, 3) {\dots};
\node[font=\Large, text=black!70] at (9.1, 2) {\dots};

\node[empty_vertex] (d91) at (10.0, 3) {};
\node[direct_commit] (d92) at (10.0, 2) {};

\node[empty_vertex] (r101) at (11.5, 3) {};
\node[empty_vertex] (r102) at (11.5, 2) {};

\draw[edge] (r101) -- (d92);
\draw[edge] (r102) -- (d92);

\node[empty_vertex] at (9.0, 5) {};
\node[empty_vertex] at (10.0, 5) {};
\node[empty_vertex] at (12.4, 6) {};
\node[empty_vertex] at (12.4, 5) {};

\draw[thick] (0.9, 0.5) -- (0.9, 0.2) node[below, align=center, font=\normalsize] {point\\of deletion};
\draw[thick] (3.0, 0.5) -- (3.0, 0.2) node[below, align=center, font=\normalsize] {$\var{ckpt_{\text{stable}}}$};
\draw[thick] (7.0, 0.5) -- (7.0, 0.2) node[below, align=center, font=\normalsize] {$\fn{earliestUnstable}{}$};
\draw[thick] (10.0, 7.55) -- (10.0, 7.85) node[above, align=center, font=\normalsize] {$\var{decidedRound}$\\($\fn{latestUnstable}{}$)};
\draw[thick] (5.5, 7.55) -- (5.5, 7.85) node[above, align=center, font=\normalsize] {GC barrier};
\draw[thick] (11.5, 7.55) -- (11.5, 7.85) node[above, align=center, font=\normalsize] {$r$};
\draw[thick] (13.05, 0.5) -- (13.05, 0.2) node[below, align=center, font=\normalsize] {wait for checkpoint\\stabilization};
\draw[thick] (15.5, 7.55) -- (15.5, 7.85) node[above, align=center, font=\normalsize] {end of receiving\\window};

\draw[<->, thick] (1.0, 0.3) -- (2.9, 0.3) node[midway, below, font=\normalsize] {$4g$};
\draw[<->, thick] (3.1, 0.3) -- (6.9, 0.3) node[midway, below, font=\normalsize] {$4G(\frac{1}{2})$};
\draw[<->, thick] (7.1, 0.3) -- (12.95, 0.3) node[midway, below, font=\normalsize] {$4g$};

\draw[<->, thick] (5.6, 7.75) -- (9.9, 7.75) node[midway, above, font=\normalsize] {$4g$};

\draw[<->, thick] (11.6, 7.75) -- (15.4, 7.75) node[midway, above, font=\normalsize] {$4g$};

\begin{scope}[shift={(1.0, 6.8)}]
    \filldraw[fill=white, draw=black!40, thick] (-0.4,-0.4) rectangle (10.2,0.4);
    \node[vertex, label=right:{\sffamily committed}] (leg1) at (0,0) {};
    \node[direct_commit, label=right:{\sffamily direct commit}] (leg2) at (2.4,0) {};
    \node[empty_vertex, label=right:{\sffamily not yet ordered}] (leg3) at (5.4,0) {};
    \node[dashed_vertex, label=right:{\sffamily missing}] (leg3) at (8.5,0) {};
\end{scope}

\end{tikzpicture}}
    \caption{Example of the different stages of a vertex. For simplicity in this example $I_\text{ckpt}$ is set to 1, so a checkpoint is created for every committed anchor in the first round of a wave. Once a valid vertex $v$ ($v.r > r$) is received by a replica, it is put into the future of the $\var{vertexBuffer}$. As soon as the round of the vertex is reached ($v.r = r$) and all its causal history is present (or irrelevant) in the DAG, $v$ is added to the DAG as undecided. When $v$ is in the causal history of a commit, $v$ gets decided, which happens in $X \sim 4G(\frac{1}{2})$ rounds ($G$ denotes the geometric distribution). This means that $v$ is either delivered if the previous committed anchor is less than $4g$ rounds ahead ($\neg \fn{ignored}{v.r} \iff \var{decidedRound}-4g< v.r$), or ignored. When the checkpoint produced at the commit that decided $v$ is stabilized, which must happen at most $4g$ rounds after the round of the checkpoint ($4\cdot \fn{earliestUnstable}{}.w +4g\leq r$), $v$ gets irrelevant. Finally, once a checkpoint with a round more than $4g$ rounds ahead of $v.r$ gets stabilized, the garbage collection takes effect and $v$ is deleted.}
    \label{fig:gcstages}
\end{figure*}
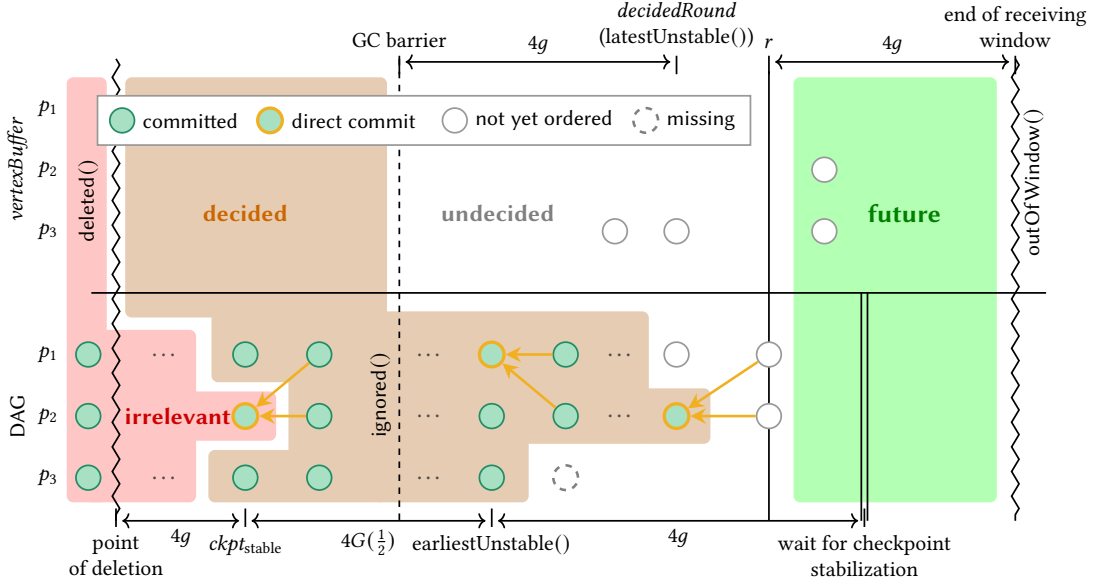
\subsection{\algoname{}'s Garbage Collection}
We now show the correctness of \algoname{}'s garbage collection (GC).
The changes \algoname{}'s GC introduces to BAB (\Cref{alg:porcupineab}) and RB (\Cref{alg:porcupinerb}) are (1) the use of a \textit{bounded receiving window}, (2) \textit{state transfers} as responses to backfilling requests for vertices that are already garbage collected, (3) ignoring vertices prior to the GC barrier during commits, and (4) actively waiting for checkpoint stabilization by checking commit rules whenever a new vertex is added to the DAG, not only when advancing to the next round.

First, we show the correctness of \algoname{}'s RB-layer (Alg.~\ref{alg:porcupinerb}) considering changes (1) and (2). In DAG-Rider the RB-layer is responsible for disseminating and agreeing on vertices that are included in the DAG and thereby delivered by the AB-layer. \algoname{}'s AB-layer does not deliver every vertex added to the DAG, only the ones that can be ordered before the GC barrier reaches them. Therefore, vertices that never reach the point of delivery do not necessarily have to reach the point of RB-Agreement. Additionally, a state transfer (2) might cause the execution of transactions from vertices that did not yet get RB-delivered, which renders receiving those vertices unnecessary. As a result, reaching RB-Agreement across all replicas might become irrelevant during runtime for the following vertices:
\begin{definition}
    [Vertex Irrelevance] A vertex $v$ and its content are considered irrelevant by a replica $p$ if one of the following conditions is met:
    \begin{enumerate}
        \item $p$ has a stabilized checkpoint covering the execution of $v$'s transactions;
        \item $p$ has a stabilized checkpoint proving that $v$'s transactions will never be executed (unless resubmitted);
    \end{enumerate}
    which is equivalent to the predicate $\fn{irrelevant}{v.r, v.p} \coloneq$\\$v.r \leq \var{ckpt_\text{stable}}.\var{delivered}[v.p] \lor v.r \leq 4\cdot (\var{ckpt_\text{stable}.w} -g)$.
\end{definition}
With this definition, we show that the TEE-based RB with changes (1) and (2) fulfills a sufficient variant of RB-Agreement -- the RB will deliver every vertex that is not already irrelevant -- which we can then use to prove the correctness of \algoname{}'s BAB-layer (Alg.~\ref{alg:porcupineab}).
An illustration of the different stages that a vertex reaches before it gets irrelevant and eventually garbage collected, is given in Figure~\ref{fig:gcstages}.

First, we demonstrate that the backfilling mechanism remains functional because replicas only garbage collect irrelevant vertices. Such vertices are inherently covered by a stable checkpoint that can itself directly serve the backfilling request.
\begin{lemma}
    [Backfilling Response]
    \label{lem:backfillingresponse}
    If a correct replica sends a vertex via message $\langle VERTEX, v\rangle$, it will always be able to respond to backfilling requests regarding the vertex and its history.
\end{lemma}
\begin{proof}
    Let $p_c$ be a correct replica that sends a vertex $v$ with $\langle VERTEX, v \rangle$. Then $v$ was either proposed by $p_c$ (Alg.~\ref{alg:porcupineab}, l.~20) or sent due to a backfilling request (Alg.~\ref{alg:porcupinerb}, l.~13). In both cases $p_c$ will directly add or has already added $v$ to its local DAG. Every vertex, including $v$, is only removed (garbage collected) from the local DAG when a stable checkpoint with $v.r \leq 4 \cdot(\var{ckpt}.w-g)$ has been obtained
    (Alg.~\ref{alg:porcupinesmr}, l.~35). Therefore, the stable checkpoint either covers the execution of $v$'s transactions ($v.r \leq \var{ckpt_\text{stable}}.\var{delivered}[v.p])$ or proves that $v$ is ignored (Alg.~\ref{alg:porcupineab}, l.~33) and its transactions are never executed ($(v.r \leq 4\cdot(\var{ckpt_\text{stable}}.w-g))$, which is all the information to render $v$ irrelevant.

    As a result, if $p_c$ needs to respond to a backfilling request for $v$, $p_c$ either still stores the vertex or can respond with a checkpoint that renders $v$ irrelevant.
    Furthermore, because $p_c$ added $v$, every vertex in $v$'s causal history is either present in the local DAG of $p_c$ or irrelevant due to a stable checkpoint (Alg.~\ref{alg:porcupineab}, l.~11).  Therefore, $p_c$ can also respond to backfilling requests regarding $v$'s history.
\end{proof}
Next, we show that although a replica might need to discard vertices that fall too far ahead of its local progress, it will eventually recover them. Before discarding such a vertex, the replica requests its causal history---including the vertex itself---via backfilling. Eventually, the replica receives sufficient history to advance its local progress, allowing the vertex to fall within its valid receiving window.
\begin{lemma}
    [Receiving Window Sufficiency]     \label{lem:slidingwindowsufficiency} If a correct replica receives a vertex from another correct replica, it will eventually add the vertex to its local DAG or reach a point in which the vertex is considered irrelevant.
\end{lemma}
\begin{proof}
    Let $p$ be a correct replica that receives a vertex $v$ from another correct replica $p_c$. Then, $p$ requests $v$'s missing causal history $H$ that is not already     irrelevant from $p_c$ (Alg.~\ref{alg:porcupinerb}, ll.~3, 14--17). If $v$ does not fit into the vertex buffer and is not already irrelevant, $p$ will also request $v$ from $p_c$ (ll.~6, 7) again. We now prove by induction that every vertex in $H \cup\{v\}$ will eventually be added to $p$'s DAG or be considered irrelevant.

    \textbf{Base Case:} Let $h\in H\cup\{v\}$ be a vertex with the smallest round $p$ requests from $p_c$. After Lemma \ref{lem:backfillingresponse} $p_c$ will respond to the backfilling request, which is why $p$ will either receive $h$ or receive a checkpoint that renders $h$ irrelevant. Once $p$ receives $h$, $h$ can directly be added to the local DAG because $h$'s history is either present or irrelevant as $h$ is the earliest missing vertex (Alg.~\ref{alg:porcupineab}, l.~11).

    \textbf{Induction Assumption:} Assume $p$ eventually adds every vertex $h\in H\cup\{v\}$ up to a fixed but arbitrary round $r < v.r$ to the local DAG or renders it irrelevant.

    \textbf{Induction Step:} Let $h \in H\cup\{v\}$ be a missing vertex with $h.r = r + 1$ that $p$ requested from $p_c$. After Lemma \ref{lem:backfillingresponse} $p_c$ will respond to the backfill request, which is why $p$ will receive $h$ or receive a checkpoint that renders $h$ irrelevant. Once $p$ receives $h$, there are three cases.\\
    \textbf{Case 1:} $h$ fits into the receiving window of $p$ and is not considered irrelevant (\ref{alg:porcupinerb}, ll. 4, 5). Then $h$ will eventually be added to the local DAG or rendered irrelevant, as by induction assumption, eventually $h$'s history will arrive or will be considered irrelevant.\\
    \textbf{Case 2:} $h$ does not fit into the receiving window of $p$ and is not considered irrelevant (\ref{alg:porcupinerb}, ll. 6, 7). Then $p$ requests $h$ from $p_c$ again until by induction assumption eventually its history arrives or $h$ or it's history are considered irrelevant.\\
    \textbf{Case 3:} $h$ is already considered irrelevant.

    By induction, $p$ will eventually add $v$ to its DAG or consider $v$ irrelevant.
\end{proof}
Combining the two Lemmata we can provide a weaker form of RB-Agreement. 
\begin{theorem}
    [Partial RB-Agreement]
    \label{the:rbagreement}
    If a correct replica adds a vertex to its local DAG, every other correct replica will either do so as well or reach a point where the vertex is considered irrelevant.
\end{theorem}
\begin{proof}
    Let $p_c$ be a correct replica that adds a vertex $v$ to its local DAG. Then $p_c$ will eventually broadcast a vertex $v'$ that references $v$.
    Let $p$ be a different correct replica. Because of perfect links, $p$ will eventually receive $v'$.
    Following Lemma \ref{lem:slidingwindowsufficiency}, $p$ will eventually add $v'$ including its history $v$ to its local DAG or consider it irrelevant.
    Because $p$ is arbitrary, every correct replica will eventually add $v$ to their local DAG or consider $v$ irrelevant.
\end{proof}
After proving that the RB-layer still fulfills a weaker variant of RB-Agreement, we now prove that the RB-Agreement is sufficient for the correctness of \algoname{}'s BAB. Furthermore, we explain why change (3), which includes ignoring vertices before the GC offset at delivery, does not affect AB-Agreement. This is because if a single correct replica decides to deliver and not ignore a vertex and thereby execute its transactions, every other correct replica will do so as well.

\begin{theorem}[AB-Agreement]
    \label{the:abagreement}
    If a correct replica calls \allowbreak$\fn{a\_deliver}{v.p, v.r, v.block}$ for a vertex $v$, every other correct replica will either do so as well or reach a state in which $v$'s transactions are already applied to the state.
\end{theorem}
\begin{proof}
    Let $p_c$ be a correct replica that a\_delivers a vertex $v$. Then $p_c$ directly committed an anchor $u$ that references $v$. Afterwards, $p_c$ will broadcast a vertex $v'$ that references every vertex that $u$ requires to fulfill the direct commit rule.
    Following partial RB-Agreement (Theorem \ref{the:rbagreement}), every correct replica will eventually add $v'$ to its local DAG or consider $v'$ irrelevant. Let $p$ be another correct replica that did not already a\_deliver $v$.\\
    \textbf{Case 1:} $p$ adds $v'$ to its local DAG.
    Then when receiving $v'$, $p$'s DAG contains the information to directly commit $u$. No commit in a round further than $u.r$ could have been triggered because otherwise, following the order properties ensured by the direct and retrospective commit rule, $v$ would have already been delivered. Therefore, $u.r > \var{decidedRound}$ which is why $u$ will be committed. Additionally, because the commit rules ensure that correct replicas commit anchors in the same order, when $p$ commits $u$, $p$'s value of $\var{decidedRound}$ equals $p_c$'s value of $\var{decidedRound}$ when $p_c$ committed $u$. Thereby, committing $u$ will a\_deliver $v$ because $v$ is undelivered and will not be ignored (Alg.~\ref{alg:porcupineab}, l.~33; Alg.~\ref{alg:porcupinerb}, ll.~22, 23). \\
    \textbf{Case 2:} $p$ considers $v'$ irrelevant due to a state transfer. Then, the state process $p$ was transferred to was valid, which means that $f+1$ replicas claimed to have encountered the same checkpoint. At least $f+1-f=1$ of these replicas is correct and has genuinely encountered the checkpoint. This correct replica $p_\sigma$ has committed an anchor $u'$ with $u.w \leq u'.w$ the checkpoint belongs to.

    \textbf{Case 2.1:} $p_\sigma$ added $v'$ to its local DAG. Then, following case 1 $p_\sigma$ did a\_deliver $v$ which means that $v$'s transactions are already applied in the state $p$ is transferred to.

    \textbf{Case 2.2:} $p_\sigma$ considers $v'$ irrelevant due to a state transfer. Then this state transfer regarded a checkpoint of an anchor $u''$ with $v'.w \leq u''.w < u'.w$. Inductively following case 2.2 down to $v'.w$, there must be a correct signer that matches case 2.1 and therefore $p$ was forwarded to a state in which $v$'s transactions are already applied to the state.\\
    Because $p$ is arbitrary, every correct replica will eventually a\_deliver $v$ or reach a state in which $v$'s transactions are already applied to the state.
\end{proof}
Ultimately, \algoname{}'s garbage collection does not compromise the safety of the BAB. This is because bounding the rounds ahead of local progress retains sufficient state for the RB and commit rule evaluation, while bounding the rounds behind local progress affects all replicas consistently.

We now provide an intuition for the liveness of \algoname{}'s AB-layer, by explaining why waiting for checkpoint stabilization (4) does not result in a deadlock. Notice that the previous scenarios sometimes required replicas to move on to the next round, which can only happen if waiting for checkpoint stabilization does not cause any deadlocks.
\begin{theorem}
    [Checkpoint Stabilization]
    \label{the:checkpointstabilization}
    A correct replica only waits for the stabilization of a checkpoint for a finite period of time.
\end{theorem}
\begin{proof}
     Let $p_c$ be a correct replica that waits for the checkpoint signature for a checkpoint $\var{ckpt}$ with wave $\var{ckpt}.w$ of a correct replica $p$ (Alg.~\ref{alg:porcupineab}, l.~14).
    Then $p_c$ has broadcast a vertex $v$ that references every vertex necessary to observe the commit of $\var{ckpt}.w$.\\
    \textbf{Case 1:} $p$ has progressed to a round $r$ with $v.r\leq r$. Then, even if $p$ does not continue to the next round because $p$ itself is waiting for checkpoint stabilization, $p$ will eventually add $v$ to its local DAG or consider $v$ irrelevant, following RB-Agreement (Theorem \ref{the:rbagreement}).\\
    \textbf{Case 2:} $p$ has only progressed to a round $r < v.r$. Then $p$ will eventually progress to a round $v.r \leq r'$ because by inductively following case 2 until the least progressed replica where only case 1 is relevant $p_c$ will not wait infinitely for checkpoint stabilization. Therefore, $p$ will eventually add $v$ to its local DAG following RB-Agreement (Theorem \ref{the:rbagreement}) or consider $v$ irrelevant.

    In both cases, $p$ eventually adds $v$ to its local DAG or considers $v$ irrelevant.\\
    \textbf{Case 1:} $p$ considers $v$ irrelevant. This means that $p$ either encountered $\var{ckpt}$ during execution and can answer the checkpoint request (Alg.~\ref{alg:porcupinesmr}, ll.~11, 12, 22--27), or has a stable checkpoint that renders $v$ and thereby $ckpt$ irrelevant and can transfer $p_c$ to a state where stabilizing $\var{ckpt}$ is no longer necessary (Alg. \ref{alg:porcupinesmr}, l. 13).\\
    \textbf{Case 2:} $p$ adds $v$ to its local DAG. Then, $p$ will commit $ckpt$ (if it has not already) and answer the checkpoint request because $v$ references all the vertices necessary to observe the commit (Alg. \ref{alg:porcupineab}, l. 38).

    Because $p$ is arbitrary, every correct replica will eventually respond to $p_c$'s checkpoint signature request. Because there are at least $2f+1-f=f+1$ correct replicas, $p_c$ can eventually stabilize $ckpt$ (Alg.~\ref{alg:porcupinesmr}, ll.~19--21, 36) and continue to the next round.
\end{proof}

\subsection{Theoretical Performance Analysis of \algoname{}'s Garbage Collection}
In the following, we show that under common WAN latencies, the proposed GC of \algoname{} does not affect the performance of the protocol. Using the single communication step of a TEE-based RB, the difference in progress (in rounds) between replicas remains below the offset of the GC and the size of the receiving window $g$, when $g$ is appropriately dimensioned.
\begin{definition}
    [Proportional Network Latencies] Let $l_{ij}$ with $i, j \in \{1, ..., n\}$ be the latency of  a message sent from replica $p_i$ to replica $p_j$. Then, $L = \{l_{ij} \in \mathbb{R}_{>0}| i,j \in \{1,...,n\}, i\neq j\}$ describes the set of latencies between replicas participating in the protocol. Let \newline $l_{\text{fastest}} = \min_{S\subset \{1, ..., n\}, |S| = f+1} \{\max\{{l_{ij} | i,j\in S, i\neq j}\}\}$ be the maximum latency between replicas in the fastest quorum and $RTT_{\text{slowest}} = \max \{l_{ij}+l_{ji}|i,j\in \{1,...,n\}, p_i, p_j \text{ are correct}\}$ the maximum round trip time (RTT) between correct replicas.
    Proportional network latencies ensure:
    \begin{enumerate}
        \item $\lfloor\frac{RTT_{\text{slowest}}}{l_{\text{fastest}}}\rfloor\leq C \in \mathbb{N}$
        \item $C$ is known to all replicas.
    \end{enumerate}
\end{definition}

\begin{proposition}
    [Progress Difference]
    \label{lem:progressdifference} Assume proportional network latencies. At all times the progress difference between replicas is below $2C$ rounds if $4g \geq 2C + 1$.
\end{proposition}
\begin{proof}
    Assume that until now the progress difference between replicas was below $2C$.
    Let $v$ be a vertex proposed by one of the at least $f+1-f=1$ correct replicas in the fastest quorum, denoted by $p_i$, at time $t_0$.
    Let $p_j$ be another correct replica. Then $p_j$ receives $v$ at $t_0 + l_{ij}$. If $p_j$ cannot directly add $v$ to its local DAG due to missing parts of $v$'s causal history, $p_j$ requests the missing history from $p_i$. Then $p_j$ will receive the missing vertices until $t_0 + l_{ij} + l_{ji} + l_{ij} < t_0 + 2RTT_{\text{slowest}}$. After receiving $v$'s causal history, which fits into the receiving window because the progress difference between replicas is below $2C<4g$, process $p_j$ will at least reach round $v.r$, which was the round of the fastest quorum when $v$ was proposed.
    Because broadcasting a vertex in the fastest quorum takes $l_{\text{fastest}}$, in that time the fastest quorum can make $\lfloor\frac{2RTT_{\text{slowest}}}{l_{\text{fastest}}}\rfloor\leq 2C$ rounds of progress.
    Therefore, $p_j$ lags at most $2C < 4g$ rounds behind the fastest quorum at all times.
\end{proof}
As a result, receiving vertices outside of the receiving window (1) as well as state transfers (2) do not take place in times of proportional network latencies. Additionally, the probability of vertices proposed by correct replicas to be ignored before the point of delivery is negligible ("AB-Validity"),
which guarantees timely fairness (3) and actively waiting for checkpoint stabilization is not required during periods of proportional network latencies (4). The last two properties will be shown in the following. Finally, combining Propositions \ref{lem:progressdifference}, \ref{the:timelyfairness}, and \ref{obs:checkpointstabilization}, GC does not affect the performance of the protocol under proportional network latencies, which we refer to as the common case.
\begin{proposition}
    [Timely Fairness]
    \label{the:timelyfairness} Assume proportional network latencies. The vertices proposed by correct replicas are never garbage collected before the point of delivery ("AB-Validity") if $4g \geq 4C + 2$.
\end{proposition}
\begin{proof}
    Let $p_i$ be a correct replica that proposes a vertex $v$ at time $t_0$. Let $p_j$ be another correct replica. Then $p_j$ adds $v$ to its local DAG at $t_0 + l_{ij} + l_{ji} + l_{ij}<t_0 + 2RTT_{\text{slowest}}$. Following Lemma \ref{lem:progressdifference}, $p_i$ lags at most $2C$ rounds behind $p_j$ at $t_0$. Because in a period of $2RTT_{\text{slowest}}$ process $p_j$ can only progress $\lfloor\frac{2RTT_{\text{slowest}}}{l_{\text{fastest}}}\rfloor\leq 2C$ further rounds, $p_j$ adds $v$ to its local DAG in round $v.r + 4C$. Therefore, $p_j$ references $v$ in round $v.r + 4C + 1$.
    Because $p_j$ was arbitrary, it follows that all $f+1$ correct replicas reference $v$ by round $v.r + 4C + 1$. By quorum intersection, this means that every vertex in round $v.r + 4C + 2$ references $v$.
    Suppose $v$ is garbage collected before the point of delivery, then there is a stable checkpoint $\var{ckpt_{stable}}$ with $v.r \leq 4\cdot(\var{ckpt_{stable}}.w - g)$. Therefore, $4\var{ckpt_{stable}}.w \geq v.r+4g$. Because we assume $4C + 2 \leq 4g$, it holds that $4\var{ckpt_{stable}.w}\geq v.r+4C+2$, which means that the anchor of $ckpt_{stable}$ references $v$, and hence, $v$ got delivered.
\end{proof}

\begin{proposition}
    [Expected Checkpoint Stabilization] \label{obs:checkpointstabilization} Assume proportional network latencies. The probability of waiting for checkpoint stabilization is negligible in $\lambda$ if $4g\geq 5C+1+4\lambda$.
\end{proposition}
\begin{proof}
    Let $p_i$ be a correct replica that creates a checkpoint $\var{ckpt}$. Then $\var{ckpt}$ was created due to a direct commit in wave $w' \geq \var{ckpt}.w$. Because a direct commit happens with probability of at least $\frac{1}{2}$, $\var{ckpt}$ is created before wave $\var{ckpt}.w + X$, where $X$ denotes a geometrically distributed random variable with mean at least 2. After creating $\var{ckpt}$, $p_i$ will broadcast a vertex $v$ in round $4\cdot (\var{ckpt}.w + X) + 1$, which references every vertex necessary to observe the commit of $\var{ckpt}$. In round $4\cdot (\var{ckpt}.w + X) + 1 + 4C$ every correct replica will have added $v$ to its local DAG (see Theorem \ref{the:timelyfairness}), which causes the commit of $\var{ckpt}$ and a response to the checkpoint signature request. Replica $p_i$ will receive the response of a correct replica $p_j$ after $l_{ji}\leq RTT_{\var{slowest}}$, in which $p_i$ has not progressed further than round $4\cdot (\var{ckpt}.w + X) + 1 + 4C + C$.
    Thus, the probability of waiting for checkpoint stabilization is $\mathbb{P}[4\cdot(\var{ckpt}.w + g) <4 \var{ckpt}.w + 4X + 1 + 5C] = \mathbb{P}[4g < 5C + 1 +4X]$.
    Therefore, the probability of process $p_i$ waiting for checkpoint stabilization scales with $\mathbb{P}[ X > \lambda] \leq \frac{1}{2^\lambda}$ which is negligible in the tolerance $\lambda$ that $g$ offers.
\end{proof}
As a result, by choosing $ g \geq \frac{5C+1}{4} +\lambda$, besides using bandwidth for checkpoint stabilization, for high enough $\lambda$ the performance of \algoname{} is unaffected by GC in the common case. During periods of network partitions, \algoname{} maintains liveness but requires state transfers to recover replicas that have fallen behind. For example, in the AWS deployment with 13 replicas (see Table~\ref{tab:latencymatrix}) we have $RTT_{\text{slowest}}\approx 199.2~\text{ms}$ and $l_{\text{fastest}} \approx 57~\text{ms}$ which results in $C = 3$, thus $g = 4 + \lambda$. 
For $\lambda \coloneq 20$ the probability of waiting for checkpoint stabilization is below $0.0001\%$. Thus, setting $g \coloneq 24$ could be sufficient.
\begin{lemma}
    [Memory Bounds] \label{lem:memorybounds} \algoname{}'s BAB and RB consume $O((n|v| + \frac{|S|}{I_{ckpt}})(g+I_{ckpt}))$ memory space with overwhelming probability, where $|v|$ denotes the size of a vertex and $|S|$ the size of a checkpoint.
\end{lemma}
\begin{proof}
The \textit{vertexBuffer} can potentially contain all vertices from the latest round that got deleted until $g$ waves in the future. Thus, the range of the vertex buffer is \(\textcolor{ColorBlue}{(}4 \cdot(\var{ckpt_{stable}}.w - g)\textcolor{ColorBlue}{,} r+4g\textcolor{ColorBlue}{]}\). With overwhelming probability, this range is included in \(\textcolor{ColorBlue}{(}r-4g-4I_{ckpt}-E-4g\textcolor{ColorBlue}{,} r+4g\textcolor{ColorBlue}{]} = \textcolor{ColorBlue}{(}r-8g-4I_{ckpt}-E\textcolor{ColorBlue}{,} r+4g\textcolor{ColorBlue}{]}\) for some constant  $E$. In conclusion, the memory consumption of the \textit{vertexBuffer} is in $O(n|v|(g + I_{ckpt}))$. A similar argumentation holds for the memory consumption of the DAG.
    In addition, \textit{pendingCkpts} stores the last stable checkpoint and all following unstable checkpoints, thus, its range is given by $\textcolor{ColorBlue}{[}4\var{ckpt_{stable}}.w\textcolor{ColorBlue}{,} r\textcolor{ColorBlue}{]}$. Again, with overwhelming probability, the range is included in $\textcolor{ColorBlue}{[}r-4g-4I_{ckpt}-E\textcolor{ColorBlue}{,} r\textcolor{ColorBlue}{]}$ for some constant $E$. Due to the checkpoint interval, \textit{pendingCkpts} contains at most $\lfloor\frac{|\textit{pendingCkpts}|}{4I_{ckpt}}\rfloor=\frac{g}{I_{ckpt}}+1$ checkpoints, yielding a memory consumption of $|S|(\frac{g}{I_{ckpt}}+1)$.\\
    In summary, the total memory consumption of \algoname{}'s BAB and RB is in $O((n|v| + \frac{|S|}{I_{ckpt}})(g+I_{ckpt}))$.
\end{proof}
\subsection{Vertex Buffer Overflow Attack}
\label{sec:vertexbufferoverflowattack}
In the following, we construct adversarial network latencies that cause a correct replica operating a DAG-based consensus instance on a Narwhal based Mempool (e.g. \cite{danezis_narwhal_2022}, \cite{spiegelman_bullshark_2022}) to receive messages ahead of its local progress. These messages will flood and exceed the replica's memory bounds.\\
Let $p_c$ be a correct replica that is cut off from all other replicas but a single faulty replica $p_f$ due to adversarial network conditions. From now on $p_f$ will refuse to answer $p_c$'s backfilling requests, but will continue to send all of its  blocks to $p_c$. Replica $p_c$ cannot add these blocks to its DAG as they are ahead of $p_c$'s local progress. Because garbage collection does only target the DAG structure, $p_f$'s blocks will, after a finite amount of time, exceed $p_c$'s memory capacity.\\
\algoname{} addresses this attack by garbage collecting the $\var{vertexBuffer}$, which results in memory bounds that are finite in expectation at all times.
\algoname{} naturally enforces garbage collection for the $\var{vertexBuffer}$ by using a bounded receiving window. By relying on backfilling and ensuring that there is a correct replica that can answer the backfilling request, blocks outside the receiving window can be safely discarded.

\section{Evaluation Appendix}

\subsection{Deployment}
\label{sec:evalapp:deployment}

\begin{table*}[t]
  \centering
  \scriptsize
  \setlength{\tabcolsep}{3.2pt}
  \caption{One-way network delays [ms] configured with \texttt{netem} between the 13 emulated AWS regions; the diagonal applies between replicas placed in the same region.
  Regions are named by billing code~\cite{awsregions}, region identifier and location:
  USE1 (\texttt{us-east-1}, N.\,Virginia),
  USW2 (\texttt{us-west-2}, Oregon),
  CAN1 (\texttt{ca-central-1}, Canada),
  EUC1 (\texttt{eu-central-1}, Frankfurt),
  EU (\texttt{eu-west-1}, Ireland),
  EUW2 (\texttt{eu-west-2}, London),
  EUW3 (\texttt{eu-west-3}, Paris),
  EUN1 (\texttt{eu-north-1}, Stockholm),
  APS1 (\texttt{ap-southeast-1}, Singapore),
  APS2 (\texttt{ap-southeast-2}, Sydney),
  APS3 (\texttt{ap-south-1}, Mumbai),
  APN1 (\texttt{ap-northeast-1}, Tokyo),
  APN2 (\texttt{ap-northeast-2}, Seoul).}
  \label{tab:latencymatrix}
  \begin{tabular}{lrrrrrrrrrrrrr}
    \toprule
     & USE1 & USW2 & CAN1 & EUC1 & EU & EUW2 & EUW3 & EUN1 & APS1 & APS2 & APS3 & APN1 & APN2 \\
    \midrule
    USE1 & 4.4 &  &  &  &  &  &  &  &  &  &  &  &  \\
    USW2 & 32.8 & 2.1 &  &  &  &  &  &  &  &  &  &  &  \\
    CAN1 & 9.4 & 31.0 & 2.3 &  &  &  &  &  &  &  &  &  &  \\
    EUC1 & 47.3 & 71.9 & 47.3 & 2.3 &  &  &  &  &  &  &  &  &  \\
    EU & 35.6 & 59.7 & 35.3 & 13.9 & 2.2 &  &  &  &  &  &  &  &  \\
    EUW2 & 39.6 & 64.9 & 40.2 & 8.8 & 7.2 & 2.1 &  &  &  &  &  &  &  \\
    EUW3 & 42.7 & 68.2 & 43.7 & 6.4 & 10.6 & 5.9 & 2.2 &  &  &  &  &  &  \\
    EUN1 & 57.0 & 77.5 & 52.9 & 11.9 & 20.2 & 15.6 & 16.3 & 1.7 &  &  &  &  &  \\
    APS1 & 110.2 & 81.7 & 110.2 & 81.0 & 89.1 & 84.9 & 83.4 & 90.6 & 2.2 &  &  &  &  \\
    APS2 & 100.6 & 70.8 & 99.6 & 126.3 & 128.2 & 133.4 & 140.4 & 136.1 & 48.0 & 2.0 &  &  &  \\
    APS3 & 96.6 & 112.4 & 96.2 & 64.3 & 63.4 & 59.0 & 54.9 & 73.9 & 32.6 & 77.8 & 2.0 &  &  \\
    APN1 & 75.6 & 50.0 & 78.4 & 114.4 & 102.4 & 107.7 & 110.9 & 123.7 & 35.9 & 53.5 & 67.7 & 2.3 &  \\
    APN2 & 91.4 & 62.2 & 89.9 & 115.9 & 118.6 & 123.8 & 127.2 & 125.4 & 37.3 & 73.2 & 67.6 & 18.4 & 2.4 \\
    \bottomrule
  \end{tabular}
\end{table*}

\Cref{tab:latencymatrix} gives the delays we configure.
They are derived from the \(p_{50}\) round-trip times reported by Cloudping~\cite{cloudping}, aggregated over a period of a year ending 2025-12-23:
for each pair we average the two measured directions and halve the result to obtain the one-way delay that \texttt{netem} applies.

\subsection{Metric Aggregation}
\label{sec:evalapp:aggregation}
Each replica records a latency histogram over the measurement window of a run.
For a given percentile we compute that percentile per replica, average it across replicas to obtain one value per run, and average across runs.
Throughput is the number of requests confirmed to clients per second, averaged over the measurement window of a run and then across runs.
Confidence intervals are computed over the per-run values.

\begin{figure*}
    \centering
    \begingroup
\pgfplotsset{compat=1.18}
\begin{tikzpicture}
\begin{axis}[
    scale only axis,
    width=0.76\linewidth, height=0.76\linewidth,
    enlargelimits=false,
    xlabel={Destination Region}, ylabel={Source Region},
    xlabel near ticks, ylabel near ticks,
    xlabel shift=5pt, ylabel shift=5pt,
    label style={font=\fontsize{9pt}{10pt}\selectfont},
    colorbar,
    colorbar style={
        width=8pt,
        title={RTT [ms]},
        title style={font=\fontsize{8pt}{9pt}\selectfont},
        ytick={0,50,...,450},
        yticklabel style={font=\fontsize{7pt}{8pt}\selectfont},
    },
    colormap/hot, point meta min=0, point meta max=450,
    xtick={0,...,33}, ytick={0,...,33},
    tick align=outside, tick style={major tick length=2pt, thin},
    xticklabel style={
        font=\fontsize{7pt}{8pt}\selectfont,
        rotate=90, anchor=near xticklabel,
    },
    yticklabel style={font=\fontsize{7pt}{8pt}\selectfont},
    xticklabels from table={data/aws-regions.dat}{region},
    yticklabels from table={data/aws-regions.dat}{region},
]
\addplot[matrix plot, mesh/cols=34, point meta=explicit]
table[x index=0, y index=1, meta index=2, col sep=space]
{data/aws-latency-matrix.dat};
\addplot[
    scatter, only marks, mark=none, point meta=explicit,
    nodes near coords={\pgfmathfloattofixed{\pgfplotspointmeta}\ifdim\pgfmathresult pt<70pt\color{white}\else\color{black}\fi
        \pgfmathprintnumber[fixed,precision=0]{\pgfplotspointmeta}},
    every node near coord/.style={
        font=\fontsize{6.5pt}{7pt}\selectfont,
        inner sep=0pt, anchor=center,
    },
]
table[x index=0, y index=1, meta index=2, col sep=space]
{data/aws-latency-matrix.dat};
\end{axis}
\end{tikzpicture}
\endgroup
    \caption{AWS round-trip-time input matrix used to sample the 100 deployments for network simulations.
    The network simulator uses half of each directed RTT as its one-way delay.}
    \label{fig:heatmap}
\end{figure*}
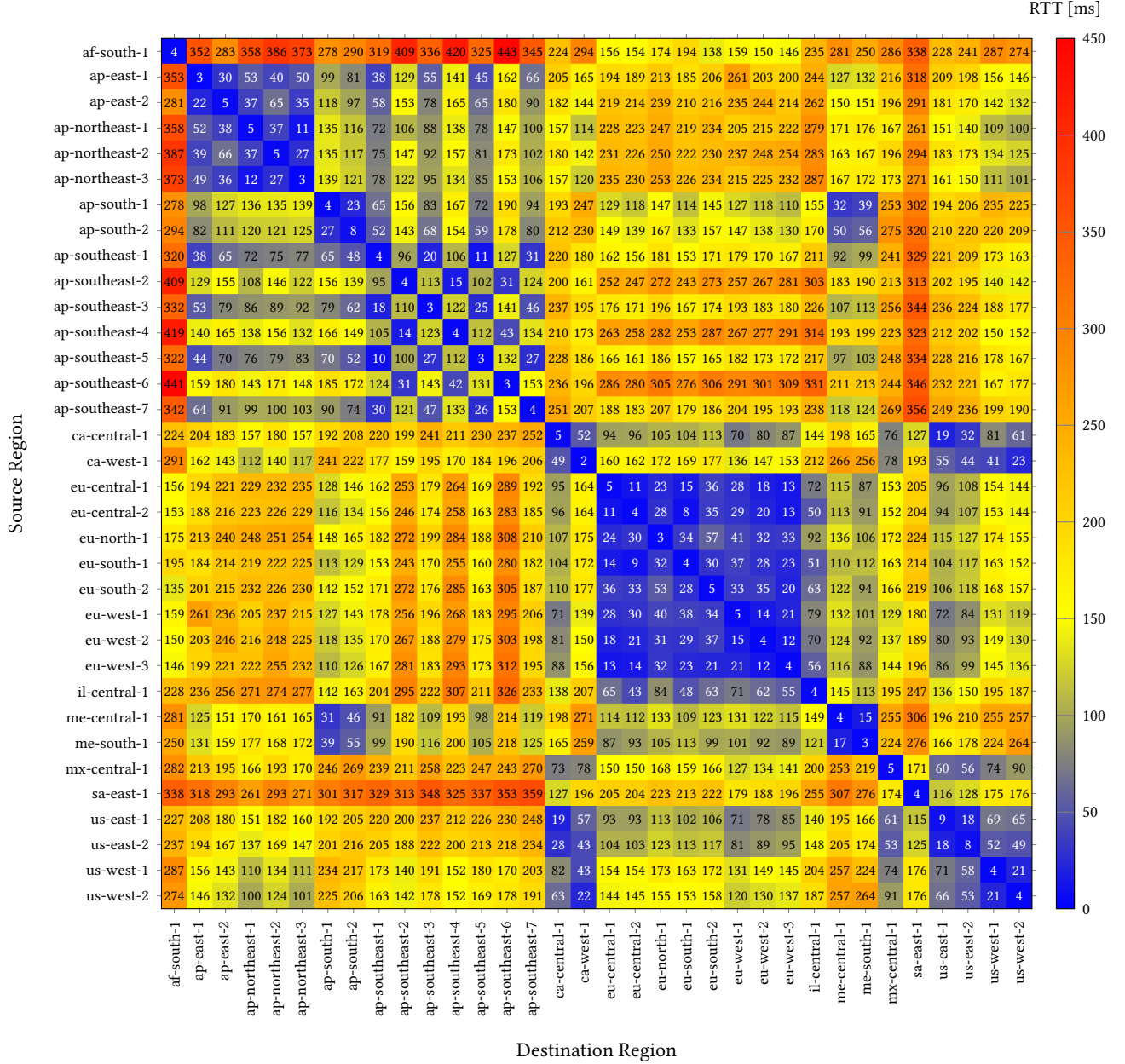

\subsection{Simulated Comparison across 100 Deployments}
\label{sec:evalapp:simulation}

\para{The simulator.} The evaluation in Sect.~\ref{sec:eval:generality} uses a discrete-event network simulator which we originally developed to compare the latency of different DAG protocols. The simulator allows users to specify region-to-region latencies and select host regions for deployed replicas and then models message delays. The broadcast protocols, commit rules and pacemaker design of existing DAG protocols can be easily re-implemented within this framework. Since cryptography is also abstracted away, differences in observed commit latency are attributable to the protocol design itself rather than separate codebases or tuning.

Replicas use reliable point-to-point links for communication whose delay is the one-way delay between their regions, which is taken as half the round-trip time in the matrix of 34 AWS
regions (see Figure~\ref{fig:heatmap}). So two replicas placed in the same region still observe that region's intra-region delay, usually between $1.1$ and $4.4$\,ms, and only a replica's message to itself is instantaneous.
Delays are also constant per region pair as the simulator models neither
jitter, nor loss, nor bandwidth and queuing.
The simulator treats \textit{timeouts} as events in the same event queue as
message deliveries, so a pacemaker is driven by the same clock as communication.
The simulated time only progresses on events. The simulator is deterministic, for instance, equal timestamps execute in insertion order, so an identical seed can reproduce a simulation run.
A DAG protocol is implemented exactly according to its specification using the same broadcast primitive and commitment rules. Every simulation run additionally checks that the committed
prefixes of all replicas agree before its measurements are recorded. As important limitations, note that all cryptographic work and execution consume no simulated time. We believe this abstraction favours no particular protocol as every system is credited with free cryptography, and the \ac{TEE}
operations \algoname{} relies on are free in the same sense as the signature verification its baselines rely on.

\para{Setup.} The comparison is paired: All eight protocols run on the same 100 deployments,
even though each uses a different fault bound. The sampler picks a
deployment's regions from the seed and the replica set size alone regardless of $f$.
Protocol parameters follow the respective papers. For instance, Shoal++ runs three DAGs staggered by $100$\,ms with a $600$\,ms round timeout started at the beginning of each round.

\para{Additional Results.} \Cref{tab:simfull} gives the per-configuration medians behind
Sect.~\ref{sec:eval:generality}. Every cell is the median over the same 100 deployments of
the replica-summary statistic defined in \cref{sec:evalapp:aggregation}.

\begin{figure*}[t]
  \centering
  \includegraphics[width=\textwidth]{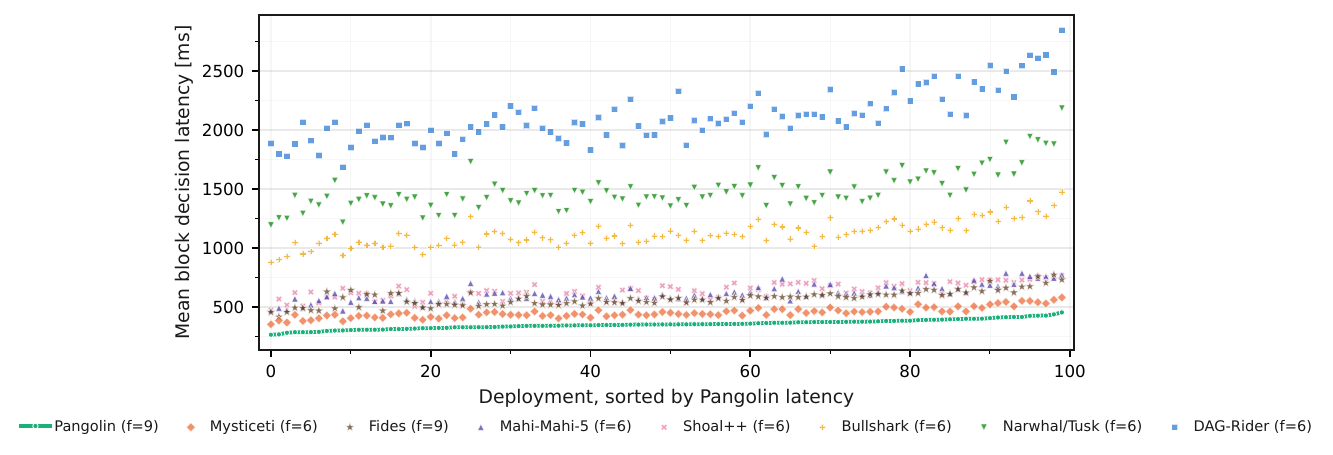}
  \caption{Per-deployment comparison, one column per deployment, columns ordered by
  \algoname{}'s own latency. \algoname{} is the connected line and is monotone by
  construction; every baseline mark lies above it in all 100 columns.}
  \label{fig:sim-paired}
\end{figure*}

\paragraph{Paired Comparison}
 \Cref{fig:sim-paired} orders the deployments by
\algoname{}'s own latency. \algoname{} wins against the baselines in all 100 deployments. The median paired difference is $+99$\,ms against Mysticeti, $+214$ ms against Fides, $+250$ ms against Mahi-Mahi-5, $+293$ ms against Shoal++, $+759$ ms against Bullshark, $+1{,}110$ ms against Narwhal/Tusk and $+1{,}731$ ms against DAG-Rider.

\begin{figure*}[t]
  \centering
  \begin{subfigure}[t]{0.49\textwidth}
    \centering
    \includegraphics[width=\textwidth]{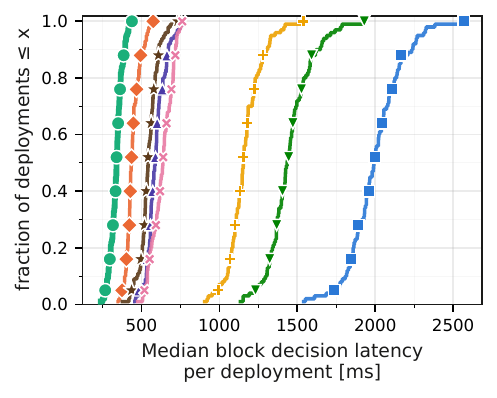}
    \caption{Median ($p_{50}$) latency.}
    \label{fig:sim-ecdf-median}
  \end{subfigure}\hfill
  \begin{subfigure}[t]{0.49\textwidth}
    \centering
    \includegraphics[width=\textwidth]{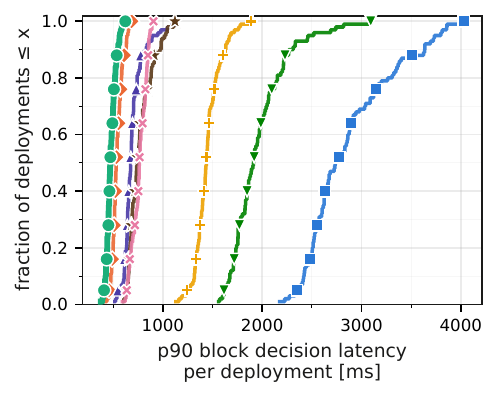}
    \caption{$p_{90}$ latency.}
    \label{fig:sim-ecdf-p90}
  \end{subfigure}
  \\[2pt]
  \includegraphics[width=\textwidth]{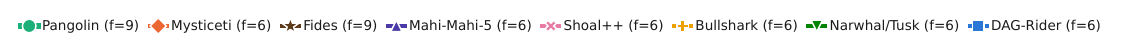}
  \caption{Cumulative distribution of block decision latency over the 100 deployments. A curve further to the left is faster on more
  deployments. In~(\subref{fig:sim-ecdf-median}) \algoname{} lies left of every baseline
  at every quantile. In (\subref{fig:sim-ecdf-p90}) it still leads, but Mysticeti has
  moved close enough to touch it.}
  \label{fig:sim-ecdf}
\end{figure*}

\paragraph{Cumulative latency distributions.} \Cref{fig:sim-ecdf} shows the same comparison
at two quantiles. At the median the eight protocols can be visually cleanly separated; only Mahi-Mahi-5 and Shoal++ touch at the top of their distributions. When looking at the tail latency ($p_{90}$) the
gap of \algoname{} to Mysticeti narrows to $1.12\times$, and at $p_{99}$ it closes to
$1.02\times$, winning on $66$ of $100$ deployments. The advantage \algoname{} holds in the median
is thus larger than the advantage it holds in the tail, which matches the behaviour \cref{sec:eval:latency} observes on the testbed for real execution.

\para{Threats to validity.} The simulator abstracts cryptography and models the network
as fixed per-pair delays, so these are block decision latencies for network-bound
configurations and are not comparable with the end-to-end request latencies of
\cref{sec:eval:client}. Also the experiment evaluates whole protocols rather than mechanism, so it
ranks protocol designs but does not attribute the ranking to individual mechanisms. All simulation  runs are fault-free.

\begin{table}[t]
  \centering
  \footnotesize
  \setlength{\tabcolsep}{3.5pt}
  \caption{Block decision latency [ms] and round rate over 100 deployments, $n=20$, fault-free.
  Each protocol runs at the largest $f$ its own resilience bound admits.}
  \label{tab:simfull}
  \begin{tabular}{lcrrrrrr}
    \toprule
    System & $f$ & quorum & mean & $p_{50}$ & $p_{90}$ & $p_{99}$ & rounds/s \\
    \midrule
    \algoname{}    & 9 & 11 & \textbf{352} & \textbf{340} & \textbf{471} & \textbf{597} & 13.1 \\
    Mysticeti      & 6 & 14 & 446 & 435 & 529 & 608 & 9.3 \\
    Fides          & 9 & 11 & 576 & 542 & 742 & 1{,}050 & 13.1 \\
    Mahi-Mahi-5    & 6 & 14 & 604 & 588 & 675 & 752 & 10.4 \\
    Shoal++        & 6 & 14 & 635 & 638 & 762 & 802 & 2.3 \\
    Bullshark      & 6 & 14 & 1{,}109 & 1{,}154 & 1{,}437 & 1{,}577 & 3.4 \\
    Narwhal/Tusk   & 6 & 14 & 1{,}444 & 1{,}435 & 1{,}912 & 2{,}643 & 3.5 \\
    DAG-Rider      & 6 & 14 & 2{,}070 & 1{,}993 & 2{,}743 & 4{,}121 & 3.5 \\
    \bottomrule
  \end{tabular}
\end{table}

\subsection{Comparison on Mysticeti's Deployment Scenario}
\label{app:sim-mysticeti-deployment}
To generate Figure~\ref{fig:motivation-latency}, we use the Method of Sect.~\ref{sec:evalapp:simulation} but with a single deployment scenario, which is the 13-region deployment, Mysticeti was originally evaluated in.
We evaluate block commit latency of selected protocols under identical network conditions.
To evaluate the average block decision latency,   we replicate the experimental setup of Mysticeti~\cite{babel_2025_mysticeti},  which covers 13 different AWS regions
across North America, Europe, and Asia-Pacific within a network simulation
to show how implemented variants of the protocols DAG-Rider~\cite{keidar_all_2021}, Narwhal~\cite{danezis_narwhal_2022}, Bullshark~\cite{spiegelman_bullshark_2022}, Shoal++~\cite{arun_2025_shoalpp}, Mahi-Mahi-5~\cite{jovanovic2025mahi}, Mysticeti~\cite{babel_2025_mysticeti}, Fides~\cite{fides_2026_arxiv} and \algoname{} would perform in an identical network environment.
Each bar is the mean over five runs that differ only in the seed of the common coin.
Fig.~\ref{fig:motivation-latency} shows \algoname{}
overtakes Mysticeti with an average block decision latency of 284~ms vs. 372~ms.

\section{Design Alternatives}

\subsection{\algoname{} without TEEs}
\label{app:sect:porcupine-without-tees}
\algoname{} can be realised  without \acp{TEE} in three steps: First, substituting \ac{TEE}-based primitives (reliable broadcast, common coin, threshold signature scheme) through standard variants. Second, using the ($n\geq 3f+1$)-resilience and standard Byzantine dissemination quorum sizes of $\lceil \frac{n+f+1}{2} \rceil$. Third, the commit rules (see Sec.~\ref{sect:design:rbc+consensus}) are not checked against the TEE, but are checked in the same way as in DAGRider~\cite{keidar_all_2021} by waiting for a wave to complete.

\clearpage

\end{document}